\documentclass{article}

\usepackage{arxiv}

\usepackage[utf8]{inputenc} 
\usepackage[T1]{fontenc}    
\usepackage{hyperref}       
\usepackage{url}            
\usepackage{booktabs}       
\usepackage{amsfonts}       
\usepackage{nicefrac}       
\usepackage{graphicx}
\usepackage{natbib}
\usepackage{doi}

\usepackage{graphicx}%
\usepackage{multirow}%
\usepackage{amsmath,amssymb,amsfonts}%
\usepackage{amsthm}%
\usepackage{mathrsfs}%
\usepackage[figuresright]{rotating}%
\usepackage[title]{appendix}%
\usepackage[dvipsnames]{xcolor}%
\usepackage{textcomp}%
\usepackage{manyfoot}%
\usepackage{booktabs}%
\usepackage{algorithm}%
\usepackage{algorithmicx}%
\usepackage{algpseudocode}%
\usepackage{program}%
\usepackage{listings}%

\newtheorem{definition}{Definition}%
\newtheorem{lemma}{Lemma}%

\title{A heuristic algorithm using tree decompositions for the maximum happy vertices problem\thanks{To appear in Journal of Heuristics}}

\date{}

\renewcommand{\headeright}{}
\renewcommand{\undertitle}{}
\renewcommand{\shorttitle}{A heuristic algorithm using tree decompositions for the maximum happy vertices problem}

\author{
    Louis Carpentier$^\text{1}$, Jorik Jooken$^\text{2}$ and Jan Goedgebeur$^\text{2,3}$ \\ \\
    $^\text{1}$Department of Computer Science, KU Leuven Brugge, Brugge, 8200, Belgium \\
	$^\text{2}$Department of Computer Science, KU Leuven KULAK, Kortrijk, 8500, Belgium \\
	$^\text{3}$Department of Applied Mathematics, Computer Science and Statistics, Ghent University, Ghent, 9000, Belgium. \\
	\{louis.carpentier, jorik.jooken, jan.goedgebeur\}@kuleuven.be
}

\begin{document}
\maketitle

\begin{abstract}
    We propose a new methodology to develop heuristic algorithms using tree decompositions. Traditionally, such algorithms construct an optimal solution of the given problem instance through a dynamic programming approach. We modify this procedure by introducing a parameter $W$ that dictates the number of dynamic programming states to consider. We drop the exactness guarantee in favour of a shorter running time. However, if $W$ is large enough such that all valid states are considered, our heuristic algorithm proves optimality of the constructed solution. In particular, we implement a heuristic algorithm for the Maximum Happy Vertices problem using this approach. Our algorithm more efficiently constructs optimal solutions compared to the exact algorithm for graphs of bounded treewidth. Furthermore, our algorithm constructs higher quality solutions than state-of-the-art heuristic algorithms Greedy-MHV and Growth-MHV for instances of which at least 40\% of the vertices are initially coloured, at the cost of a larger running time.
\end{abstract}


\section{Introduction}

Many combinatorial graph optimisation problems are $\mathcal{NP}$-hard. This implies that no polynomial time algorithm exists to solve these problems for every graph, unless $\mathcal{P} = \mathcal{NP}$ \citep{karp2010reducibility}. One way to deal with this, is to exploit special structures in the input graph. Researchers have developed algorithms that solve certain graph optimisation problems, if the graphs are restricted to a certain class, e.g., trees \citep{aravind2016linear,jooken2020multi}, claw-free graphs \citep{minty1980maximal}, planar graphs \citep{baker1994approximation} or bipartite graphs \citep{kHonig1931grafok}. While these algorithms have polynomial time complexity, they cannot handle graphs of other classes, thus limiting their applicability. 

To avoid this limitation, we can resort to algorithms using the \textit{tree decomposition} of a graph \citep{robertson1984graph}, which decomposes the graph into a tree-like structure (we will formally define this concept in Section \ref{sec:tree_decomposition}). The \textit{treewidth} is a parameter related to a tree decomposition and measures how tree-like a graph is. A tree decomposition allows for a dynamic programming approach, by computing tables of partial solutions at each node of the tree decomposition in a bottom-up fashion. This results in a fixed-parameter tractable algorithm. Such algorithms have a polynomial dependence on the size of the input graph, but potentially have an exponential dependence on some relevant, secondary measurement \citep{cygan2015parameterized}. For algorithms using tree decompositions, this secondary parameter is the treewidth. 

The current paper proposes to develop algorithms using the tree decomposition of a graph, which removes the exponential dependence on the treewidth. This dependency originates from the size of the dynamic programming tables at each node in the tree decomposition. Our solution proposes to include a parameter $W$, which we call the width, which dictates how many entries of the dynamic programming table should be computed. By setting $W$ to a small value and only partially computing the dynamic programming table, we remove the exponential dependency in the time complexity, at the cost of losing the exactness guarantee. This allows to dynamically trade off between the exactness of the solution and the algorithm runtime (i.e., if the entire dynamic programming table is computed the solution is guaranteed to be exact, but this comes at the cost of a longer runtime).

In particular, we illustrate this methodology on the \textit{Maximum Happy Vertices Problem} (MHV). This special graph colouring problem was recently proposed by \cite{zhang2015algorithmic}. Typically, graph colouring problems require adjacent vertices to have a different colour. In the MHV problem, however, the goal is to maximise the number of happy vertices, given a set of initially coloured vertices. A vertex is \textit{happy} if it has the same colour as all of its neighbours. Several algorithms for the MHV problem have been proposed, including an exact, polynomial time algorithm for graphs of bounded treewidth \citep{agrawal2017parameterized, agrawal2020parameterized}. The algorithm described in these papers yields exact solutions, but is impractical when the treewidth of the graph is too large. 

The main contributions of this paper are as follows:
\begin{enumerate}
    \item We propose a new methodology to use tree decompositions in algorithm design. Traditionally, tree decompositions have only been used in exact algorithms. We go beyond this approach by dropping the exactness constraint in favour of algorithm speed. Our methodology allows to dynamically convert the algorithm into either an exact or a heuristic algorithm, solely by setting the value of a single parameter. 
    
    \item We develop a heuristic algorithm for solving the MHV problem and tuned its hyperparameters with SMAC \citep{hutter2011sequential}, an automatic algorithm configurator. We further show that our algorithm constructs higher quality solutions than Greedy-MHV and Growth-MHV -- two state-of-the-art heuristic algorithms proposed by \cite{zhang2015algorithmic} -- if at least 40\% of the vertices are initially coloured, at the cost of a larger running time. 

    \item We implemented the exact algorithm of \cite{agrawal2017parameterized} and \cite{agrawal2020parameterized} for solving the MHV problem on graphs of bounded treewidth. We further show that our algorithm more efficiently computes optimal colourings than this exact algorithm for graphs of bounded treewidth. 
\end{enumerate}

The results in this paper were obtained through a large-scale experiment, requiring more than 2000 CPU-hours. We refer to \cite{carpentier2022developing} for a more elaborate presentation of the work in this study. 

The remainder of this paper is structured as follows: we provide the necessary background knowledge with regard to the MHV problem and tree decompositions in Sections \ref{sec:mhv} and \ref{sec:tree_decomposition}, respectively. In Section \ref{sec:exactAlgoTd}, we describe the exact algorithm for graphs of bounded treewidth from \cite{agrawal2017parameterized} and \cite{agrawal2020parameterized}. Next, we discuss our proposed algorithm in Section \ref{sec:my-algo}. We introduce the experimental setup and the datasets that we use in Section \ref{sec:experimental-setup}, after which we compare the results obtained by the different algorithms in Section \ref{sec:results}. We recapitulate the main conclusions of this paper in Section \ref{sec:conclusions}, and give some pointers for future work. 

\section{The Maximum Happy Vertices problem} \label{sec:mhv}

\subsection{Problem definition} \label{sec:MHVDefiniton}

We first clarify the notation related to graph theory used in this work. A graph $G$ consists of a set of vertices $V(G)$ and a set of edges $E(G)$. The set $N_G(v) = \{u \in V(G) \mid \{u, v\} \in E(G)\}$ are the neighbours of $v$ in $G$. The degree of a vertex $v$, denoted as $\text{deg}_G(v)$, equals the number of neighbours of $v$, thus $\lvert N_G(v) \rvert = \text{deg}_G(v)$. We denote $\Delta_G = \max_{v \in V(G)} \text{deg}_G(v)$ as the largest degree in $G$. We simply write $N(v)$, $\text{deg}(v)$ and $\Delta$ if the graph is clear from context. A subgraph $G'$ of $G$ has $V(G') \subseteq V(G)$ and $E(G') \subseteq \{\{u, v\} \in E(G) \mid u, v \in V(G')\}$. We say that $G'$ is an induced subgraph of $G$ if $E(G')$ is exactly equal to the latter set.

The MHV problem, introduced by \cite{zhang2015algorithmic}, is a vertex colouring problem based on the \textit{happiness} of a vertex.
\begin{definition}[Happiness of a vertex]\label{def:happiness}
	Given a graph $G$ and a colour function $c: V(G) \rightarrow \{1, \ldots, k\}$. A vertex $v \in V(G)$ is said to be \textit{happy} if and only if $\forall v' \in N(v): c(v) = c(v')$. Otherwise, vertex $v$ is said to be \textit{unhappy}. 
\end{definition}

\begin{figure}
    \centering
    \includegraphics[scale=0.7]{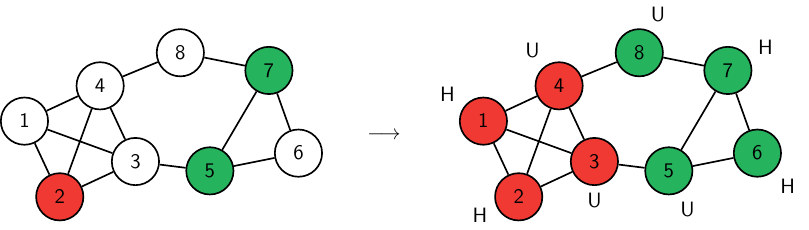}
    \caption[A visualisation of the MHV Problem.]{A visualisation of the MHV problem. On the left a partial colour function is given for some graph $G$, which is extended to a complete colour function of $G$. Vertices 1, 2, 6 and 7 are happy in the extension, which is marked with an 'H'. The other vertices, marked with a 'U', are unhappy.}
    \label{fig:MHV_example}
\end{figure}

The MHV problem follows straightforwardly from the concept of happy vertices. 
\begin{definition}[Maximum Happy Vertices Problem -- MHV \citep{zhang2015algorithmic}] \label{def:mhv}
	Given a graph $G$ and a partial colour function $c: V(G) \rightarrow \{1, \ldots, k\}$. The goal is to extend $c$ to a full colouring $c'$ such that the number of happy vertices is maximised.
\end{definition}
Fig. \ref{fig:MHV_example} visualises this problem. The solution consists of 4 happy vertices, which is the maximum number of happy vertices for this particular instance. Note that other extensions exist with 4 happy vertices, for example by colouring all uncoloured vertices green.

The MHV problem has applications in a variety of areas. \cite{zhang2015algorithmic} proposed it when studying social networks, which are governed by \textit{homophily}: one's friends are typically not a random sample of the population, but have similar characteristics such as the place they live, occupation, interests and race \citep{easley2010networks}. Other applications involve partitioning a group of people, such that many people are in the same group as their friends, for example in creating team building groups \citep{lewis2019finding} or assigning wedding table seats \citep{PeetersFlorian2020OvhM}. In general, any clustering application can be converted into an instance of the MHV problem, with the graph dictating which data points are strongly related and the colours indicating the cluster of the vertex.

\subsection{Constructive algorithms by Zhang and Li}

Besides introducing the MHV problem, \cite{zhang2015algorithmic} additionally proposed two heuristic algorithms for solving it: Greedy-MHV and Growth-MHV. 

The most simple of the two is Greedy-MHV: the algorithm assigns the same colour to all uncoloured vertices. This results in $k$ colour functions, one for each colour. Each of these is evaluated by counting the number of happy vertices. The colour function with the most happy vertices is returned as the solution. \cite{zhang2015algorithmic} proved that Greedy-MHV is a $1/k$-approximation algorithm. Each pair of neighbours must be considered for evaluating a colouring. Doing this for all $k$ colour functions results in a time complexity of $\mathcal{O}(k \cdot \lvert E(G) \rvert)$

Growth-MHV starts from a graph $G$ and a partial colour function $c$ of $G$, and greedily assigns a colour to some uncoloured vertex. These greedy choices are made through a labelling of $V(G)$.

\begin{definition}[Growth-MHV Labels] \label{def:growth-mhv-labels}
    Given a graph $G$ and a fixed partial colour function of $G$. Every coloured vertex $v \in V(G)$ has one of the following labels:
    \begin{itemize}
        \item $v$ is a $H$-vertex if all of its neighbours are assigned the same colour.  
        \item $v$ is a $U$-vertex if any of its neighbours has a different colour. 
        \item $v$ is a $P$-vertex if all of its coloured neighbours are assigned the same colour as $v$ and some neighbours are not coloured. 
    \end{itemize}
    Every $v \in V(G)$ that is not coloured yet has one of the following labels:
    \begin{itemize}
        \item $v$ is a $L_P$-vertex if any of its neighbours is a $P$-vertex.  
        \item $v$ is a $L_H$-vertex if none of its neighbours are $P$-vertices and $v$ has the potential to become happy, that is: $v$ has at least one coloured neighbour and all of its neighbouring $U$-vertices have the same colour. 
        \item $v$ is a $L_U$-vertex if none of its neighbours are $P$-vertices and $v$ can not become happy, that is: at least two of its neighbouring $U$-vertices are coloured differently.  
        \item $v$ is a $L_F$-vertex if none of its neighbours are coloured. 
    \end{itemize}
\end{definition}

Algorithm \ref{alg:growth-mhv} shows how Growth-MHV executes, following the definition in \cite{lewis2019finding} and \cite{PeetersFlorian2020OvhM}. \cite{zhang2015algorithmic} excluded the else-statement on line \ref{alg:growth-mhv-lfcheck} by assuming that the graphs are connected. By colouring vertex $v$, its label changes. Furthermore, the label of all neighbours of $v$ can change if $v$ is a $L_U$- or $L_F$-vertex. If $v$ is a $P$- or $L_H$-vertex on the other hand, then the labels of vertices within a distance of three edges can change. \cite{zhang2015algorithmic} only considered vertices within a distance of two, but this has been noted and rectified by \cite{lewis2019finding}. Possibly multiple vertices have the same label during selection. Ties are broken based on the degree of the vertices. This strategy was proposed by \cite{lewis2019finding}, and its effectiveness was experimentally analysed by \cite{PeetersFlorian2020OvhM}. The while loop iterates at most $\lvert V(G) \rvert$ times, and retrieving the neighbourhood has a complexity of $\mathcal{O}(\lvert E(G) \rvert)$ through a breadth-first search starting from the coloured vertex. The overall time complexity is $\mathcal{O}(\lvert V(G) \rvert \cdot \lvert E(G) \rvert)$.

\begin{algorithm}
    \small
	\caption{Growth-MHV} \label{alg:growth-mhv}
    \textbf{Input:} A graph $G$ and a partial colouring $c$ of $G$ with $k$ colours \newline
    \textbf{Output:} A colouring $c'$ of the vertices of $G$
	\begin{algorithmic}[1]
	    \State $c'$ $\leftarrow$ $c$
        \While{$c'$ is not a complete colouring of $G$}
            \If{there exists a $P$-vertex $v$}
                \State $c'(n)$ $\leftarrow$ $c'(v)$ for all $L_P$-vertices $n \in N(v)$
            \ElsIf{there exists a $L_H$-vertex $v$}
                \State $c'(v')$ $\leftarrow c'(u)$ for all $v' \in \{v\} \cup N(v)$ and some $U$-vertex $u \in N(v)$
            \ElsIf{there exists a $L_U$-vertex $v$}
                \State $c'(v)$ $\leftarrow$ $c'(u)$ for some $U$-vertex $u \in N(v)$
            \Else \hspace{1mm}some uncoloured $L_F$-vertex $v$ exists \label{alg:growth-mhv-lfcheck}
                \State $c'(v) \leftarrow$ a random colour
            \EndIf
            \State Recompute the labels of all relevant vertices
        \EndWhile
        \State \Return $c'$
	\end{algorithmic}
\end{algorithm}

\subsection{Metaheuristic algorithms}

Since the introduction of the MHV problem, several alternative heuristic methods to Greedy-MHV and Growth-MHV have been proposed. Besides analysing the MHV problem, \cite{lewis2019finding} develop an algorithm based on the Construct, Merge, Solve \& Adapt (CMSA) methodology of \cite{blum2016construct}. The authors improved upon this solution with a Tabu Search based approach, which is particularly suited for larger problem instances \citep{thiruvady2020tackling}. \cite{PeetersFlorian2020OvhM} proposes to use simulated annealing to tackle the MHV problem. More recently, \cite{ghirardi2021simple} propose a simple but effective matheuristic improvement approach. Because the algorithm of \cite{ghirardi2021simple} generally outperforms the alternative approaches, we compare it to our method in Section~\ref{sec:results}. Below we provide a high-level overview of their matheuristic. 

The algorithm of \cite{ghirardi2021simple} first generates an initial solution using Greedy-MHV (the authors also experimented with Growth-MHV, but did not notice a significant difference). Second, the solution is restricted such that each vertex $v$ can only take a colour of another vertex $v'$ within distance $L$ of $v$. $L$ is a parameter of the algorithm and their experiments indicated that 2 seems to be the best value for $L$ in most cases. The restricted instance is then solved optimally using integer programming. Next, this newly found solution is used to restrict the possible colours of each vertex, and the process continues until no further improvements are made. This method is improved by using a multi-start setting, in which all $k$ colourings generated by Greedy-MHV are optimised, instead of only using the best one.

\section{Tree decompositions} \label{sec:tree_decomposition}

\cite{robertson1984graph} introduced \textit{tree decompositions} in their series of papers on graph minors:

\begin{definition}[Tree decomposition \citep{cygan2015parameterized, robertson1984graph}] \label{def:td}
    A \textit{tree decomposition} of graph $G$ is a pair $\mathcal{T}=(T, \mathcal{X})$, where $T$ is a tree and $\mathcal{X}=\{X_t \subseteq V(G) \mid t \in V(T)\}$ is a family of subsets of $V(G)$ with $X_t$ called the bag corresponding to node $t$, satisfying the following properties:
    \begin{enumerate}
        \item $\bigcup_{t \in V(T)} X_t = V(G)$.
        
        \item For every edge $e \in E(G)$, there exists a $t \in V(T)$ such that $e$ has both ends in $X_t$.
        
        \item For every $v \in V(G)$, the set $T_v = \{t \in V(T) \mid v \in X_t\}$, i.e., the set of nodes whose corresponding bags contain $v$, induces a connected subtree of $T$.
    \end{enumerate}
\end{definition}

\begin{figure}
    \centering
    \includegraphics[scale=0.7]{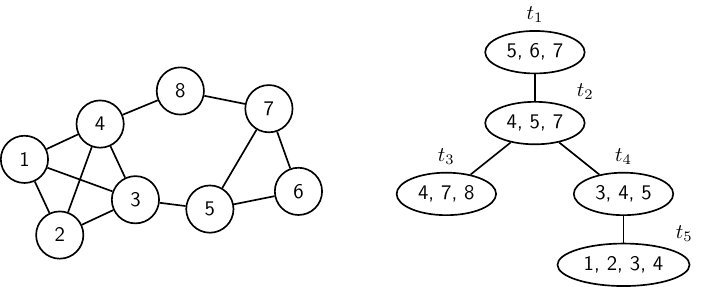}
    \caption{An example graph (left) and some tree decomposition (right) for that graph. The integers within a node of a tree decomposition refer to the identifiers of the vertices that are contained in the corresponding bag. For example, bag $X_{t_4} = \{3,4,5\}$.}
    \label{fig:tree_decomposition}
\end{figure}

The vertices of $T$ are referred to as \textit{nodes} in order to distinguish them from the vertices of $G$. For a rooted tree $T$, by $G_t$ with $t \in V(T)$ we denote the subgraph of $G$ induced by the vertices $X_t \cup \{v \in X_{t'} \mid t' \in \text{desc}(t)\}$, with $\text{desc}(t)$ the descendants of $t$ in $T$. A tree decomposition is shown Fig.  \ref{fig:tree_decomposition}.

The width $w$ of a tree decomposition $\mathcal{T}=(T, \mathcal{X})$ is equal to the cardinality of the largest bag in the tree minus 1, or $w = \max_{X \in \mathcal{X}} \lvert X \rvert - 1$. The tree decomposition in Fig. \ref{fig:tree_decomposition} has width $w = 3$ due to node $t_5$. A graph $G$ usually has multiple tree decompositions of varying width. One could create a tree decomposition $\mathcal{T}=(T, \mathcal{X})$, in which $T$ exists of only one node $t$ with $X_t = V(G)$, satisfying all properties above. This leads us to the \textit{treewidth} of a graph $G$.

\begin{definition}[Treewidth of a graph]
    The \textit{treewidth} of a graph $G$, denoted by $\text{tw}(G)$, is the minimum possible width of a tree decomposition of $G$.
\end{definition}

A tree decomposition can be converted into a normalized form: the \textit{nice tree decomposition}, defined below. Generally, nice tree decompositions give no algorithmic advantages over normal tree decompositions. They do, however, simplify algorithm design, since the relations of connected nodes in $T$ are restricted to only a few simple types. 

\begin{definition}[Nice Tree Decomposition]
    A tree decomposition $\mathcal{T}=(T, \mathcal{X})$ of graph $G$ is said to be \textit{nice} if $T$ is rooted and the following conditions hold:
    \begin{enumerate}
        \item $X_r = \emptyset$ with $r$ the root of $T$.
        \item $X_l = \emptyset$ for every leaf $l$ in $T$.
        \item Every non-leaf node $t \in V(T)$ has one of the following types:
        \begin{itemize}
            \item Introduce node: $t$ has exactly 1 child $t'$ such that $X_t = X_{t'} \cup \{v\}$ for some vertex $v \not\in X_{t'}$. Vertex $v$ is said to be introduced in node $t$.
            \item Forget node: $t$ has exactly 1 child $t'$ such that $X_t = X_{t'} \setminus \{v\}$ for some vertex $v \in X_{t'}$. Vertex $v$ is said to be forgotten in node $t$.
            \item Join node: $t$ has exactly 2 children $t_1$ and $t_2$, such that $X_t = X_{t_1} = X_{t_2}$.
        \end{itemize}
    \end{enumerate}
    \label{def:nicetd}
\end{definition}

Lemma \ref{lem:construct_nice_tree_decomposition} shows that tree decompositions can efficiently be converted into nice tree decompositions of equal width.

\begin{lemma}[Constructing Nice Tree Decompositions \citep{cygan2015parameterized}]
    \label{lem:construct_nice_tree_decomposition}
    If a graph $G$ admits a tree decomposition of width at most $w$, then it also admits a nice tree decomposition of width at most $w$. Moreover, given a tree decomposition $\mathcal{T}=(T, \mathcal{X})$ of $G$ of width at most $w$, one can in time $\mathcal{O}\big(w^2 \cdot max(\lvert V(T) \rvert, \lvert V(G) \rvert\big)$ compute a nice tree decomposition of $G$ of width at most $w$ that has at most $\mathcal{O}(w\lvert V(G) \rvert)$ nodes.
\end{lemma}

A \textit{separation} $(A,B)$ of a graph $G$ with separator $A \cap B$ is a pair such that $A \cup B = V(G)$ and there is no edge between $A \setminus B$ and $B \setminus A$. Separators are interesting, because they lead to a number of subproblems for which the optimal solutions only interact with each other through the vertices of the separator. The following lemma shows that the nodes of a tree decomposition form a sequence of separations. This lemma is often implicitly applied in algorithms using tree decompositions.

\begin{lemma}[Separations in a Tree Decomposition \cite{cygan2015parameterized}] \label{lemma:separation}
    Let $\mathcal{T} = (T, \mathcal{X})$ be a tree decomposition of graph $G$ and $\{a,b\}$ be an edge of $E(T)$ connecting nodes $a$ and $b$. The forest $T-\{a,b\}$ obtained from $T$ by deleting edge $\{a,b\}$ consists of two connected components $T_a$ (containing node $a$) and $T_b$ (containing node $b$). Let $A = \bigcup_{t \in V(T_a)} X_t$ and $B = \bigcup_{t \in V(T_b)} X_t$. Then $(A, B)$ is a separation of $G$ with separator $X_a \cap X_b$.
\end{lemma}

Algorithms using tree decompositions often have a polynomial dependence on the size of the graph, but an exponential dependence on the width of the used tree decomposition. It is thus desirable to compute a tree decomposition of minimal width. Unfortunately, deciding if the treewidth of a graph $G$ is at most $w$ is $\mathcal{NP}$-complete \citep{arnborg1987complexity}. Nonetheless, practical algorithms computing tree decompositions of minimal (or small) width exist. There was a track for computing tree decompositions as part of the second edition of the Parameterized Algorithms and Computational Experiments Challenge \citep{dell2018pace} (PACE 2017\footnote{\url{https://pacechallenge.org/2017/}}). This challenge aims to bridge the gap between the theory of algorithm design and analysis, and the practice of algorithm engineering. Both exact and heuristic algorithms could be submitted for the challenge. PACE 2017 resulted in multiple highly optimised, open source algorithms for computing a tree decomposition of a graph \citep{tamaki2019positive,hamann2018graph,bannach2019practical}, which can be used to develop algorithms using tree decompositions. 

\section{Exact algorithm for graphs of bounded treewidth} \label{sec:exactAlgoTd}

Our heuristic algorithm is based on the exact algorithm by \cite{agrawal2017parameterized} and \cite{agrawal2020parameterized} for solving the MHV problem in polynomial time on graphs of bounded treewidth. In this section, we briefly describe this algorithm, since it forms the basis of our heuristic algorithm and we use similar terminology for describing it in the following section. We refer to \cite{agrawal2020parameterized} for the correctness proofs of the recursive formulas. Note that we use colour functions, whereas \cite{agrawal2020parameterized} used the notion of a colouring as a partition of the vertices. 

Given a graph $G$ and partial colour function $c: V(G) \rightarrow \{1, \ldots, k\}$. The algorithm first selects for each $i \in \{1, \ldots, k\}$ an arbitrary vertex $v_i^* \in V(G)$ and constructs the set $S^* = \{v_i^* \mid i \in \{1, \ldots, k\}\}$, a set containing one coloured vertex for each colour. Next a tree decomposition is constructed, and converted into a nice tree decomposition $\mathcal{T}'=(T', \mathcal{X}')$ using Lemma \ref{lem:construct_nice_tree_decomposition}, of width $w$ with root $r' \in V(T')$. Tree decomposition $\mathcal{T}=(T, \mathcal{X})$ is obtained by further processing $\mathcal{T}'$ such that $T = T'$ with root $r = r'$, and $\mathcal{X} = \{X_t' \cup S^* \mid X_t' \in \mathcal{X}'\}$. This ensures that $X_t$ contains at least one coloured vertex for every colour. The width of $\mathcal{T}$ is at most $w + k$. The notion of introduce, forget and join nodes naturally extend this more processed tree decomposition $\mathcal{T}$.

Consider a partial colour function $c': X_t \rightarrow \{1, \ldots, k\}$ of $G$ with $t \in V(T)$, which colours all vertices in bag $X_t$. Let $\mathcal{H} = \{(H_i, U_i) \mid i \in \{1, \dots, k\} \land H_i \cup U_i = c'^{-1}(i) \land H_i \cap U_i = \emptyset\}$ be a partition of $X_t$. The vertices in $H_i$ will be happy in the final colouring and those in $U_i$ are assumed to be unhappy. For every $c'$, multiple $\mathcal{H}$ exist. A tuple $\tau = [t, c', \mathcal{H}]$ is valid if $c'$ extends $c$. A colour function $c_\tau: V(G_t) \rightarrow \{1, \dots, k\}$ is said to be a \textit{$\tau$-good colouring} if $\tau$ is valid and following conditions are met:
\begin{itemize}
    \item $\forall v \in X_t: c'(v) = c_\tau(v)$: $c_\tau$ extends $c'$
    \item $\forall v \in V(G_t): c(v) = c_\tau(v)$ if $v$ is coloured by $c$: $c_\tau$ extends $c$ in $G_t$.
    \item $\forall v \in H_i, i \in \{1, \dots, k\} : v$ is happy in $G_t$ with respect to $c_\tau$.
\end{itemize}

The algorithm builds a table $\Pi(\tau)$ of valid tuples $\tau$ for each node $t \in V(T)$, which is set to the maximum number of happy vertices in $V(G_t) \setminus \big(\bigcup_{i \in \{1, \dots, k\}} U_i\big)$ over all $\tau$-good colour functions. If no $\tau$-good colouring exists, then $\Pi(\tau) = -\infty$. These values can be computed in a bottom-up fashion over the structure of the tree decomposition $\mathcal{T}$. Because $G = G_r$ with $r$ the root of $T$, the maximum number of happy vertices in $G$ can be read from the table $\Pi(\tau)$ at root $r$. We now describe how $\Pi(\tau)$ can be computed for each type of node in $\mathcal{T}$.

\textbf{Leaf node.} Suppose that $t$ is a leaf node. We know that $X_t = S^*$ and $c'(v) = c(v)$ for all $v \in S^*$. There can be at most one $\tau$-good colouring for every tuple $\tau = [t, c', \mathcal{H}]$, namely $c_\tau = c'$. Furthermore, the happy vertices $H$ in $S^*$ can be found by checking the adjacencies in $S^*$. If some $H_i$ contains a vertex that is not in $H$, then $c_\tau$ is no $\tau$-good colouring, and $\Pi(\tau) = -\infty$. Otherwise, $\Pi(\tau) = \lvert H \setminus \big(\bigcup_{i \in \{1, \dots, k\}} U_i\big) \rvert$, that is the number of happy vertices that are not in any $U_i$. This is justified by the uniqueness of $c_\tau$.

\textbf{Introduce node.} Suppose that $t$ is an introduce node with unique child $\bar{t}$ and introduced vertex $\tilde{v}$. We must compute $\Pi(\tau)$ for some $\tau = [t, c', \mathcal{H}]$. Assume that $c'(\tilde{v}) = i$. Two cases immediately arise in which $\Pi(\tau) = -\infty$: 
\begin{enumerate}
    \item If $\tilde{v} \in H_i$ and $c'(n) \neq i$ for some $n \in N_{G_t}(\tilde{v})$. The introduced vertex can not be happy if it has a differently coloured neighbour in $G_t$.
    
    \item If $H_j \setminus N_{G_t}(\tilde{v}) \neq \emptyset$ for some $j \in \{1, \dots, k\} \setminus \{i\}$, if some neighbour of $\tilde{v}$ is happy but has a different colour than $\tilde{v}$.
\end{enumerate} %
Otherwise we construct tuple $\bar{\tau} = [\bar{t}, \bar{c'}, \bar{\mathcal{H}}]$, with $\bar{c'} = c'$ but in which the colour assignment of $\tilde{v}$ is removed and $\bar{\mathcal{H}} = \big(\mathcal{H} \setminus \{(H_i, U_i)\}\big) \cup \{(H_i \setminus \{\tilde{v}\}, U_i \setminus \{\tilde{v}\}\}$. Then $\bar{\tau}$ is a valid tuple for child $\bar{t}$. We recursively compute $\Pi(\tau)$ as follows:
\begin{align}
    \Pi(\tau) = 
    \begin{cases} 
      1 + \Pi(\bar{\tau}) & \text{if } \tilde{v} \in H_i  \\ 
      \Pi(\bar{\tau}) & \text{if } \tilde{v} \in U_i
    \end{cases}
\end{align}

\textbf{Forget node.} Suppose that $t$ is a forget node with unique child $\bar{t}$ and forgotten vertex $\tilde{v}$. Given tuple $\tau = [t, c', \mathcal{H}]$. For every $i \in \{1, \dots, k\}$, let $c'_i$ be equal to $c'$ with the additional colour assignment $c'_i(\tilde{v}) = i$. If $\tilde{v}$ is coloured in initial colouring $c$, then $\mathcal{C} = \{c'_{c(\tilde{v})}\}$. Otherwise, $\mathcal{C} = \{c'_i \mid i \in \{1, \dots, k\}\}$. Define partition $\mathcal{H}_{i+} = \big(\mathcal{H} \setminus \{(H_i, U_i)\}\big) \cup \{(H_i \cup \{\tilde{v}\}, U_i)\}$ and $\mathcal{H}_{i-} = \big(\mathcal{H} \setminus \{(H_i, U_i)\}\big) \cup \{(H_i, U_i \cup \{\tilde{v}\})\}$. $\mathcal{H}_{i+}$ and $\mathcal{H}_{i-}$ are obtained from $\mathcal{H}$ by adding $\tilde{v}$ to $H_i$ and $U_i$ respectively. We can now compute $\Pi(\tau)$ as:
\begin{align}
    \Pi(\tau) = \displaystyle\max_{\bar{c'_i} \in \mathcal{C}, \# \in \{+, -\}} \Pi(\bar{t}, \bar{c'_i},\mathcal{H}_{i\#})
    \label{eq:exact_forget_node}
\end{align}

\textbf{Join node.} Suppose that $t$ is a join node with children $t_1$ and $t_2$. Recall from Definition \ref{def:nicetd} of nice tree decompositions that $X_t = X_{t_1} = X_{t_2}$. For $\tau = [t, c', \mathcal{H}]$, we can compute $\Pi(\tau)$ as 
\begin{align}
    \Pi(t, c', \mathcal{H}) = \Pi(t_1, c', \mathcal{H}) + \Pi(t_2, c', \mathcal{H}) - \sum_{i \in \{1, \dots, k\}} \lvert H_i \rvert
    \label{eq:exact_td_join_node}
\end{align}

\textbf{Runtime analysis.} Given a graph $G$ with $n = \lvert V(G)\rvert$ and initial colouring $c$ with $k$ colours. A tree decomposition of $G$ of width $w \leq 5 \cdot \text{tw}(G) + 4$ can be constructed in time $\mathcal{O}(2^{\text{tw}(G)}n)$ due to \cite{bodlaender2016c}, which can be converted into a nice tree decomposition in time $\mathcal{O}(wn)$ of equal width using Lemma \ref{lem:construct_nice_tree_decomposition}. Table $\Pi(\tau)$ for any node $t$ contains at most $k^{w+1}2^{k+w+1}$ entries. The total running time of the algorithm is $\mathcal{O}\big(k^{w+1}2^{k+w+1} n^{\mathcal{O}(1)}\big) = \mathcal{O}\big(2^{\mathcal{O}(k + w \log(k))} n^{\mathcal{O}(1)}\big)$. To obtain a linear running time in $n$, \cite{bodlaender2013fine} proposes a data structure -- that can be constructed in time $\mathcal{O}(wn)$ -- that allows for adjacency checking in time $\mathcal{O}(w)$. Using this data structure results in a running time of $\mathcal{O}\big(2^{\mathcal{O}(k + w \log(k))} n\big)$ to solve the MHV problem, which is linear in $n$.

\section{Heuristic algorithm using tree decompositions} \label{sec:my-algo}

\begin{table}[b]
    \centering
    \caption{An overview of the symbols used for describing the algorithm, and their semantic meaning. Several of the subscripts are left out in order to keep this table manageable.}
    \label{tab:algorithm_symbols}
    \begin{tabular}{p{0.15\textwidth}p{0.75\textwidth}}
        \toprule
        \textbf{Symbol} & \textbf{Description} \\
        \midrule
        $G$ & A graph with vertices $V(G)$ and edges $E(G)$ \\ \midrule
        $\text{deg}(v)$ & The degree of a vertex $v$, the number of neighbours of $v$ \\ \midrule
        $\Delta$ & The maximum degree over all vertices in the graph \\ \midrule
        
        $k$ & The number of colours to use \\ \midrule
        $c$ & The initial colour function of the MHV instance \\ \midrule
        
        $t$ & a node of the tree decomposition \\ \midrule
        $X_t$ & the set of vertices, the bag, corresponding to node $t$ \\ \midrule
        $G_t$ & The graph induced by the vertices in $X_t$ and all bags corresponding to the descendants of $t$ \\ \midrule
        
        $\tau = [t, c', \tilde{\mathcal{H}}]$ & The tuples constructed for each node $t$ in the tree decomposition, which dictate the solution of $G_t$ \\ \midrule
        $c'$ & The colouring corresponding to a tuple $\tau$ for node $t$, which dictates the colours of the vertices in $G_t$ \\ \midrule
        $\tilde{\mathcal{H}}$ & The labelling corresponding to a tuple $\tau$, which assigns a label from Definition \ref{def:heuristic_labels} to all vertices in $G$ \\ \midrule
        
        $W$ & The width of the algorithm, the number of tuples $\tau$ to construct at each node of the tree decomposition \\ \midrule
        $\Pi(\tau)$ & The evaluation of a tuple $\tau$, computed according to Equation \ref{eq:heuristic_value} \\ \midrule
        $\mathcal{L}$ & The list of tuples $\tau$ that is constructed at each node of the tree decomposition \\ \midrule
        
        $\mathtt{H}, \mathtt{U}, \mathtt{P_H}, \mathtt{P_U}, \mathtt{UK}$ & The labels from Definition \ref{def:heuristic_labels} \\ \midrule
        $W_\mathtt{H}$ & The weight of $\mathtt{H}$-vertices when computing $\Pi(\tau)$ (similar counts exist for the other labels) \\ \midrule
        $C_\mathtt{H}(\tau)$ & The number of $\mathtt{H}$-vertices in a tuple $\tau$  (similar weights exist for the other labels)\\
        \bottomrule
    \end{tabular}
\end{table}

While the algorithm of \cite{agrawal2017parameterized} enumerates all possible $\tau$-tuples for a given node $t$ in the tree decomposition, our algorithm constructs the tuples given the tuples constructed in the children of $t$. By doing so, we avoid enumerating all possible tuples in $t$, including those for which no $\tau$-good colouring exists. Furthermore, we can choose to only generate a specified number of tuples for each node in the tree decomposition. This is achieved through a user defined parameter $W$, which we call the \textit{width} of the algorithm. We lose the exactness guarantee of our algorithm if not all tuples are constructed, in favour of shorter running times. However, if $W$ is large enough such that all tuples are constructed, the algorithm constructs an optimal solution. Our algorithm -- and more precisely $W$ -- offers a trade-off between optimality and running time. 

The remainder of this section is organized as follows. First we discuss how the algorithm computes the most promising tuples. Afterwards we explain how each type of node in the nice tree decomposition is handled. We conclude this section with an analysis of the algorithm. The description of our algorithm introduces a fair amount of symbolic notation. An overview of all symbols is given in Table \ref{tab:algorithm_symbols}. The source code of this algorithm is available on \url{https://github.com/LouisCarpentier42/HeuristicAlgorithmsUsingTreeDecompositions}.

\subsection{Detecting promising colour functions: evaluating partial colourings} \label{sec:labeling}

\begin{figure}[h]
    \centering
    \includegraphics[scale=0.7]{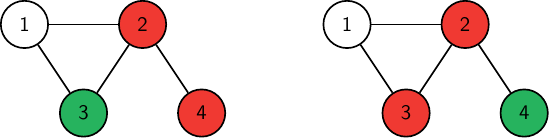}
    \caption{Problems with naively computing the most promising colour functions. The left partial colouring has more happy vertices (only vertex 4), but ultimately the right colouring can have two happy vertices (vertices 1 and 3, if 1 is coloured red).}
    \label{fig:heuristic_naive_labeling}
\end{figure}

For some node $t \in V(T)$, the algorithm decides which tuples are most promising and should be passed on to the parent of $t$. A naive solution is to simply count the number of happy vertices. However, this can lead to non-optimal decisions, as is shown in Fig. \ref{fig:heuristic_naive_labeling}. The algorithm distinguishes between several types of vertices in order to better evaluate a partial colouring.

\begin{definition}[Heuristic Algorithm Labels]
    \label{def:heuristic_labels}
    For a tree decomposition $\mathcal{T} = (T, \mathcal{X})$ of graph $G$, given a node $t \in V(T)$ and partial colour function $c' : V(G_t) \rightarrow \{1, \dots k\}$. Each vertex $v \in V(G)$ receives one of the following labels:
    \begin{itemize}
        \item $v$ is a $\mathtt{H}$-vertex if $v$ is coloured, all of its coloured neighbours share the same colour, and $v$ must be happy in the complete colouring of $G$ extending $c'$.
        
        \item $v$ is a $\mathtt{U}$-vertex if $v$ is coloured and has a differently coloured neighbour, or if $v$ is uncoloured but destined to be unhappy. 
        
        \item $v$ is a $\mathtt{P_H}$-vertex if $v$ is not coloured by $c'$, but has coloured neighbours of only one colour, $v$ could potentially become happy.
    
        \item $v$ is a $\mathtt{P_U}$-vertex if $v \in X_t$ and all of its coloured neighbours share the same colour, but $v$ is assumed to be unhappy in the complete colouring of $G$ extending $c'$.
    
        \item In any other case $v$ is a $\mathtt{UK}$-vertex, the state of $v$ is unknown.
    \end{itemize}
\end{definition}

We now modify the definition of $\tau$ such that $\mathcal{H}$ coincides with the labels above. We define the complete function $\tilde{\mathcal{H}} : V(G) \rightarrow \{\mathtt{H}, \mathtt{U}, \mathtt{P_H}, \mathtt{P_U}, \mathtt{UK}\}$, which maps all vertices onto a label of Definition \ref{def:heuristic_labels}. Our approach will generate tuples $\tau = [t, c', \tilde{\mathcal{H}}]$. Note that $\mathcal{H}$ in the exact algorithm partitions only the vertices in the bag. The $\mathtt{P_U}$-label may seem redundant, but this also occurs in the exact algorithm from \cite{agrawal2017parameterized}. A vertex can be part of some $U_i$, while all its neighbours share the same colour. We want to differentiate between these vertices since they could potentially become happy, in contrast to $\mathtt{U}$-vertices which are effectively unhappy. 

\begin{figure}
    \centering
    \includegraphics[scale=0.7]{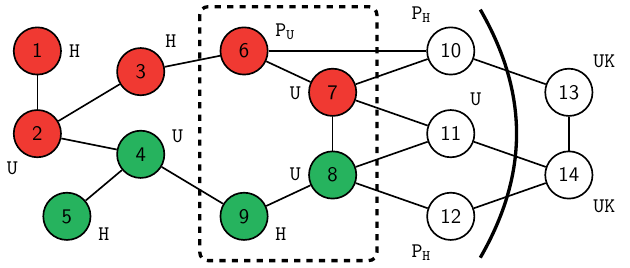}
    \caption{An example of the labels assigned by the heuristic algorithm for some colouring $c'$ and some node $t$. The labels of $\tilde{\mathcal{H}}$ are shown besides the vertices. Bag $X_t$ is marked with the dashed box. Exactly the vertices of $G_t$ are coloured. The black line separates the border from the other uncoloured vertices.}
    \label{fig:heuristic_labels}
\end{figure}

Fig. \ref{fig:heuristic_labels} shows an example of the labels. The uncoloured vertices are separated into two sets: the vertices adjacent to a vertex in the bag, and the remaining vertices. We call this first set the border. These are either $\mathtt{U}$- or $\mathtt{P_H}$-vertices. The algorithm will iteratively colour vertices from the border (more precisely, in the introduce node), after which the border shifts to include more vertices. 

The labels are directly used for computing the value $\Pi(\tau)$ of a tuple $\tau = [t, c', \tilde{\mathcal{H}}]$. In the exact algorithm, this equals to the maximal number of happy vertices in a $\tau$-good colouring. Our algorithm can exploit additional information through the labels. We assign a weight to each label, denoted as $W_\mathtt{X}$, with $\mathtt{X}$ any of the labels in Definition \ref{def:heuristic_labels}. The $\mathtt{UK}$-vertices are fixed for some node $t$, and give no additional information. For that reason we set $W_\mathtt{UK} = 0$. The other weights are parameters of the algorithm and can be chosen by the user. The evaluation of a tuple $\tau = [t, c', \tilde{\mathcal{H}}]$ is equal to
\begin{align}
	\Pi(\tau) = W_\mathtt{H} \cdot C_\mathtt{H}(\tau) + W_\mathtt{U} \cdot C_\mathtt{U}(\tau) + W_\mathtt{P_H} \cdot C_\mathtt{P_H}(\tau) + W_\mathtt{P_U} \cdot C_\mathtt{P_U}(\tau)
	\label{eq:heuristic_value}
\end{align}
with $C_\mathtt{X}(\tau)$ the number of vertices in $\tilde{\mathcal{H}}$ with label $\mathtt{X}$. The algorithm maintains and dynamically updates these counts when a vertex is coloured to ensure that Equation \ref{eq:heuristic_value} can be computed in $\mathcal{O}(1)$ time.

The algorithm generates a list $\mathcal{L}$ of tuples $\tau$ for each node $t \in V(T)$. If a new tuple $\tau'$ is constructed and $\mathcal{L}$ already contains $W$ tuples, then the algorithm must efficiently decide which $\tau \in \mathcal{L}$ to discard. Practically, we do this by representing $\mathcal{L}$ as a C++ map (i.e., a red-black tree) with as key $\Pi(\tau)$ from Equation \ref{eq:heuristic_value} and as value the tuple $\tau$ itself, sorted in ascending order of $\Pi(\tau)$. The tuples with low value are pushed to the front of $\mathcal{L}$. The worst value is possibly not unique. In this case the algorithm discards a random tuple with worst value. While sorting ensures a quicker lookup, adding a tuple to $\mathcal{L}$ requires logarithmic time. 

We follow a standard approach for dynamic programming algorithms that allows us to obtain a complete colouring of $V(G)$ by keeping track of backpointers. Such mechanism is necessary since the tuples $\tau$ only dictate the colours of the vertices in the bag. This has also been mentioned from \cite{agrawal2017parameterized}. In order to simplify further explanation, we say that a vertex $v$ is coloured by partial colouring $c'$ if $v$ is effectively coloured by $c'$ or if any colouring which can be reached through these pointers assigns a colour to $v$.

Note that the algorithm of \cite{agrawal2017parameterized} constructs a more structured tree decomposition by adding set $S^* = \{v_i^* \mid i \in \{1, \dots, k\}\}$ to $X_t$ for all $t \in V(T)$. Their main argument for doing so is to simplify the proofs. However, $S^*$ has no influence on a practical algorithm and only increases the running time. We have chosen to not construct this more structured nice tree decomposition and use the \textit{plain} nice tree decomposition.

We now proceed to describe how the tuples $\tau = [t, c', \tilde{\mathcal{H}}]$ are constructed in each type of node of a nice tree decomposition.

\subsection{Handling leaf nodes}

Handling a leaf node is straightforward because each leaf node $t \in V(T)$ has $X_t = \emptyset$ and $V(G_t) = \emptyset$. There exists exactly one tuple $\tau = [t, c', \tilde{\mathcal{H}}]$, namely with $c'$ a colour function with empty domain and complete function $\tilde{\mathcal{H}}$ with $\tilde{\mathcal{H}}(v) = \mathtt{UK}$ for all $v \in V(G)$. The algorithm constructs this tuple, and passes it on to the parent of $t$.

\subsection{Handling introduce nodes}

An introduce node $t \in V(G)$ with child node $t_\text{child}$ must assign a colour to the introduced vertex $\tilde{v}$. By doing so, the border moves and includes all neighbours $v \in N(\tilde{v})$ of $\tilde{v}$. 

The procedure for handling introduce nodes is given in Algorithm \ref{alg:heuristic_introduce}. For a tuple $\tau_\text{child} = [t_\text{child}, c'_\text{child}, \tilde{\mathcal{H}}_\text{child}]$, the procedure constructs tuples $\tau_{i,\mathtt{X}} = [t, c', \tilde{\mathcal{H}}]$ as follows: $c'$ extends $c'_\text{child}$ by colouring $\tilde{v}$ with colour $i \in \{1, \dots, k\}$ and $\tilde{\mathcal{H}}$ extends $\tilde{\mathcal{H}}_\text{child}$ by setting $\tilde{\mathcal{H}}(\tilde{v}) = \mathtt{X}$ with $\mathtt{X} \in \{\mathtt{H}, \mathtt{U}, \mathtt{P_U}\}$ and adjusting the labels of $n \in N(\tilde{v})$.

If $t_\text{child}$ is a leaf node, then $\tilde{v}$ will be a $\mathtt{UK}$-vertex. Its happiness only depends on the vertices coloured by $c$. All neighbours $v \in N(\tilde{v})$ are added to the border, and their labels change by definition. For tuple $\tau_{i,\mathtt{X}}$, if $v$ has a neighbour $n \in N(v)$ with $c(n) \neq i$, then $v$ is a $\mathtt{U}$-vertex, otherwise $v$ is a $\mathtt{P_H}$-vertex. 

If $t_\text{child}$ is not a leaf node, then $\tilde{v}$ is either a $\mathtt{U}$- or $\mathtt{P_H}$-vertex. If some vertex $v \in N(\tilde{v})$ with $c'_\text{child}(v) \neq i$ and $\tilde{\mathcal{H}}_\text{child}(v) = \mathtt{H}$ exists, then no valid tuple $\tau_{i,\mathtt{X}}$ exists, because $v$ can never be happy. Moreover, if two vertices $v_1, v_2 \in N(\tilde{v})$ exist such that $\tilde{\mathcal{H}}_\text{child}(v_1) = \tilde{\mathcal{H}}_\text{child}(v_2) = \mathtt{H}$ and $c'_\text{child}(v_1) \neq c'_\text{child}(v_2)$, then no valid $\tau$ can be constructed. In this case we construct tuple $\tau_{i,backup}$, in which we set $c'(\tilde{v}) = i$ and $\tilde{\mathcal{H}}(\tilde{v}) = \mathtt{U}$, but additionally set $\tilde{\mathcal{H}}(v) = \mathtt{U}$ for all $v \in N(\tilde{v})$ with $\tilde{\mathcal{H}}_\text{child}(v) = \mathtt{H}$ and $c'_\text{child}(v) \neq i$. These neighbours are exactly the conflicting neighbours, and all conflicts are resolved by assigning the $\mathtt{U}$-label. This backup tuple is added to a backup list $\mathcal{L}_\text{backup}$. We only return this list if no valid extensions exist for any $\tau_\text{child} \in \mathcal{L}_\text{child}$. Thus we only have to compute $\tau_{i, \text{backup}}$ if $\mathcal{L}$ is empty.

If no problematic vertices exist, then the happiness of $\tilde{v}$ can easily be set. If $\tilde{v}$ is a $\mathtt{U}$-vertex, we construct $\tau_{i, \mathtt{U}}$ because $\tilde{v}$ already has two differently coloured neighbours and will remain unhappy independent of its colour. Otherwise, $\tilde{v}$ is a $\mathtt{P_H}$-vertex, and all neighbours $v \in N(\tilde{v}) \cap X_t$ share the same colour $i$. We construct the tuples $\tau_{i, \mathtt{H}}$, $\tau_{i, \mathtt{P_U}}$ and $\tau_{j, \mathtt{U}}$ for all $j \in \{1, \dots, k\} \setminus \{i\}$.

The labels of vertices in the border are recomputed on line \ref{alg:line:recompute_labels}. Assigning a colour to $\tilde{v}$ only affects the labels of vertices $v \in N(\tilde{v})$, which limits the number of labels to recompute. If $v$ is a $\mathtt{U}$-vertex, then $v$ will remain unhappy after colouring $\tilde{v}$. When $v$ is a $\mathtt{P_H}$- or $\mathtt{P_U}$-vertex with coloured neighbours of colour $j$, then it becomes a $\mathtt{U}$-vertex only if $\tilde{v}$ is assigned colour $i \neq j$. Lastly, it is possible that $v$ is a $\mathtt{UK}$-vertex, because the border has shifted by colouring $\tilde{v}$. If vertex $v$ has a neighbour $v' \in N(v)$ with $c(v') \neq i$, then $v$ receives a $\mathtt{U}$-label. Otherwise, $v$ is a $\mathtt{P_H}$-vertex.

\begin{algorithm}
    \small
	\caption{Handle an introduce node} \label{alg:heuristic_introduce}
	\textbf{Input:} Introduce node $t$ with introduced vertex $\tilde{v}$ and child $t_\text{child}$, graph $G$, partial colouring $c$ of $G$ with $k$ colours, and $\mathcal{L}_\text{child}$ the tuples constructed in $t_\text{child}$ \newline
	\textbf{Output:} A list of tuples $\tau = [t, c', \tilde{\mathcal{H}}]$
	\begin{algorithmic}[1]
    	\State $\mathcal{L}$, $\mathcal{L}_\text{backup} \leftarrow$ two empty lists for the newly constructed tuples
        \ForAll{$\tau_\text{child} = [t_\text{child}, c'_\text{child}, \tilde{\mathcal{H}}_\text{child}] \in \mathcal{L}_\text{child}$}
            \ForAll{$i \in \{1, \dots, k\}$}
                \If{$\tilde{v}$ is a $\mathtt{UK}$-vertex}
                    \If{$\exists v \in N(\tilde{v}): c(v) \neq i$}
                        \State Add $\tau_{i, \mathtt{U}}$ to $\mathcal{L}$
                    \Else
                        \State Add $\tau_{i, \mathtt{H}}$ and $\tau_{i, \mathtt{P_U}}$ to $\mathcal{L}$
                    \EndIf
                \ElsIf{$\exists v \in N(\tilde{v}): c'_\text{child}(v) \neq i \land \tilde{\mathcal{H}}_\text{child}(v) = \mathtt{H}$}
                    \If{$\mathcal{L}$ is empty}
                        \State add $\tau_{i, \text{backup}}$ to $\mathcal{L}_\text{backup}$ 
                    \EndIf
                \ElsIf{$\tilde{v}$ is a $\mathtt{P_H}$-vertex with neighbours coloured in $i$}
                    \State add $\tau_{i,\mathtt{H}}$ and $\tau_{i,\mathtt{P_U}}$ to $\mathcal{L}$
                \Else
                    \State add $\tau_{i,\mathtt{U}}$ to $\mathcal{L}$
                \EndIf
                \State recompute $\tilde{\mathcal{H}}(v)$ for every $v \in N(\tilde{v})$ and constructed $\tau_{i,\mathtt{X}}$ \label{alg:line:recompute_labels}
            \EndFor
        \EndFor
        
    	\If{$\mathcal{L}$ is empty}
    	    \State \Return $\mathcal{L}_\text{backup}$
    	\Else
    		\State \Return $\mathcal{L}$
    	\EndIf
	\end{algorithmic}
\end{algorithm}

Algorithm \ref{alg:heuristic_introduce} is equal to the procedure for handling introduce nodes in the exact algorithm for graphs of bounded treewidth, if $\mathcal{L}$ never reaches its capacity $W$. Our approach distinguishes between $\mathtt{U}$- and $\mathtt{P_U}$-vertices. However, both labels have equivalent semantics: the vertex will be unhappy in the completed colour function.

The procedure constructs at most $2k$ tuples $\tau_{i, \mathtt{X}}$ for each tuple $\tau_\text{child} \in \mathcal{L}_\text{child}$. Because $\mathcal{L}$ is sorted, adding a tuple has logarithmic time complexity. The labels of all $v \in N(\tilde{v})$ must be recomputed. The algorithm checks the colour of $n$ with respect to $c$ for all $n \in N(v)$ in order to decide whether $v$ is a $\mathtt{U}$- or $\mathtt{P_H}$-vertex, which results in checking the colour of at most $\Delta$ vertices for each $v$, with $\Delta$ the maximum degree of a vertex in G. All together, Algorithm~\ref{alg:heuristic_introduce} has a running time of $\mathcal{O}\big(2kW (\log(W) + \text{deg}(\tilde{v}) \Delta)\big)$.

\subsection{Handling forget nodes}

Given a forget node $t \in V(T)$ with forgotten vertex $\tilde{v}$. The exact algorithm iterates over all $\tau = [t, c', \mathcal{H}]$ and maximises $\Pi(\tau)$ over all possible decisions for $\tilde{v}$, as is shown in Equation \ref{eq:exact_forget_node}. The happiness of $\tilde{v}$ is fixed for some $\tau$, due to Lemma \ref{lemma:separation} and because $\tilde{v}$ is forgotten in $t$. In our approach, not all tuples $\tau_\text{child}$ are available in $\mathcal{L}_\text{child}$. For that reason we maximise for every colour and happiness assignment of $\tilde{v}$ over all the tuples that have been constructed. The pseudocode for this procedure is shown in Algorithm \ref{alg:heuristic_forget}. Note that $\tilde{v}$ potentially is a $\mathtt{P_U}$-vertex, but once forgotten it must be either a $\mathtt{H}$- or $\mathtt{U}$-vertex. Due to Definition \ref{def:heuristic_labels} and Lemma \ref{lemma:separation}, we know that $\tilde{v}$ shares the same colour as all its neighbours, and thus we can safely set $\tilde{\mathcal{H}}(\tilde{v}) = \mathtt{H}$.

\begin{algorithm}
    \small
	\caption{Handle a forget node} \label{alg:heuristic_forget}
	\textbf{Input:} Forget node $t$ with forgotten vertex $\tilde{v}$ and child $t_\text{child}$, graph $G$, partial colouring $c$ of $G$ with $k$ colours, and $\mathcal{L}_\text{child}$ the tuples constructed in $t_\text{child}$ \newline
	\textbf{Output:} A list of tuples $\tau = [t, c', \tilde{\mathcal{H}}]$
	\begin{algorithmic}[1]
    	\State $\mathcal{L}' \leftarrow$ an empty list for the merged tuples, sorted according to $c'$ and $\tilde{\mathcal{H}}$
        \ForAll{$\tau_\text{child} = [t_\text{child}, c'_\text{child}, \tilde{\mathcal{H}}_\text{child}] \in \mathcal{L}_\text{child}$}
            \If{$\tilde{\mathcal{H}}(\tilde{v}) = \mathtt{P_U}$}
                \State $\tilde{\mathcal{H}}(\tilde{v}) \leftarrow \mathtt{H}$
            \EndIf
            \If{$\exists \tau = [t, c', \tilde{\mathcal{H}}] \in \mathcal{L}' : \forall v \in X_t: c'(v) = c'_\text{child}(v) \land \tilde{\mathcal{H}}(v) = \tilde{\mathcal{H}}_\text{child}(v)$}
                \State replace $\tau$ in $\mathcal{L}'$ by $\tau_\text{child}$ if $\Pi(\tau) < \Pi(\tau_\text{child})$
            \Else
                \State add $\tau_\text{child}$ to $\mathcal{L}'$
            \EndIf
        \EndFor
        \State $\mathcal{L} \leftarrow$ the tuples in $\mathcal{L}'$, but sorted according to $\Pi(\tau)$
    	\State \Return $\mathcal{L}$
	\end{algorithmic}
\end{algorithm}

If parameter $W$ is taken large enough, then all possible colours and happiness values of $\tilde{v}$ are present in $\mathcal{L}$. In this case, Algorithm \ref{alg:heuristic_forget} exactly computes Equation \ref{eq:exact_forget_node}, which guarantees the exactness of the procedure for large enough $W$.

Algorithm \ref{alg:heuristic_forget} uses an intermediate list $\mathcal{L}'$, which allows for logarithmic lookup. At most $W$ tuples are added to $\mathcal{L}'$, and thus no $\tau$ ever needs to be discarded. Finally, this list must be sorted according to $\Pi(\tau)$. The total time complexity of Algorithm \ref{alg:heuristic_forget} equals $\mathcal{O}(W\log(W))$.

\subsection{Handling join nodes}

\begin{figure}[h]
    \centering
    \includegraphics[scale=0.7]{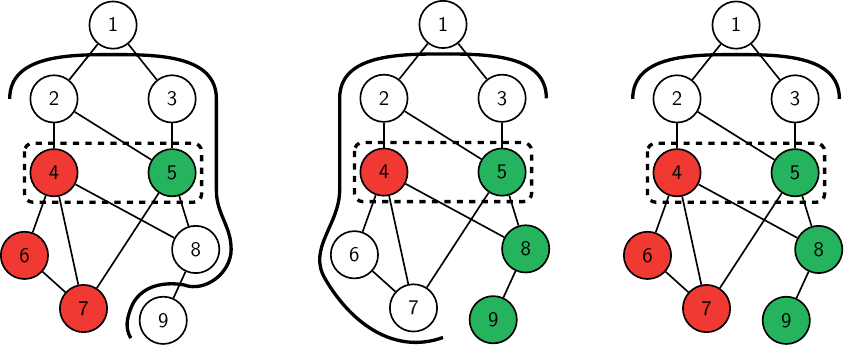}
    \caption{An example of how two tuples are merged in a join node $t$ with children $t_1$ and $t_2$. Bag $X_t$ is marked with the dashed box. The labels are left out for simplification. The black curve separates the border from the other uncoloured $\mathtt{UK}$-vertices. The left and center figure show the tuples of the children, the right figure shows their combination.}
    \label{fig:heuristic_join_border}
\end{figure}

Given a join node $t \in V(T)$, we know that $t$ has exactly two children $t_1$ and $t_2$ such that $X_t =X_{t_1} = X_{t_2}$. The exact algorithm iterates over every $\tau = [t, c', \mathcal{H}]$ in the join node and computes $\Pi(\tau)$ by combining its evaluation in $t_1$ and $t_2$. We can interpret this as combining two tuples $\tau_1$ and $\tau_2$ for child nodes $t_1$ and $t_2$ to construct the tuple $\tau$ in $t$. This is exactly how our heuristic approach works. For every $\tau_1 \in \mathcal{L}_1$, all $\tau_2 \in \mathcal{L}_2$ are compared to see if there exists a match, in which case $\tau_1$ and $\tau_2$ are combined into a full colouring of $G_t$. An example is given in Fig. \ref{fig:heuristic_join_border}. The tuples partition $V(G)$ into four sets:
\begin{enumerate}
    \item The vertices in $X_t$, which is equal to $\{4,5\}$ in Fig. \ref{fig:heuristic_join_border}. 
    
    \item The vertices in $V(G_{t_1}) \setminus X_t$. This is set $\{6,7\}$ in Fig. \ref{fig:heuristic_join_border}.
    
    \item The vertices in $V(G_{t_2}) \setminus X_t$. This is set $\{8,9\}$ in Fig. \ref{fig:heuristic_join_border}.
    
    \item The vertices in $V(G) \setminus V(G_t)$, set $\{1,2,3\}$ in Fig. \ref{fig:heuristic_join_border}. These are the vertices that are not coloured.
\end{enumerate}
Because of Lemma \ref{lemma:separation}, the solution for vertices in the last three sets do not influence each other, for a fixed colouring of the vertices in $X_t$. The pseudocode of the procedure for handling join nodes is given in Algorithm \ref{alg:heuristic_join}.

\begin{algorithm}
    \small
	\caption{Handle a join node} \label{alg:heuristic_join}
	\textbf{Input:} Join node $t$ with children $t_1$ and $t_2$, graph $G$, partial colouring $c$ of $G$ with $k$ colours, and $\mathcal{L}_1$ and $\mathcal{L}_2$ the lists of tuples constructed in $t_1$ and $t_2$ \newline
	\textbf{Output:} A list of tuples $\tau = [t, c', \tilde{\mathcal{H}}]$
	\begin{algorithmic}[1]
    	\State $\mathcal{L}_o$, $\mathcal{L}_i \leftarrow $ decide which of $\mathcal{L}_1$ and $\mathcal{L}_2$ should be used for the outer and inner loop
    	\State $\mathcal{L}$, $\mathcal{L}_\text{backup} \leftarrow$ two empty lists for the constructed tuples
    	\ForAll{$\tau_o = [t_o, c'_o, \tilde{\mathcal{H}}_o] \in \mathcal{L}_o$}
            \State $\tau_\text{best} \leftarrow$ the first tuple $\tau_i \in \mathcal{L}_i$\;
            \State $d_\text{best} \leftarrow +\infty$\;
            \State $\text{match\_found} \leftarrow false$\;
        	\ForAll{$\tau_i = [t_i, c'_i, \tilde{\mathcal{H}}_i] \in \mathcal{L}_i$}
        	    \State $d \leftarrow$ the distance between $\tau_o$ and $\tau_i$
        	    \If{$d=0$}
        	        \State $\text{match\_found} \leftarrow true$
                    \State merge $\tau_o$ and $\tau_i$, and add the result to $\mathcal{L}$
                    \State \textbf{break}
                \ElsIf{$d < d_\text{best}$}
                    \State $\tau_\text{best} \leftarrow \tau_i$\;
                    \State $d_\text{best} \leftarrow d$\;
                \EndIf
        	\EndFor
    	    \If{not $\text{match\_found}$ and  $\mathcal{L}$ is empty}
    	        \State heuristically merge $\tau_o$ and $\tau_\text{best}$, and add the result to $\mathcal{L}_\text{backup}$\;
    	    \EndIf
    	\EndFor
    	
    	\If{$\mathcal{L}$ is empty}
    	    \State \Return $\mathcal{L}_\text{backup}$
    	\Else
    		\State \Return $\mathcal{L}$
    	\EndIf
	\end{algorithmic}
\end{algorithm}

The procedure consists of two loops, one for each child.  We will denote $\mathcal{L}_o$ and $\mathcal{L}_i$ as the list used for the outer and inner loop respectively. We implemented three different mechanisms for selecting $\mathcal{L}_o$: randomly either $\mathcal{L}_1$ or $\mathcal{L}_2$, the list with the largest cardinality, or to the list with smallest cardinality. For each tuple in $\mathcal{L}_o$, at most one new tuple is constructed. If we select the largest of $\mathcal{L}_1$ and $\mathcal{L}_2$ in the outer loop, then more tuples will be constructed, which makes it more likely to construct the optimal $\tau$. Conversely, if we select the smallest list, fewer tuples will be constructed, resulting in shorter running times. 

In the exact algorithm, the colour and happiness for all $v \in X_t$ must exactly match. Because our heuristic approach only generates $W$ tuples, it is likely that for some tuples no exact match exists. Furthermore, it is possible that no match exists for any of the tuples, in which case no new tuples would be constructed. A list $\mathcal{L}_\text{backup}$ is maintained, similarly as we did for introduce nodes. For every $\tau_o \in \mathcal{L}_o$, if no match in $\mathcal{L}_i$ exists, then the most similar $\tau_i$ is heuristically merged with $\tau_o$. 

The similarity of a $\tau_i$ is measured through a distance metric with a weighting scheme. Each vertex $v \in X_t$ receives a weight, and this weight is added to the total distance if $c'_o(v) \neq c'_i(v)$. Similarly, this weight is added if $\tilde{\mathcal{H}}_o(v) \neq \tilde{\mathcal{H}}_i(v)$. A naive approach is to assign each vertex a weight of 1. However, vertices with neighbour $n \in N(v)$ such that $n \not\in X_t$ can have more influence when we need to heuristically combine two tuples. We have implemented several approaches to insert this knowledge into the distance computation.
\begin{enumerate}
    \item Vertex $v$ has a weight 1 if there exists some $n \in N(v) \setminus X_t$.
    \item Vertex $v$ has a weight 1 if there exists some $n \in  N(v)$ in the border. 
    \item Vertex $v$ has a weight 1 if there exists some $n \in N(v) \setminus X_t$ not in the border.
\end{enumerate}
For each of these strategies, we add an additional approach that sets the weight of $v$ to the number of vertices in the corresponding set. For example, for the first strategy, a similar weighting mechanism is implemented in which the weight of $v$ is equal to $\lvert N(v) \setminus X_t \rvert$. 

We developed two procedures for heuristically merging $\tau_o = [t_o, c'_o, \tilde{\mathcal{H}}_o]$ and $\tau_i = [t_i, c'_i, \tilde{\mathcal{H}}_i]$ into $\tau = [t, c', \tilde{\mathcal{H}}]$:
\begin{enumerate}
    \item Copy all colours and labels from $\tau_o$ to construct $\tau$. Next, the colours and labels of $\tau_i$ are copied into $\tau$ for all $v \in V(G_{t_i}) \setminus X_t$, that is the vertices in the subtree of $T$ rooted at $t_i$. This is visualised in Fig. \ref{fig:heuristic_join_copy_bag}. It shows that $\tau$ can be invalid, since $\tau_i$ and $\tau_o$ do not perfectly match. This only occurs for a pair of connected vertices $v_o \in X_t$ and $v_i \in V(G_{t_i}) \setminus X_t$, because only for these vertices the colours of some neighbour can change. This approach constructs two tuples, one as described and another one by swapping the role of $\tau_o$ and $\tau_i$.
    
    Copying $c'$ and $\tilde{\mathcal{H}}$ requires constant time by using pointers. In worst case, the labels of $\lvert X_t \rvert \Delta$ vertices must be verified, for which the colour of all their neighbours must be checked. The total running time is $\mathcal{O}(\lvert X_t \rvert \Delta^2)$.
    
    \begin{figure}[b]
        \centering
        \includegraphics[scale=0.7]{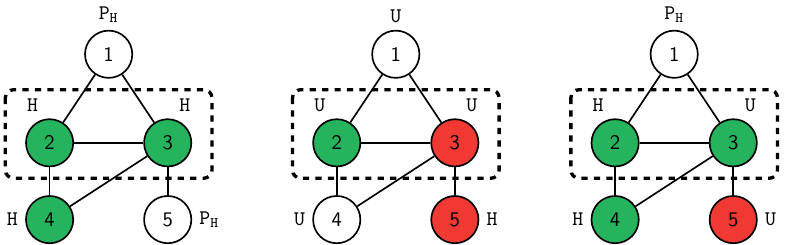}
        \caption{An example of merging two tuples $\tau_o$ (left) and $\tau_i$ (center) into a new tuple $\tau$ (right) by copying the decisions in the bag from $\tau_o$. Bag $X_t$ is marked with the dashed box. The label of vertex 3 changed in $\tau$ because it is adjacent to a differently coloured vertex.}
        \label{fig:heuristic_join_copy_bag}
    \end{figure}
    
    \item The second approach tries to maintain as many equivalent colour and label assignments. First set $c'(v) = c'_o(v)$ for all $v \in X_t$ with $c'_o(v) = c'_o(v)$. We set $\tilde{\mathcal{H}}(v)$ to either $\tilde{\mathcal{H}}_o(v)$ or $\tilde{\mathcal{H}}_i(v)$, with priority for the $\mathtt{H}$-label, then the $\mathtt{P_U}$-label, and lastly the $\mathtt{U}$-label. If $v$ is a $\mathtt{H}$- or $\mathtt{P_U}$-vertex, then all neighbours $n \in N(v)$ are assigned colour $c'(v)$, to ensure validity of the label. Next, the labels of coloured, unlabelled vertices are set, in which a vertex is assigned a $\mathtt{H}$-label whenever possible. Again, in this case the neighbours of the vertex are assigned the same colour. Some vertex is possibly not assigned a colour if $X_t$ consists of multiple components. Such a vertex $v$ is randomly assigned a colour $c'_o(v)$ or $c'_i(v)$, and next its label is recomputed. Lastly, the labels of the vertices in the border must be recomputed. 
    
    Every vertex $v \in X_t$ must be considered. In worst case, none of the vertices have a matching colour and all their neighbours must be checked to decide $\tilde{\mathcal{H}}(v)$, which results in checking $\mathcal{O}(\lvert X_t \rvert \Delta)$ vertices. Afterwards, the labels of all vertices in the border must be recomputed. The border contains at most $\lvert X_t \rvert \Delta$ vertices, and for every vertex in the border we must check the colour of all its neighbours. This last step results in a time complexity of $\mathcal{O}(\lvert X_t \rvert \Delta^2)$. 
\end{enumerate}
    
As for handling introduce and forget nodes, the algorithm for handling a join node converges to the exact algorithm of \cite{agrawal2017parameterized}, when $W$ is taken large enough. All valid tuples are constructed in the child nodes of $t$ and thus a match will always be found. In this case no heuristic approach will be executed. 

For computing the distance between tuples, the algorithm needs to iterate over all vertices in $X_t$. If the neighbours of a vertex are taken into account, then the distance computation has a running time of $\mathcal{O}(\lvert X_t \rvert \Delta)$. Both heuristic combination algorithms have a running time of $\mathcal{O}(\lvert X_t \rvert \Delta^2)$. However, this step is not executed in the inner for-loop. Because $\mathcal{L}_1$ and $\mathcal{L}_2$ contain at most $W$ tuples,  Algorithm \ref{alg:heuristic_join} has a total time complexity of $\mathcal{O}(W^2 \lvert X_t \rvert \Delta)$.

\subsection{Analysis}

If $W$ is taken large enough, all valid tuples $\tau$ will be constructed and our heuristic algorithm converges to the exact algorithm of \cite{agrawal2017parameterized}. Given a tree decomposition $\mathcal{T} = (T, \mathcal{X})$ of width $w$, our heuristic approach becomes an exact algorithm if
\begin{align}
    W \geq (2k)^{w+1} 
    \label{eq:exactness_guarantee}
\end{align}
with $k$ the number of colours. For every vertex $v \in X_t$ for some $t \in T$, there are $k$ possible colours and $v$ can be either happy (a $\mathtt{H}$-vertex) or unhappy (a $\mathtt{U}$- or $\mathtt{P_U}$-vertex). We have $\lvert X_t \rvert \leq w + 1$, which gives the above bound.

Many of those $\tau = [t, c', \tilde{\mathcal{H}}]$ are invalid: if two adjacent vertices are coloured differently, but one of them is a $\mathtt{H}$-vertex. Generally, we can not reduce above upper bound. If none of the $\mathcal{L}$ lists reaches its capacity $W$ however, the algorithm has constructed all valid $\tau$ tuples and thus returns an exact solution. This is detected through a simple check. When a list $\mathcal{L}$ is generated, the algorithm checks if $\lvert \mathcal{L} \rvert < W$. Our heuristic algorithm can guarantee an exact solution if this check succeeds for all $t \in V(T)$, even if $W < (2k)^{w+1}$.

Above we have discussed the time complexity of each individual procedure. Given a graph $G$ and a tree decomposition $\mathcal{T} = (T, \mathcal{X})$ of $G$ of width $w$. Denote the number of leaf nodes with $\#\mathtt{L}$, the number of introduce nodes with $\#\mathtt{I}$, the number of forget nodes with $\#\mathtt{F}$, and the number of join nodes with $\#\mathtt{J}$. The total time complexity of our heuristic algorithm is equal to
\begin{align}
    \mathcal{O}\Big(\#\mathtt{L} + \#\mathtt{I} kW (\log(W) + \Delta^2) + \#\mathtt{F} W\log(W) + \#\mathtt{J} W^2 w \Delta\Big)\text{.}
\end{align}

This time complexity does not include computing the tree decomposition. Unfortunately, constructing a tree decomposition of minimal width is $\mathcal{NP}$-hard. This is a drawback in every approach that uses tree decompositions. But if the tree decomposition is available for some graph $G$, then it can be used for solving the MHV problem on $G$ for multiple initial colour functions. 

\section{Experimental setup} \label{sec:experimental-setup}

\subsection{Constructing tree decompositions} \label{sec:constructing-td}

We used FlowCutter \citep{hamann2018graph} for constructing a tree decomposition of the graphs. This solver finished second in the PACE 2017 challenge in the heuristic track for computing tree decompositions. It was the only solver that could construct a tree decomposition for all instances within 30 minutes. The algorithm by \cite{tamaki2019positive} sometimes constructed a better tree decomposition for smaller instances, and won the challenge. However, we want to solve larger instances of the MHV problem, for which FlowCutter more reliably constructs a tree decomposition. The tree decomposition is converted into a nice tree decomposition, using the implementation of \cite{bannach2019practical}.

FlowCutter is not an integral part of our algorithm. We use it as a preprocessing step for constructing a tree decomposition of a graph, after which our algorithm takes this tree decomposition as argument. We execute FlowCutter only once for a given graph, and use the tree decomposition for multiple initial colourings and hyperparameter configurations. This allows us to replace FlowCutter with any other solver, in case better algorithms for constructing tree decompositions would become available in the future. FlowCutter required 2.5 seconds to construct a tree decomposition for graphs with 1000 vertices, and smaller graphs required less than a second. For these reasons we decided to not include FlowCutter's computing time in the runtime of our algorithm in the results shown in Section \ref{sec:results}.

\subsection{Generating Instances} \label{sec:problem-instances}

An instance of the MHV problem consists of two parts: a graph $G$ and partial colour function $c$. \cite{lewis2019finding} developed a generator which requires four parameters: the number of vertices $n$, the graph density $p$, the number of colours $k$ and the percent of the vertices to be coloured $q$. If $\lfloor q n \rfloor < k$, then no instance will be constructed because there are more colours than vertices to colour. Otherwise, a graph will be constructed according to the Erd{\H{o}}s-Rényi model \citep{erdHos1960evolution}: a graph with $n$ vertices, and every possible edge is included with fixed probability $p$. Next, the vertices of $G$ are shuffled, and the first $k$ vertices of this permutation are set to the colour 1 to $k$ respectively. The next $(qn - k)$ vertices are each assigned a random colour from $\{1,\dots,k\}$.

\cite{lewis2019finding} identified the difficult-to-solve instances of the MHV problem. They generated 72 000 instances with 1000 vertices, $k \in \{10, 50, 100, 250\}$, and varying $p$ and $q$, which they solved using an integer programming formulation and time limit of 600 seconds. The model was able to produce one solution for every graph within this time, but could not necessarily prove its optimality. From this experiment, \cite{lewis2019finding} concluded that instances become easier if $k$, $p$ and $q$ are large, because these instances have many vertices that are destined to be unhappy. The hardest instances of the MHV problem in their study have $p = 5/(n-1)$ and $q = 0.1$.  

For our first experiment, we manually generated difficult-to-solve instances using the generator of \cite{lewis2019finding}. The constructed graphs consisted of 30, 50, 100, 250, 500, 750 or 1000 vertices, and we chose $k \in \{3,10,30\}$. We also generated \textit{easier} instances, with $p \in \{0.05, 0.5, 0.9, 5/(n-1)\}$ and $q \in \{0.05, 0.1, 0.2, 0.3, 0.4, 0.5, 0.6, 0.7, 0.8, 0.9\}$. This leads to a total of 840 configurations. Only 716 of these configurations are valid, because an instance must satisfy $\lfloor q n \rfloor \geq k$. For each valid configuration 5 instances are generated, resulting in 3~580 MHV-instances.

On top of this, we also include the instances generated by \cite{ghirardi2021simple} for experimentation, which were also generated using the generator of \cite{lewis2019finding}. The difference with our instances is their size. \cite{ghirardi2021simple} set the number of colors $k$ to 10 and 50, the percentage of precoloured vertices $q = 0.1$, the number of vertices $n$ to 250, 500, 750, 1000, 2000, 3000, 4000, 5000, 7500 and 10000, and graph density $p = 5/(n-1)$. Notice that the combination $k=50$, $q=0.1$ and $n=250$ is invalid, because of the constraint $\lfloor q n \rfloor \geq k$. This results in 19 valid combinations, for which 20 instances were generated each, or 380 instances in total.

The House of Graphs database \citep{coolsaet2023house} (which can be accessed at \url{https://houseofgraphs.org/}) has a meta-directory containing complete lists of all graphs of a certain class, up to a limited number of vertices. We only considered connected graphs with at least 5 vertices. For each instance, 10\% of the vertices are precoloured using 3 colours. Graphs with $\lfloor 0.1 \cdot n \rfloor < 3$ result in invalid configurations, but in these cases we colour 3 random vertices in each of the 3 colours. House of Graphs also offers a database of \textit{interesting} graphs. There is no formal definition of \textit{interesting} graphs, but House of Graphs aims to find a workable definition of \textit{interesting}. Because researchers can add new \textit{interesting} graphs to the House of Graphs database, we mention the date on which the graphs were downloaded. Next we discuss the graph classes used for experimentation. 
\begin{enumerate}
    \item \textbf{Bipartite graphs.} A bipartite graph is a graph in which the vertices can be divided into two disjoint and independent sets. No full dataset of smaller graphs is available on House of Graphs, but 4~113 \textit{interesting} graphs are available (24 April 2022).
     
    \item \textbf{Claw-Free Graphs.} A claw is a graph with 4 vertices and 3 edges, connecting 1 vertex to all other vertices. A graph is claw-free if it does not have a claw as induced subgraph. House Of Graphs has 462 \textit{interesting} claw-free graphs available (22 April 2022). No full dataset of small claw-free graphs is available. 
    
    \item \textbf{Cubic Graphs.} A cubic graph is a graph where all vertices have degree 3. House of graphs has a dataset of all cubic graphs with up to 24 vertices. We used all cubic graphs having 6 to 14 vertices for experiments, and 1000 randomly selected cubic graphs with 16, 18, 20 and 22 vertices, resulting in a total of 4~620 graphs. Additionally, the 8~243 \textit{interesting} graphs have been used for experimentation (24 April 2022). 
    
    \item \textbf{Eulerian Graphs.} An Eulerian graph is a graph in which an Eulerian cycle exists, a cycle that visits every edge of the graph exactly once. No full dataset for small Eulerian graphs is available on House of Graphs. However, Brendan McKay made such dataset for Eulerian graphs with up to 12 vertices publicly available\footnote{\url{http://users.cecs.anu.edu.au/~bdm/data/graphs.html}}. All Eulerian graphs with 6 to 9 vertices were considered, as well as 3~000 randomly selected graphs with 10 and 11 vertices, totalling in 8~011 graphs. Additionally 498 \textit{interesting} Eulerian graphs were available in the House of Graphs database (24 April 2022). 
    
    \item \textbf{Non-Hamiltonian Graphs}. A Hamiltonian cycle of a graph visits all vertices exactly once. A non-Hamiltonian graph does not contain such a cycle. No dataset for small non-Hamiltonian graphs is available on House of Graphs, but there are 7~459 \textit{interesting} graphs available (24 April 2022).
    
    \item \textbf{Planar Graphs.} A planar graph can be embedded in the plane with the edges only intersecting in the vertices. The House of Graphs database contains all connected planar graphs up to 11 vertices. All planar graphs with 6 or 7 vertices were used, and 2~000 randomly selected graphs with 8, 9, 10 and 11 vertices. This results in 8~745 graphs. In addition, the 3~468 \textit{interesting} connected planar graphs available on House of Graphs were used for experimentation (22 April 2022). 
    
    \item \textbf{Trees.} House of Graphs contains all trees with up to 20 vertices. All trees with at most 11 vertices were used and 500 random trees with 12 to 20 vertices, giving a total of 4~928 trees. Additionally 259 \textit{interesting} trees have been used for experimentation (22 April 2022). 
\end{enumerate}

We will denote the set of graphs taken from a complete list of graphs as \texttt{small}-graphs. The interesting graphs available on House of Graphs are denoted by \texttt{interesting}-graphs. Table \ref{tab:graph_classes} gives an overview of the structure of the different graph classes.

We made all instances, including the manually generated Erd{\H{o}}s-Rényi graphs, the Erd{\H{o}}s-Rényi graphs used in \cite{ghirardi2021simple}\footnote{We thank Prof. Marco Ghirardi for providing us the instances.}, and the graphs of certain classes from Table~\ref{tab:graph_classes}, publicly available on \url{https://github.com/LouisCarpentier42/HeuristicAlgorithmsUsingTreeDecompositions/tree/main/instances}.

\begin{table}
    \centering
    \caption[Overview of the sizes of the graphs of different classes.]{Overview of the sizes of the graphs of different classes, and the width of the used tree decompositions for these graphs. The \texttt{small} and \texttt{interesting} graphs are separated if both types were used for the class, otherwise only \texttt{interesting} graphs were used. The last two columns refer to the width of the tree decomposition constructed by FlowCutter.}
    \label{tab:graph_classes}
    \begin{tabular}{ll *{5}{c}}
        \toprule
        
        & & \textbf{\#graphs} & \textbf{\shortstack{average\\\#vertices}} & \textbf{\shortstack{max\\\#vertices}} & \textbf{\shortstack{average\\width}} & \textbf{\shortstack{max\\width}} \\
        \midrule
        
        \multicolumn{2}{l}{\textbf{Bipartite}}
        & 4 113 & 143.95 & 250 &  17.41 & 104 \\
        \midrule
        
        \multicolumn{2}{l}{\textbf{Claw-Free}}
        & 462 & 17.66 & 243 & 6.73 & 128 \\
        \midrule
        
        \multirow{2}{*}{\textbf{Cubic}}
        & {\ttfamily\fontseries{b}\selectfont interesting} & 8 243 & 96.47 & 250 & 12.26 & 54 \\
        & {\ttfamily\fontseries{b}\selectfont small} & 4 620 & 18.27 & 22 & 4.30 & 6 \\
        \midrule
        
        \multirow{2}{*}{\textbf{Eulerian}}
        & {\ttfamily\fontseries{b}\selectfont interesting} & 498 & 23.88 & 208 & 11.46 & 156 \\
        & {\ttfamily\fontseries{b}\selectfont small} & 8 011 & 10.09 & 11 & 4.47 & 8 \\
        \midrule
        
        \multicolumn{2}{l}{\textbf{Non-Hamiltonian}}
        & 7 459 & 29.07 & 231 & 4.84 & 9 \\
        \midrule
        
        \multirow{2}{*}{\textbf{Planar}}
        & {\ttfamily\fontseries{b}\selectfont interesting} & 3 468 & 27.45 & 231 & 4.53 & 14 \\
        & {\ttfamily\fontseries{b}\selectfont small} & 8 745 & 9.28 & 11 & 2.94 & 4 \\
        \midrule
        
        \multirow{2}{*}{\textbf{Trees}}
        & {\ttfamily\fontseries{b}\selectfont interesting} & 259 & 21.94 & 181 & 1 & 1  \\
        & {\ttfamily\fontseries{b}\selectfont small} & 4 928 & 15.50 & 20 & 1 & 1  \\
        \bottomrule
    \end{tabular}
\end{table}

\subsection{Tuning the hyperparameters}

Our heuristic algorithm uses several hyperparameters, which were tuned using SMAC (Sequential Model-based Algorithm Configuration) \citep{hutter2011sequential}. Given a parameterised algorithm $A$ with parameters $\theta_1, \dots, \theta_n$ and corresponding domains $\Theta_1, \dots, \Theta_n$, a set of instances $S$ and a performance metric $m$, SMAC searches for a parameter configuration $\langle \theta_1, \dots, \theta_n \rangle \in \Theta_1 \times \dots \times \Theta_n$ that minimises the performance $m$ on instances $S$. 

The generator of \cite{lewis2019finding} was used for constructing $S$, with $n \in \{50, 100, 250, 500, 750, 1000\}$, $p = 5/(n-1)$, $k \in \{3, 10, 50\}$ and $q \in \{0.05, 0.1, 0.25, 0.5, 0.75\}$. This gives a total of 91 valid configurations (recall that $\lfloor q n \rfloor < k$ is forbidden). For each configuration we generated 15 train instances and 1 test instance. We chose performance metric $m$ as 1 minus the percent of vertices that are happy, because SMAC minimises the objective function. Using the total number of happy vertices adds a bias towards larger graphs. We permitted SMAC to run for 20 hours total, and each configuration had a maximum runtime of 150 seconds. 

We must specify the domains of each parameter before running SMAC. For the categorical parameters in the join node procedure, we can simply enumerate all possible values. The domains of integer parameters must be manually constrained. Denote $W_\mathtt{X}$ as the weight for a $\mathtt{X}$-vertex. We restricted the domains of the label weights as follows:
\begin{align}
    W_\mathtt{H}, W_\mathtt{P_H} \in [-5,20] \quad \text{and} \quad W_\mathtt{U}, W_\mathtt{P_U} \in [-10,10] 
\end{align}
This ensures that a priori happy vertices have a higher weight than unhappy vertices, reducing the search space for SMAC. We additionally added the following constraints: 
\begin{align}
    W_\mathtt{H} \geq W_\mathtt{P_H} \quad \text{and} \quad 
    W_\mathtt{H} \geq W_\mathtt{P_U} \quad \text{and} \quad 
    W_\mathtt{U} \leq W_\mathtt{P_H} \quad \text{and} \quad 
    W_\mathtt{U} \leq W_\mathtt{P_U}
\end{align}
Lastly we restricted the domain of $W$ to $[1, 2048]$, with a logarithmic scale because for larger values of $W$ a small deviation will have less influence. The tuned parameters are shown in Table \ref{tab:tuned_parameters}. These are used throughout the experiments in Section \ref{sec:results}, with the exception of $W$, which we vary in order to analyse its influence on the algorithm. 

\begin{table}
    \centering
    \caption{An overview of the tuned hyperparameters using SMAC.}
    \label{tab:tuned_parameters}
    \begin{tabular}{p{0.45\textwidth}p{0.45\textwidth}}
        \toprule
        \multicolumn{1}{c}{\textbf{Hyperparameter}} & \multicolumn{1}{c}{\textbf{Tuned value}} \\
        \midrule
        
        $W$ & 67 \\ 
        \midrule
        $\mathtt{H}$ & 15 \\ 
        \midrule
        $W_\mathtt{U}$ & -9 \\ 
        \midrule
        $W_\mathtt{P_H}$ & 4 \\ 
        \midrule 
        $W_\mathtt{P_U}$ & -8 \\ 
        \midrule
        Deciding which $\mathcal{L}$ is used in the outer loop in Algorithm \ref{alg:heuristic_join} & Use the $\mathcal{L}$ with fewest partial solutions in the outer loop  \\ 
        \midrule
        The weights for distance computation in Algorithm \ref{alg:heuristic_join} & The number of neighbours that are not in the bag \\ 
        \midrule
        Procedure to heuristically combine two partial solutions in Algorithm \ref{alg:heuristic_join} & Copy all the decisions of the vertices in the bag \\
        
        \bottomrule
    \end{tabular}
\end{table}

\subsection{Correctness} \label{sec:practical_correctness}

Our heuristic algorithm is complex and bugs can easily sneak in unnoticed. We implemented the following sanity checks to ensure the correctness of our implementation.
\begin{enumerate}
    \item The final colouring may only consist of $\mathtt{H}$- and $\mathtt{U}$-vertices, which must effectively be happy and unhappy, respectively. 
    \item If our algorithm indicates having found an exact solution, then this solution must effectively be optimal. We verified this by comparing our results to those of the exact algorithm for graphs of bounded treewidth. 
    \item We verified that our algorithm optimally solves all instances satisfying Equation \ref{eq:exactness_guarantee}.
\end{enumerate}

We used the tuned hyperparameters from Table \ref{tab:tuned_parameters}, but instead set $W~=~10~000$, to ensure that many instances were solved exactly. Initial colouring $c$ colours $10\%$ of the vertices, using $k=3$ colours. The dataset used for verifying these sanity checks is a subset of the instances of certain graph classes described above in Section \ref{sec:problem-instances}. We included all \texttt{small}-graphs. Larger, \texttt{interesting}-graphs require a lot of resources to solve exactly, thus we only considered a subset of these. Using Equation \ref{eq:exactness_guarantee}, our algorithm guarantees an optimal solution if $w+1 < \log_{2\cdot3}10000 = 5.14$. We included all \texttt{interesting}-graphs with $w < 9$, because Equation \ref{eq:exactness_guarantee} is an upper bound. This bound on the width of the tree decomposition is also advantageous for the exact algorithm for graphs of bounded treewidth. In total, we use 43~507 instances for the sanity checks. 

We implemented the exact algorithm for graphs of bounded treewidth ourselves, because \cite{agrawal2017parameterized} and \cite{agrawal2020parameterized} did not provide any implementation. We empirically show that our implementation is correct by comparing it to a naive brute force algorithm which iterates over all possible colour functions. Both algorithms were tested on 5~000 random instances with graphs of degree 2 up to 9. The algorithms returned equal number of happy vertices for every instance, which increases our confidence of a correct implementation.

\section{Results} \label{sec:results}

Below we discuss the performance of our algorithm. All experiments used the tuned parameters in Table \ref{tab:tuned_parameters}, but $W$ was varied depending on the experiment. The experiments were conducted on the Genius cluster of the Flemish Supercomputer Center (VSC), using Xeon Gold 6140 CPUs with 5GB RAM. Executing all the experiments in this work required more than 2000 CPU-hours. An overview of the numerical results is given in Table \ref{tab:results}.

First we clarify the results shown in the figures throughout this section. The likelihood of finding an exact solution is computed as the percentage of instances of which our algorithm proved optimality. We measure the quality of a solution as the percentage of vertices that are happy, rather than the total number of happy vertices to reduce bias towards large graphs. We average this percentage over all instances with a certain property (e.g., the number of vertices) to retrieve a final quality measure. Similarly the running time is averaged over all algorithm executions. Lastly, we average the width and number of nodes in the tree decomposition when visualising the size of the tree decomposition in function of the number of vertices of the graph. 

\begin{sidewaystable}
    \small
    \centering
    \caption{The numerical results of our heuristic algorithm using tuned hyperparameters compared to Greedy-MHV and Growth-MHV. We show the average percent of happy vertices and average running time in milliseconds, and the percentage of instances our algorithm solved exactly. The last two columns show the percentage of instances for which the quality of the solution constructed by our algorithm is strictly better or equivalent to the best solution of Greedy-MHV and Growth-MHV.}
    \label{tab:results}
    \begin{tabular}{ll *{9}{c}}
        \toprule
        & & \multicolumn{3}{c}{\textbf{Our Algorithm}} & \multicolumn{2}{c}{\textbf{Greedy-MHV}} & \multicolumn{2}{c}{\textbf{Growth-MHV}} & & \\
        & & \textbf{quality} & \textbf{time ms} & \textbf{exact} & \textbf{quality} & \textbf{time ms} & \textbf{quality} & \textbf{time ms} & \textbf{Improved} & \textbf{Equal} \\
        \midrule
        
        \multicolumn{2}{l}{\textbf{Sanity Checks}}
        & 0.58 & 19902.24 & 84.48\% & / & / & / & / & / & / \\
        \midrule
        
        \multicolumn{2}{l}{\textbf{Erd{\H{o}}s-Rényi}}
        & 0.09 & 11817.98 & 3.61\% & 0.12 & 0.631 & 0.12 & 63.357 & 4.99\% & 68.19\% \\
        \midrule
        
        \multicolumn{2}{l}{\textbf{Bipartite}}
        & 0.6 & 194.27 & 8.75\% & 0.78 & 0.008 & 0.7 & 0.471 & 5.88\% & 8.87\% \\
        \midrule
        
        \multicolumn{2}{l}{\textbf{Claw-Free}}
        & 0.24 & 8.8 & 29.65\% & 0.24 & 0.002 & 0.24 & 0.055 & 1.52\% & 91.99\% \\
        \midrule
        
        \multirow{2}{*}{\textbf{Cubic}}
        & {\ttfamily\fontseries{b}\selectfont interesting} & 0.65 & 116.43 & 0.1\% & 0.78 & 0.005 & 0.72 & 0.309 & 11.15\% & 8.55\% \\
        & {\ttfamily\fontseries{b}\selectfont small} & 0.62 & 7.21 & 0.35\% & 0.64 & 0.001 & 0.62 & 0.049 & 4.81\% & 74.37\% \\
        \midrule
        
        \multirow{2}{*}{\textbf{Eulerian}}
        & {\ttfamily\fontseries{b}\selectfont interesting} & 0.39 & 31.32 & 12.25\% & 0.39 & 0.003 & 0.38 & 0.16 & 2.41\% & 92.37\% \\
        & {\ttfamily\fontseries{b}\selectfont small} & 0.29 & 1.21 & 20.42\% & 0.28 & 0.001 & 0.27 & 0.017 & 1.35\% & 98.65\% \\
        \midrule
        
        \multicolumn{2}{l}{\textbf{Non-Hamiltonian}}
        & 0.67 & 17.09 & 9.18\% & 0.71 & 0.002 & 0.67 & 0.105 & 4.5\% & 57.82\% \\
        \midrule
        
        \multirow{2}{*}{\textbf{Planar}}
        & {\ttfamily\fontseries{b}\selectfont interesting} & 0.6 & 14.75 & 24.45\% & 0.63 & 0.002 & 0.6 & 0.1 & 5.25\% & 70.36\% \\
        & {\ttfamily\fontseries{b}\selectfont small} & 0.4 & 0.48 & 69.07\% & 0.38 & 0.001 & 0.38 & 0.016 & 3.82\% & 96.18\% \\
        \midrule
        
        \multirow{2}{*}{\textbf{Trees}}
        & {\ttfamily\fontseries{b}\selectfont interesting} & 0.76 & 0.79 & 100.0\% & 0.68 & 0.001 & 0.75 & 0.043 & 26.25\% & 73.75\% \\
        & {\ttfamily\fontseries{b}\selectfont small} & 0.8 & 0.46 & 100.0\% & 0.71 & 0.001 & 0.78 & 0.029 & 28.35\% & 71.65\% \\
        \bottomrule
    \end{tabular}
\end{sidewaystable}

\subsection{Sanity Checks} \label{sec:sanity_checks}

\begin{figure}
     \centering
     \includegraphics[width=0.48\textwidth]{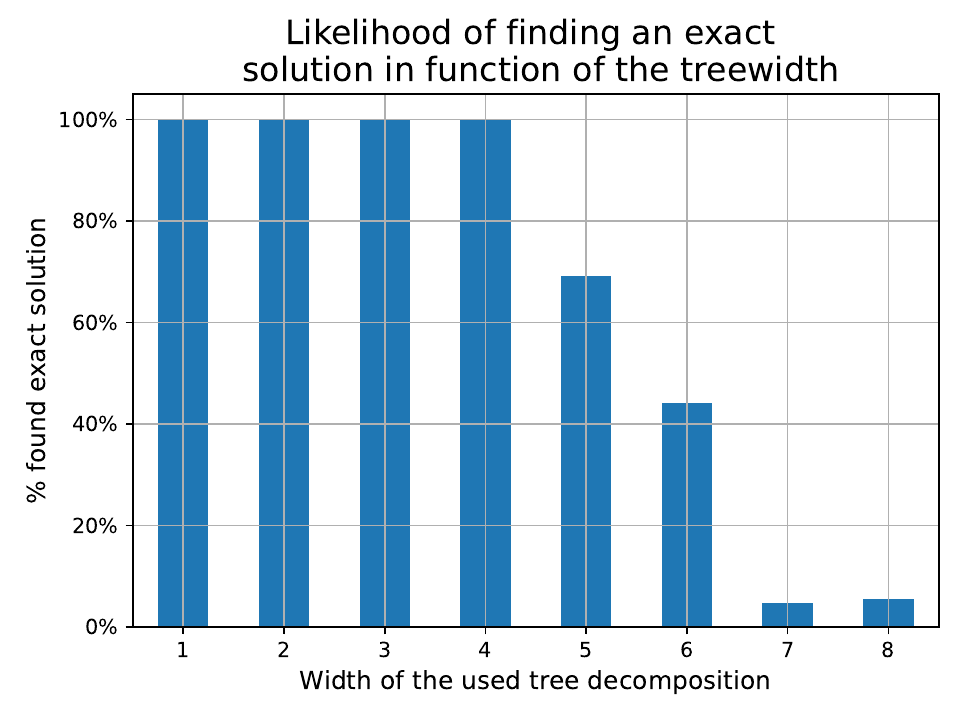}
     \\
     \includegraphics[width=0.48\textwidth]{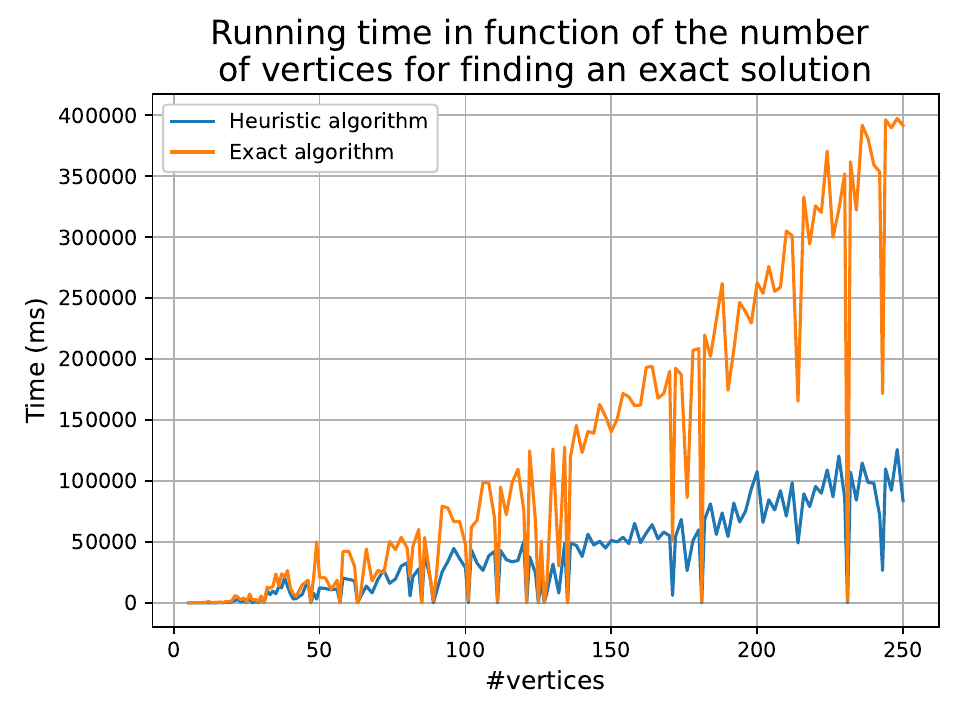}
     \hfil
     \includegraphics[width=0.48\textwidth]{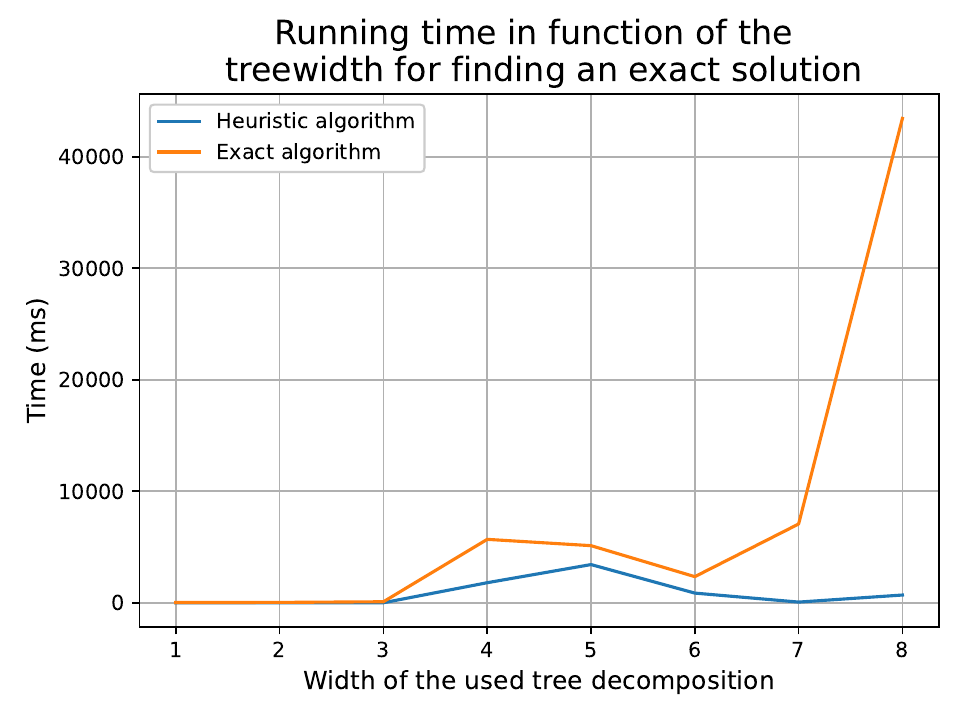}
    \caption{The performance of our algorithm compared to the exact algorithm for graphs of bounded treewidth by \cite{agrawal2017parameterized}. The comparisons at the bottom only include the results for graphs that have been solved exactly by our algorithm.}
    \label{fig:sanity_checks}
\end{figure}

We discussed three different Sanity Checks in Subsection \ref{sec:practical_correctness}, which all succeeded on all 43~507 instances. The first sanity check verifies that the labelling is correctly maintained by the algorithm, and applies to all instances. However, the second check only applies to instances for which our algorithm indicated to have found an exact solution. In total 36~754 instances, or $84.48\%$, have been solved exactly by our algorithm. The solution of the remaining 6~753 instances were not proven to be exact, but could still be optimal. The results are visualised in Fig. \ref{fig:sanity_checks}.

A large number of instances were solved exactly, including instances which do not satisfy Equation \ref{eq:exactness_guarantee}. Due to this equation, all instances with $w + 1 < 5.14$ or $w \leq 4$ are solved exactly using $W~=~10~000$. Additionally, instances using a tree decomposition with larger width were solved optimally because only valid partial solutions are considered. As the width of the tree decomposition increases, more partial solutions are possible, and the likelihood of finding an exact solution decreases.

In Fig. \ref{fig:sanity_checks}, we show the running time of our heuristic algorithm and the exact algorithm by \cite{agrawal2017parameterized}, only for the 36~754 instances that have been solved exactly by our algorithm. The running time is similar for smaller graphs, but our algorithm outperforms the exact algorithm for larger graphs or when using a tree decomposition with larger width. Our algorithm only considers valid partial solutions, while the exact algorithm iterates over all possible (valid and invalid) partial solutions. Our heuristic algorithm thus has two advantages over the exact algorithm for graphs of bounded treewidth. Firstly, it has a smaller running time for solving instances exactly. And secondly, we can choose to make our algorithm heuristic to further decrease the running time.

\subsection{Performance on Erd{\H{o}}s-Rényi Graphs} \label{sec:random_graphs}

The discussion on the results on Erd{\H{o}}s-Rényi graphs are split in two parts. First, we describe the results on the manually generated instances in Section~\ref{sec:small-erdos-renyi}, and compare to Greedy-MHV and Growth-MHV. This provides a first analysis of the algorithm behavior. Next, our heuristic is validated on the larger instances generated by \cite{ghirardi2021simple} in Section~\ref{sec:large-erdos-renyi}, allowing to quantify the scalability. For this experiment, we compare to the matheuristic by \cite{ghirardi2021simple}\footnote{We thank Prof. Marco Ghirardi for providing us with the source code of the matheuristic.}. The first experiment is referred to as \textit{small} while the second is referred to as \textit{large}.

\subsubsection{Small Erd{\H{o}}s-Rényi Graphs} \label{sec:small-erdos-renyi}

\begin{figure}[h]
    \centering
    \includegraphics[width=0.48\textwidth]{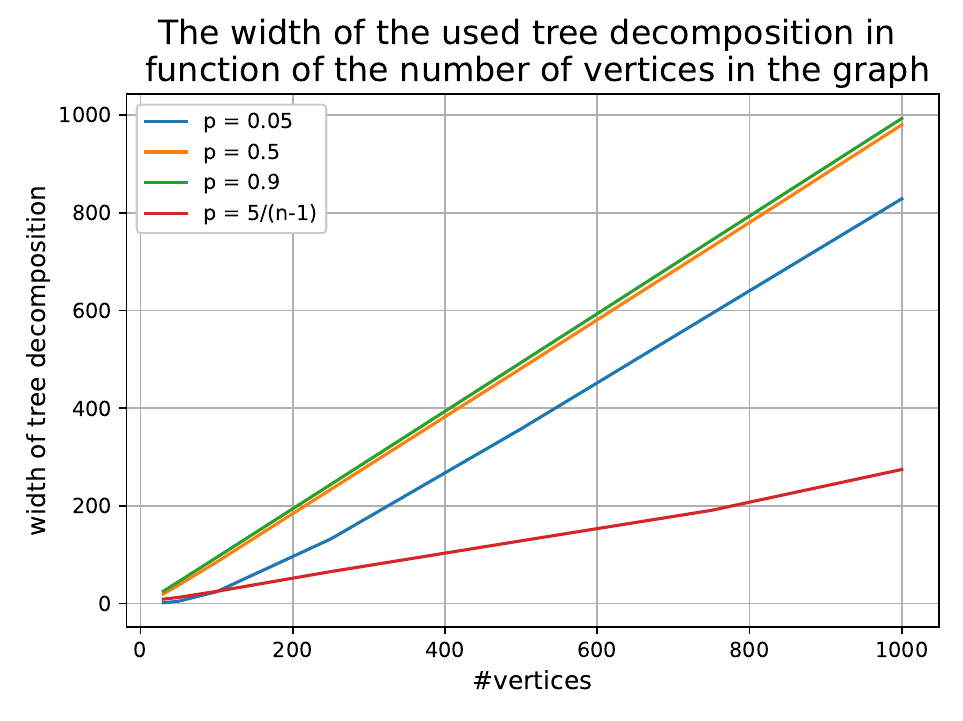} 
    \hfil
    \includegraphics[width=0.48\textwidth]{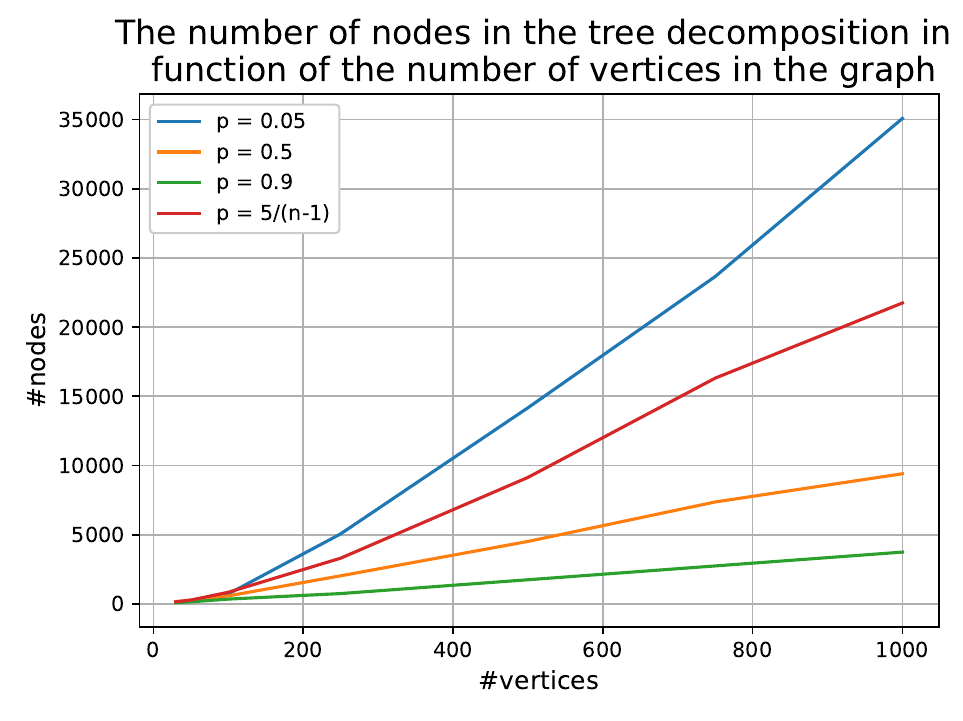} 
    \caption{The size of the tree decomposition of small Erd{\H{o}}s-Rényi graphs.}
    \label{fig:er_data}
\end{figure}

We generated 3~580 MHV instances using the Erd{\H{o}}s-Rényi model. We first analyse the size of the tree decompositions constructed by FlowCutter in Fig.~\ref{fig:er_data}. A tree decomposition converts a graph into a tree-like structure. The more a graph resembles a tree, the smaller the width of this structure. A tree has a low density $p$, i.e., it has only few edges. For that reason we separate the data over the different densities used for constructing graphs. Recall that density $p = 5/(n-1)$ of graph $G$ with $n = \lvert V(G) \rvert$ results in the most difficult instances~\citep{lewis2019finding}.

Given a graph $G$ and tree decomposition $\mathcal{T} = (T, \mathcal{X})$ of $G$ of width $w$, $w$ and $\lvert V(G) \rvert$ are positively correlated. Adjacent vertices must both end up in the same bag of the tree decomposition. If many vertices are connected, if $G$ is dense, possibly all vertices end up in the same bag, resulting in a large width. On the other hand, a larger density implies fewer nodes in the tree decomposition. If, for some vertex $v \in V(G)$, there exists a node $t \in V(T)$ such that $v \in X_t$ and $N(v) \subseteq X_t$, then $v$ does not need to be in any other bag. If this happens for many vertices, then all these vertices can be contained into a single bag, reducing the total number of nodes. If only few neighbours of $v$ are not in $X_t$, then $v$ needs to share a bag with only few other vertices. 

We have solved all 3~580 instances using multiple values of $W$. For large $W$, graphs with many vertices and large treewidth require a lot of computational resources to solve. For that reason we chose $W \in \{8, 16, 32, 64, 128, 256, 512, 1024, 2048\}$ for graphs with at most 250 vertices, $W \in \{8, 16, 32, 64, 128, 256, 512, 1024\}$ for graphs with the number of vertices between 250  and 500, and $W \in \{8, 16, 32, 64, 128, 256\}$ for graphs with more than 500 vertices. We additionally solved every instance using both Greedy-MHV and Growth-MHV.

\begin{figure}[h]
    \centering
    \includegraphics[width=0.48\textwidth]{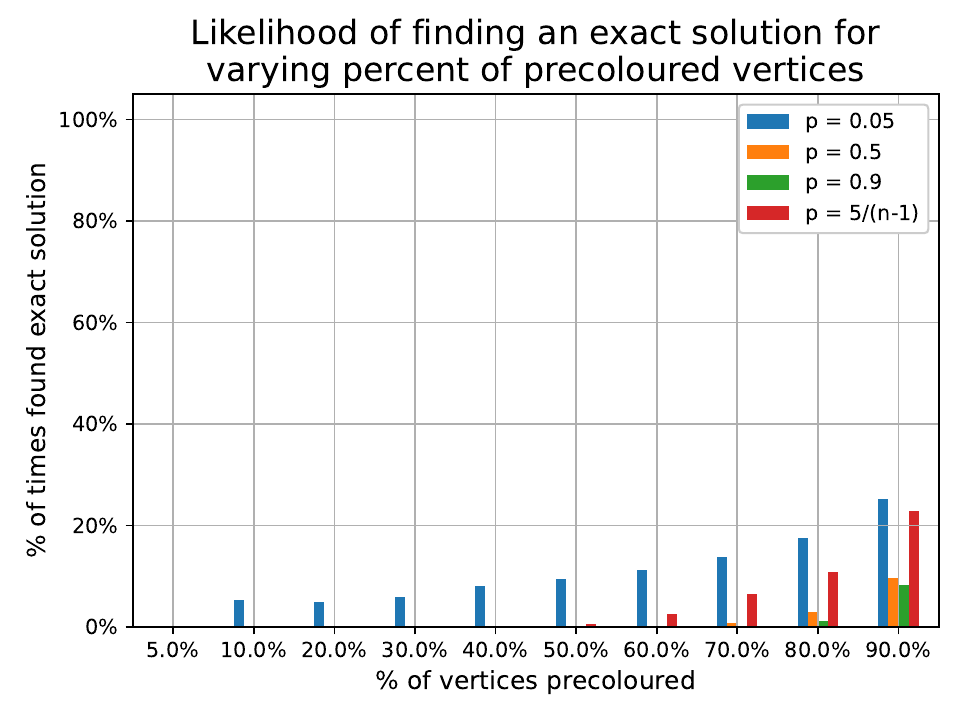}
    \hfil
    \includegraphics[width=0.48\textwidth]{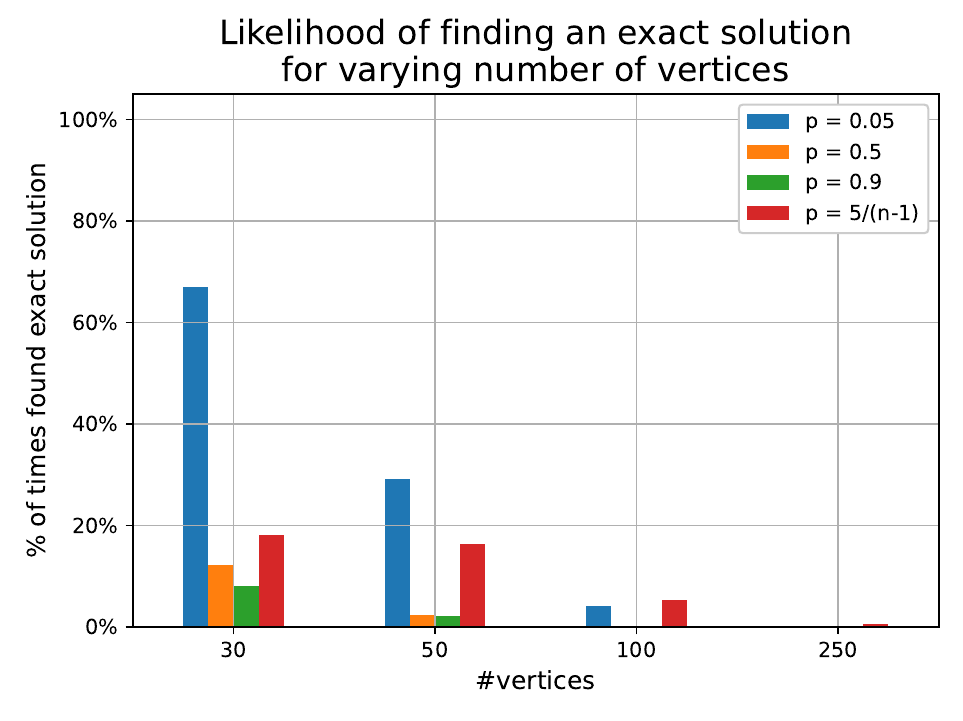}
    \caption{The probability that our heuristic algorithm proves optimality of the constructed solution for small Erd{\H{o}}s-Rényi graphs.}
    \label{fig:er_exactness}
\end{figure}

Our algorithm solved 28~040 instances using varying values of $W$. For 1~304 runs optimality was proven. Fig.~\ref{fig:er_exactness} shows a more detailed overview of which instances were solved optimally. We only show the results for graphs with at most 250 vertices, in order not to introduce a bias towards larger values of $W$. Most optimally solved instances consist of graphs with fewer vertices, low density, and many initially coloured vertices. This mostly corresponds to the easy-to-solve instances discussed by \cite{lewis2019finding}, because such instances have fewer valid tuples to consider. However, \cite{lewis2019finding} discusses that a lower density results in more difficult instances, since the happiness of many vertices is not yet fixed. Sparser graphs are more tree-like, and tend to have a smaller treewidth, for which our algorithm is more appropriate due to Equation~\ref{eq:exactness_guarantee}. Greedy-MHV and Growth-MHV are not able to exploit the sparseness of such instances. 

\begin{figure} [h]
    \centering
    \includegraphics[width=0.48\textwidth]{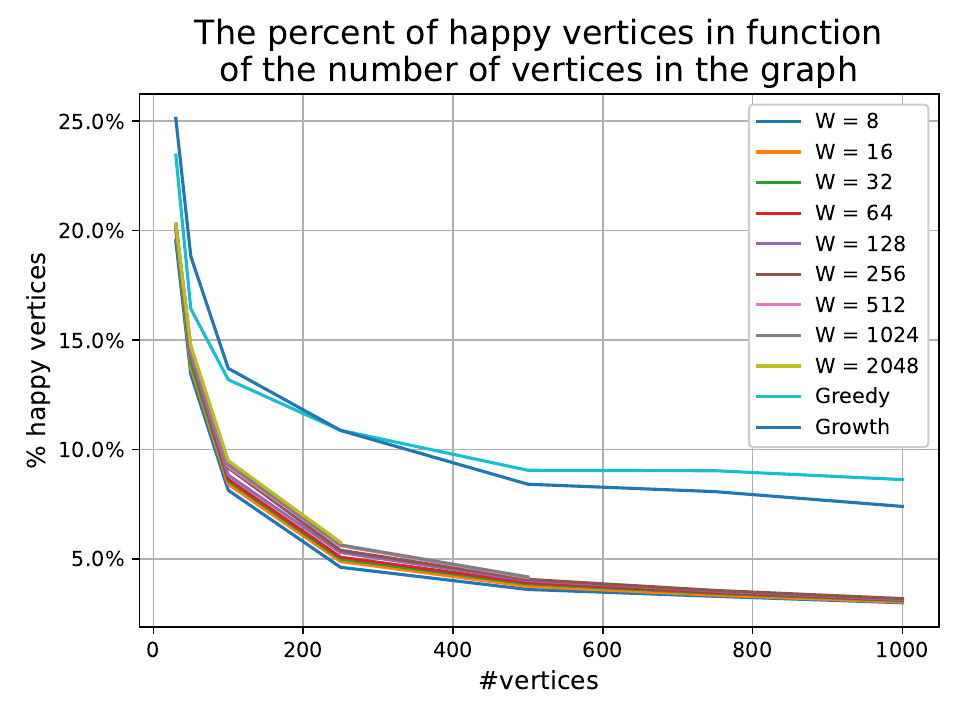} 
    \hfil
    \includegraphics[width=0.48\textwidth]{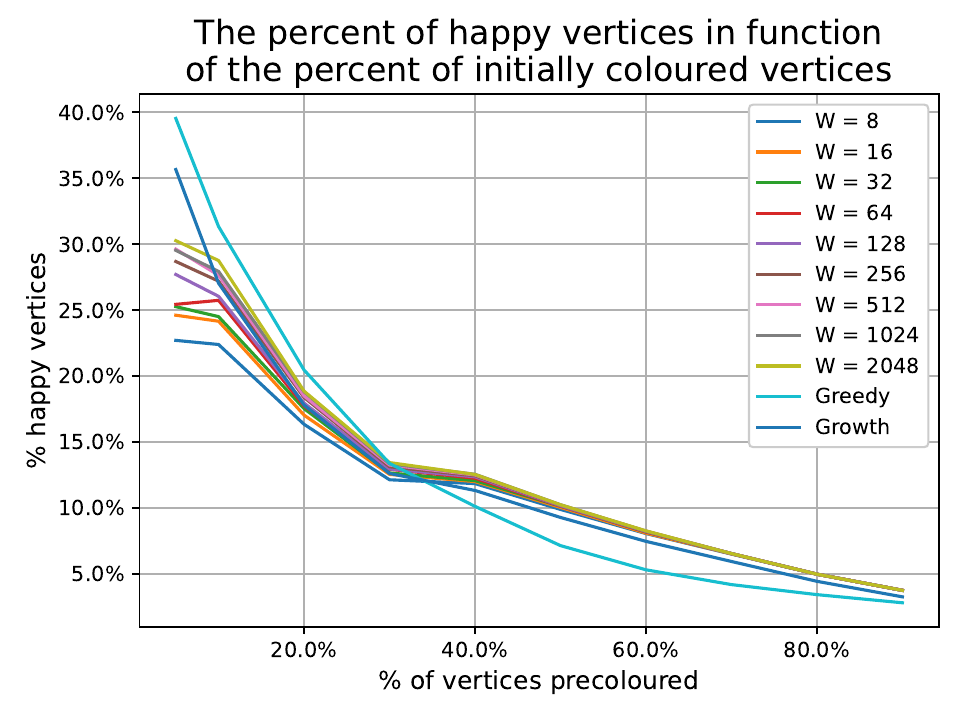} 
    \caption{The quality of the final solution of our heuristic algorithm compared to Greedy-MHV and Growth-MHV for small Erd{\H{o}}s-Rényi graphs.}
    \label{fig:er_quality}
\end{figure}

Fig. \ref{fig:er_quality} shows the quality of the colouring constructed by our algorithm, Greedy-MHV, and Growth-MHV. Our algorithm constructs higher quality solutions if at least 40\% of the vertices are initially coloured. Such instances have many vertices that are destined to be unhappy. A fine grained colouring is necessary in order to find those vertices that can be happy, which our algorithm constructs. More initially coloured vertices results in fewer valid tuples, which in turn indicates that varying $W$ has a smaller influence, and explains why the curves of our algorithm converge. 

Our algorithm does not scale well to larger instances. Large Erd{\H{o}}s-Rényi Graphs tend to have a large treewidth, which implies a huge number of tuples for every node in the tree decomposition. Maintaining 2~048 tuples (the largest $W$ used in our experiment) is only a fraction of the complete space of possible tuples. This also explains why varying $W$ has a small influence on the quality for larger instances. 

\begin{figure} [h]
    \centering
    \includegraphics[width=0.48\textwidth]{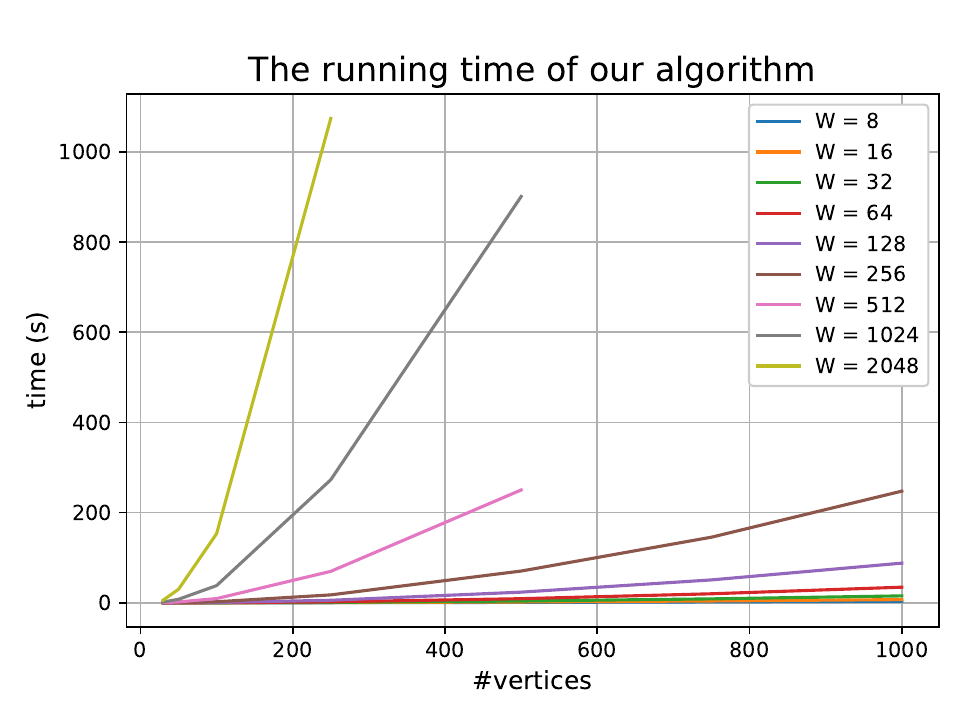}
    \hfil
    \includegraphics[width=0.48\textwidth]{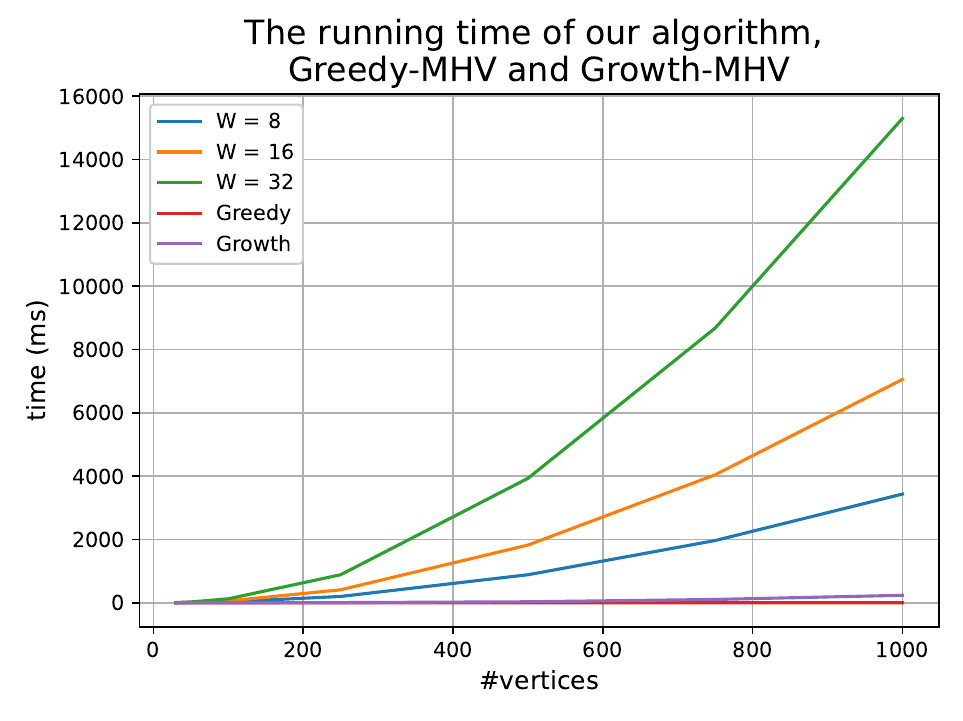}
    \hfil
    \includegraphics[width=0.48\textwidth]{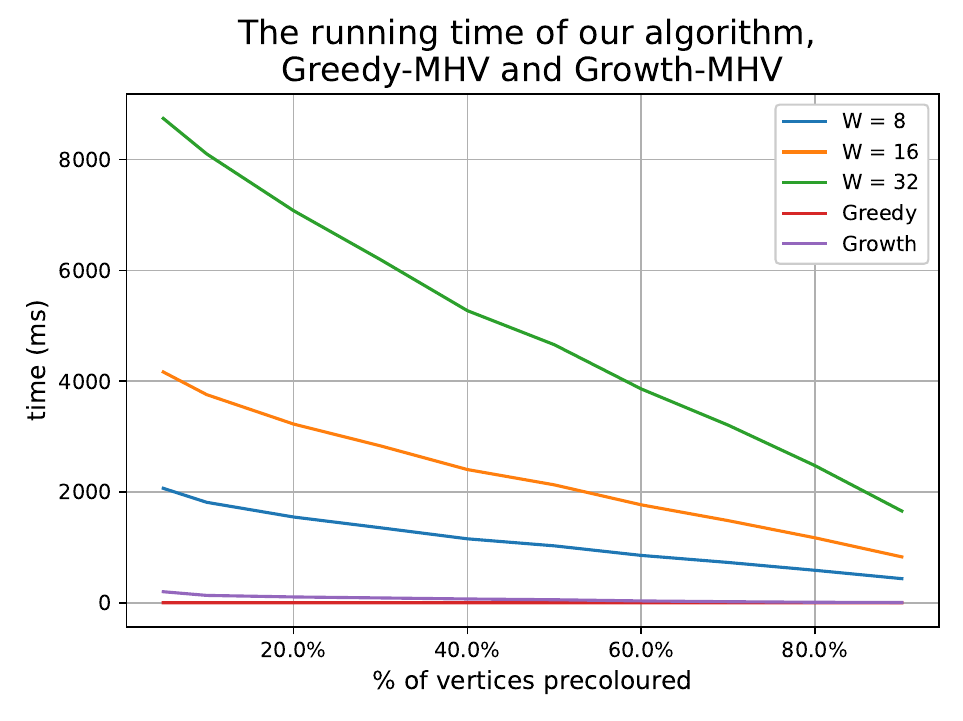}
    \hfil
    \includegraphics[width=0.48\textwidth]{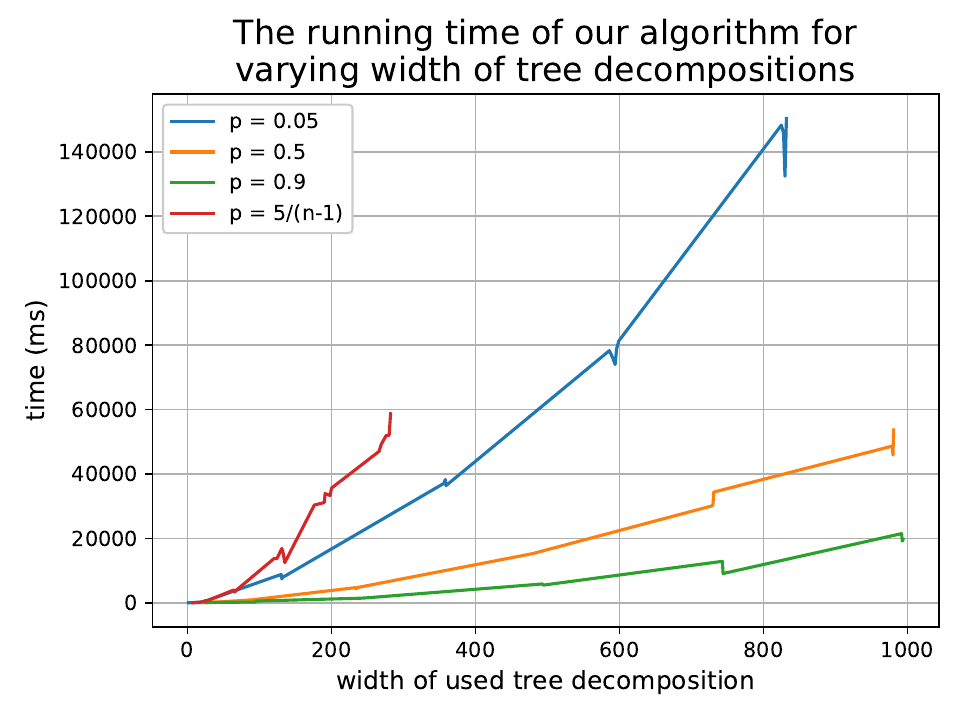}
    \caption{The running time of our heuristic algorithm, Greedy-MHV and Growth-MHV for small Erd{\H{o}}s-Rényi graphs.}
    \label{fig:er_time}
\end{figure}

Fig. \ref{fig:er_time} shows the running time of our heuristic algorithm. It is interesting to see that the running time decreases as the number of initially coloured vertices increases. These are exactly the instances for which our algorithm is more appropriate, as discussed above. Our algorithm has to consider fewer tuples in every node because fewer valid tuples exist.

Unfortunately, it requires a lot of computational resources, especially for larger $W$. The reason is that our algorithm performs a reasonably complex operation at every node of the tree decomposition. Exact algorithms for graphs of bounded treewidth also suffer from this problem. While having nice theoretical properties, a polynomial complexity in the number of vertices but exponential in the treewidth, they are time consuming in practice due to large hidden constants in the big-O notation.

\subsubsection{Large Erd{\H{o}}s-Rényi Graphs} \label{sec:large-erdos-renyi}

\cite{ghirardi2021simple} generated 380 instances of the MHV problem, with graphs containing up to 10~000 vertices. However, the implementation of \cite{bannach2019practical}, used to convert a tree decomposition into a nice tree decomposition, requires a lot of memory for larger graphs. We limited the RAM usage to 60GB. While this is more memory than available in a standard desktop, we set the limit to 60GB to assess algorithm performance on larger instances. All 40 graphs with 10~000 vertices, 36 graphs with 7~500 vertices, and 5 graphs with 5~000 vertices exceeded this memory limit. Our heuristic algorithm was executed on the remaining 299 instances. We used tuned hyperparameters as shown in Table~\ref{tab:tuned_parameters} and chose $W \in \{8, 16, 32, 64, 128, 265\}$ to measure its effect. This resulted in 1794 algorithm executions. 

\begin{figure}[h]
    \centering
    \includegraphics[width=0.48\textwidth]{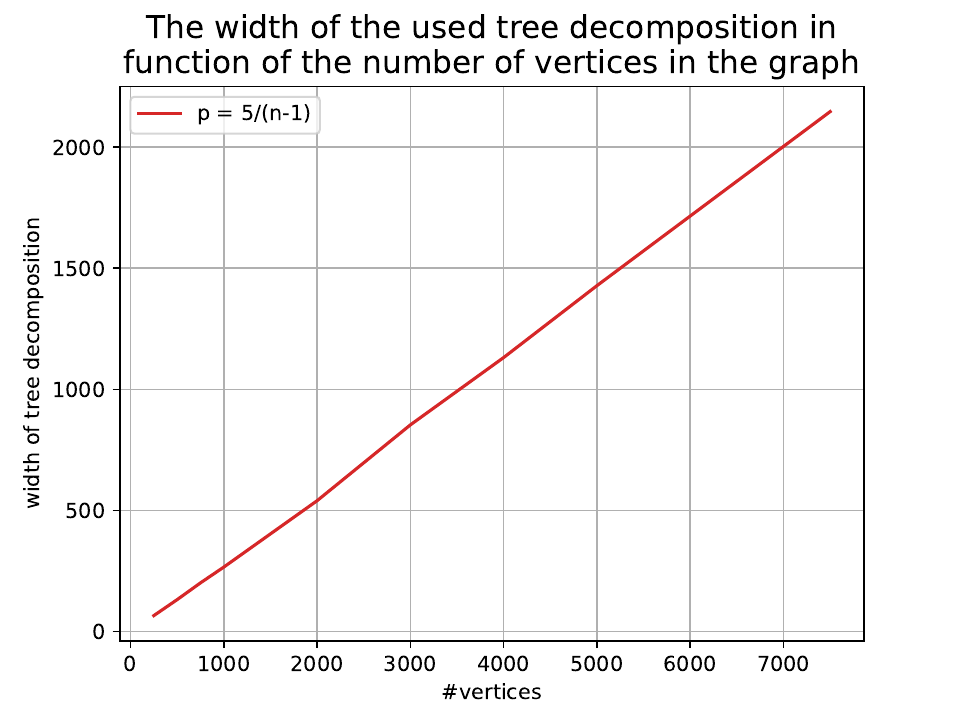} 
    \hfil
    \includegraphics[width=0.48\textwidth]{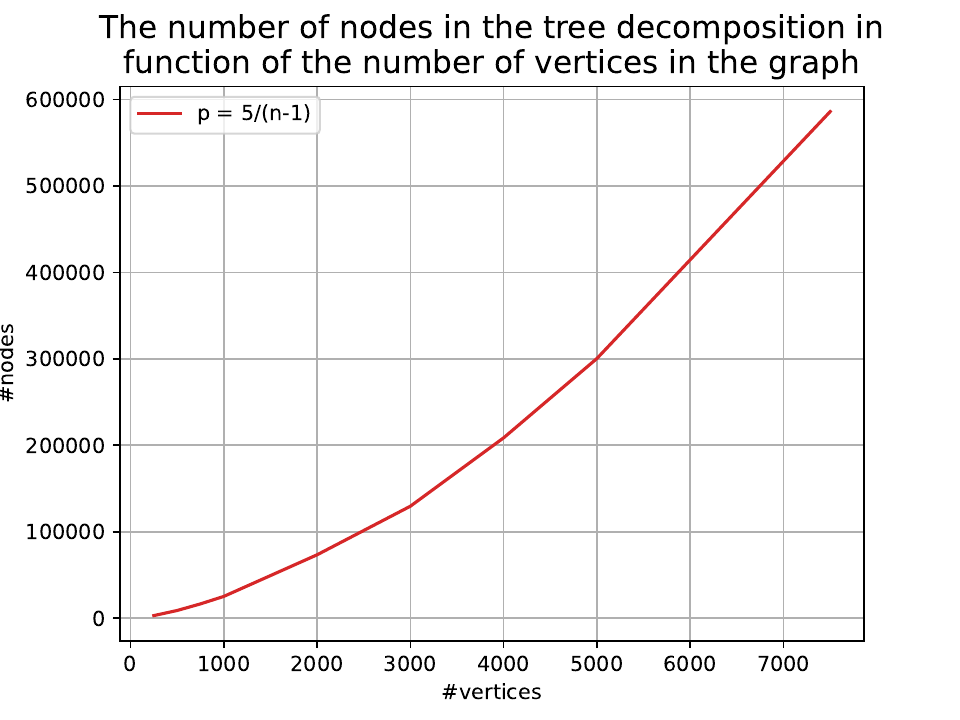} 
    \caption{The size of the tree decomposition of large Erd{\H{o}}s-Rényi graphs.}
    \label{fig:er_large_data}
\end{figure}

Fig.~\ref{fig:er_large_data} shows the size of the tree decomposition in function of the size of the graphs, including the graphs for which a nice tree decomposition could be constructed. The figures indicate that the conclusions from the previous section also hold for larger graphs. The size of the tree decomposition (both the treewidth and number of nodes) is positively correlated with the number of vertices in the graph. This indicates that the running time will also follow a similar trend as the previous section. To limited the required resources, we limit execution time to 8 hours. Out of the 1794 algorithm executions, 1701 were successfully completed. In the results below, we only include the graphs for which at least one configuration was completed, for both our algorithm and the matheuristic of \cite{ghirardi2021simple}.

\begin{figure} [h]
    \centering
    \includegraphics[width=0.48\textwidth]{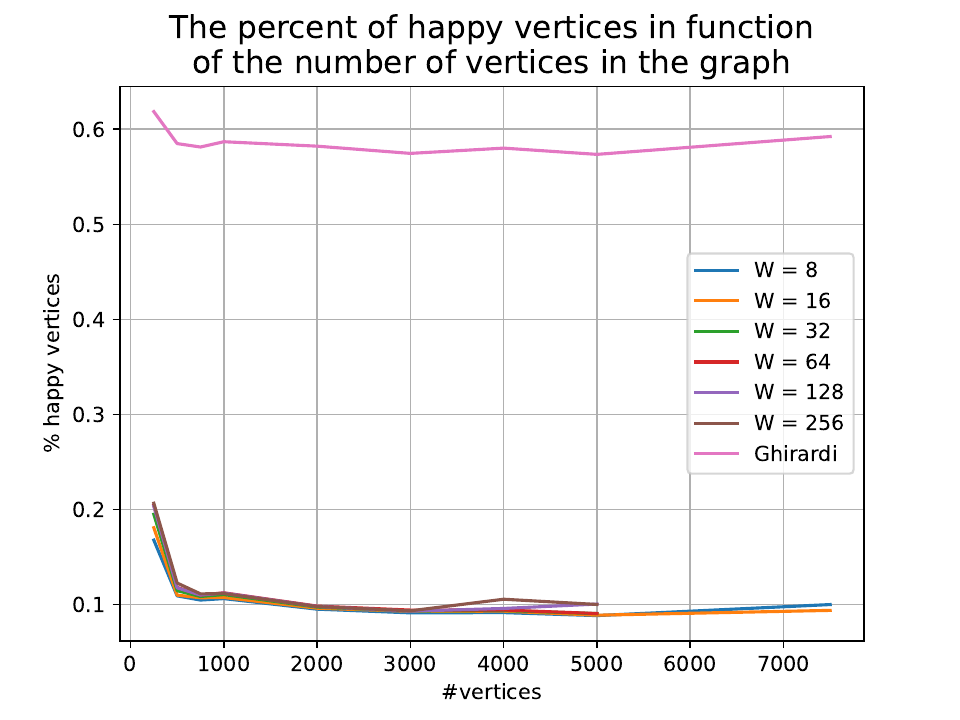}
    \hfil
    \includegraphics[width=0.48\textwidth]{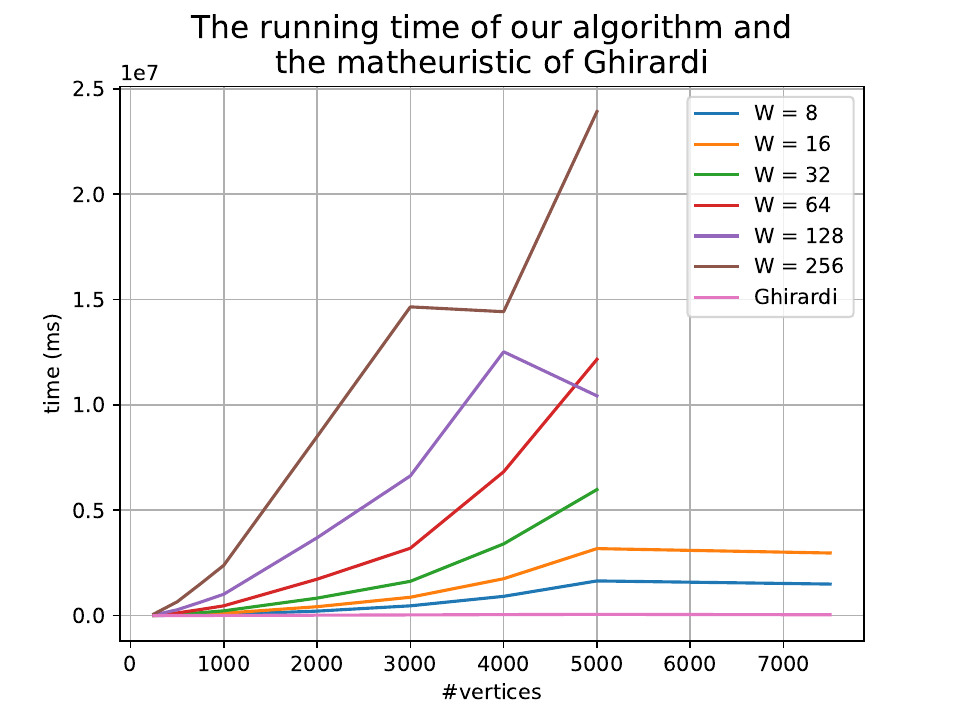}
    \caption{The quality and running time of our heuristic algorithm and the matheuristic of Ghirardi for large Erd{\H{o}}s-Rényi graphs.}
    \label{fig:er_large_quality_time}
\end{figure}

Fig.~\ref{fig:er_large_quality_time} shows the percentage of happy vertices and the runtime of our algorithm in function of the size of the graphs. These figures continue the trend from Figs.~\ref{fig:er_quality} and~\ref{fig:er_time}, indicating that our algorithm has trouble scaling to larger instances. The matheuristic of \cite{ghirardi2021simple}, however, quickly provides a high quality solution. The large running time can be explained through the size of the tree decomposition. First, the tree decomposition consists of many nodes. For each node, a complex operation is performed. If we limit the runtime to 8 hours, each node should be handled within 1 second if the tree decomposition consists of 28~000 nodes. The graphs with 1~000 vertices in this dataset have nice tree decompositions with approximately 28~000 nodes. The nice tree decompositions of larger graphs consist of even more nodes. Graphs with 5~000 vertices have tree decompositions with approximately 300~000 nodes, which means that each node should be handled within 0.093 milliseconds. Nevertheless, for some instances, the heuristic algorithm is able to meet this time constraint. 

The degradation in solution quality can be explained by combining the treewidth of the graphs with Equation~\ref{eq:exactness_guarantee}. In the current setup, at most 256 partial solutions were computed in each node. For graphs with 1~000 vertices (which are relatively small in the current setup), the treewidth averaged to 250. Given the width parameter of the algorithm $W=256$, treewidth $w=250$ and the number of colours $k=10$, and using  Equation~\ref{eq:exact_forget_node}, we find that $W/(2k)^{w+1} = 256/(2\cdot10)^{250+1} \approx 10^{-322}\%$ of the partial solutions were computed. Thus, only an extremely small fraction of the search space is considered. Because of this, it is difficult for the algorithm to find high quality solutions. 

\subsection{Performance on graph classes}

\begin{figure}
    \centering
    \includegraphics[width=0.48\textwidth]{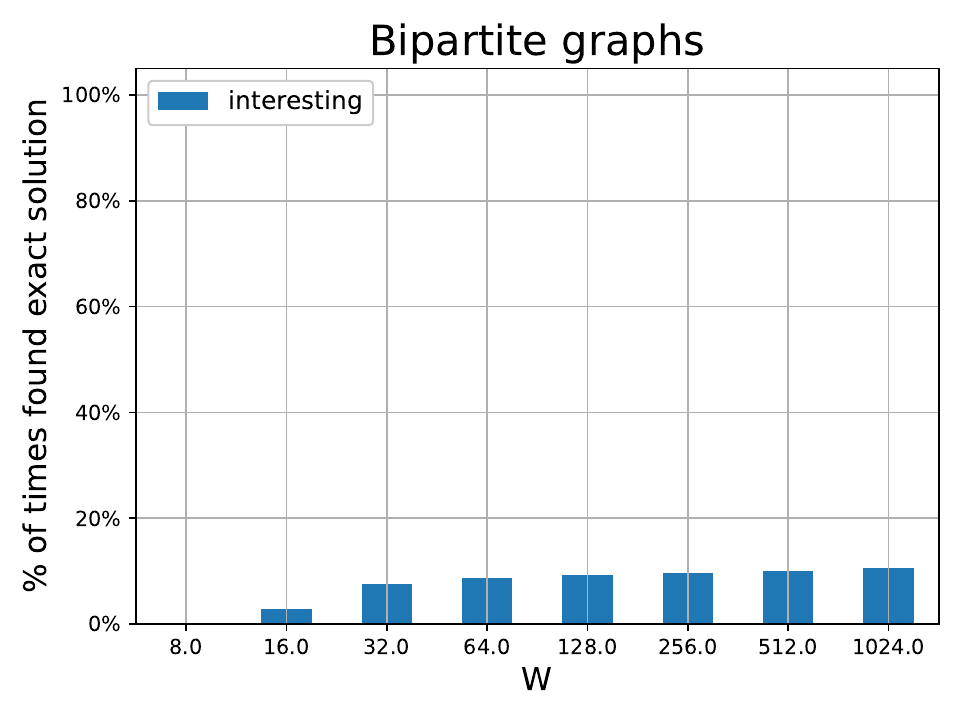}
    \hfil
    \includegraphics[width=0.48\textwidth]{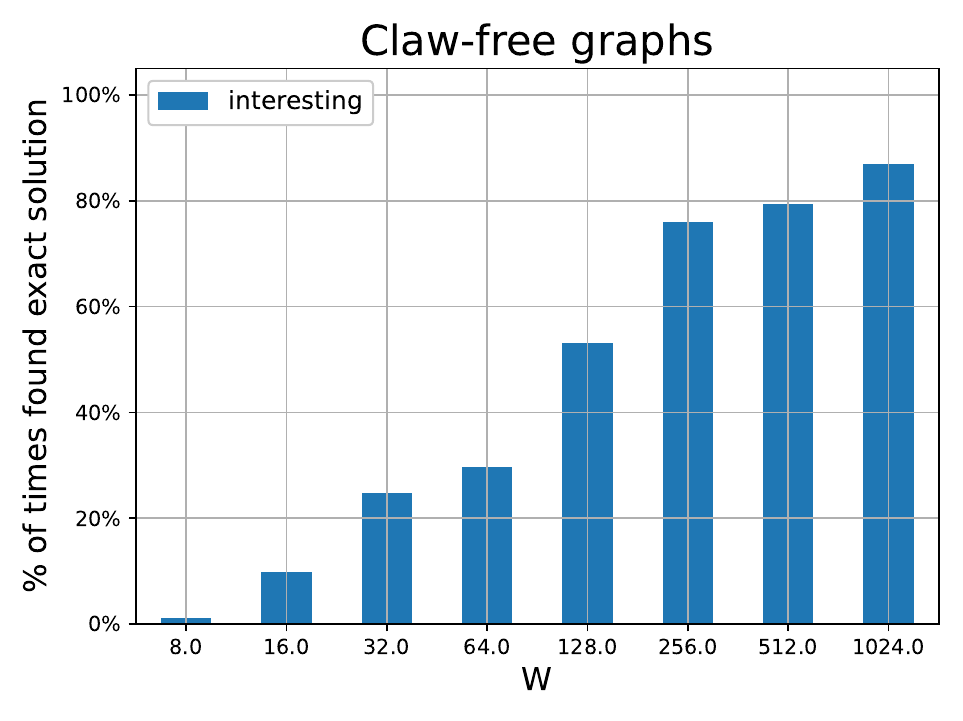}
    \hfil
    \includegraphics[width=0.48\textwidth]{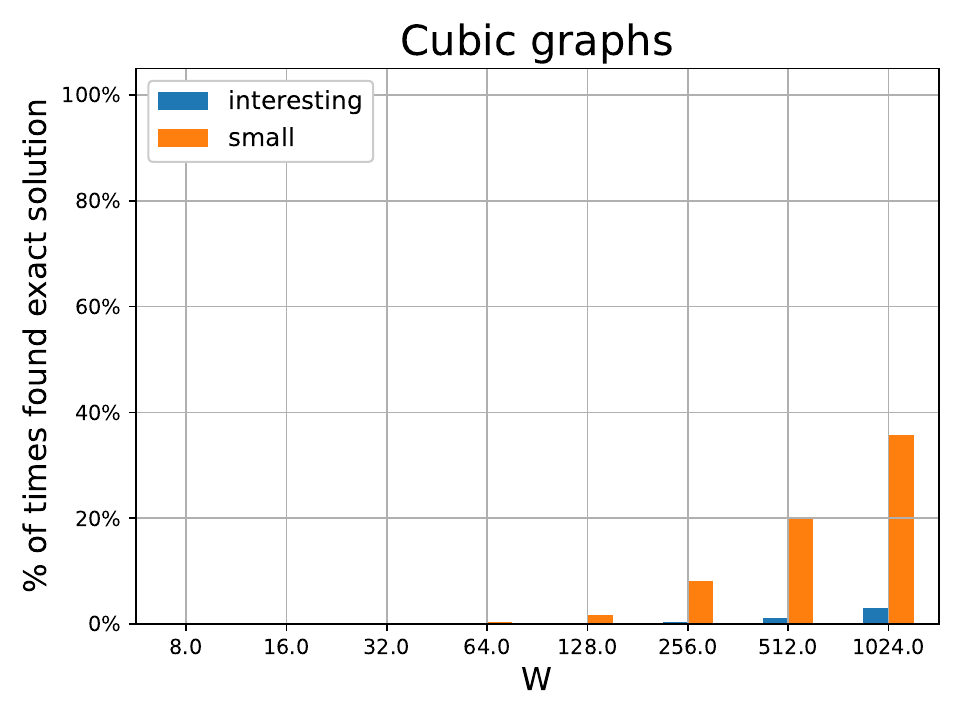}
    \hfil
    \includegraphics[width=0.48\textwidth]{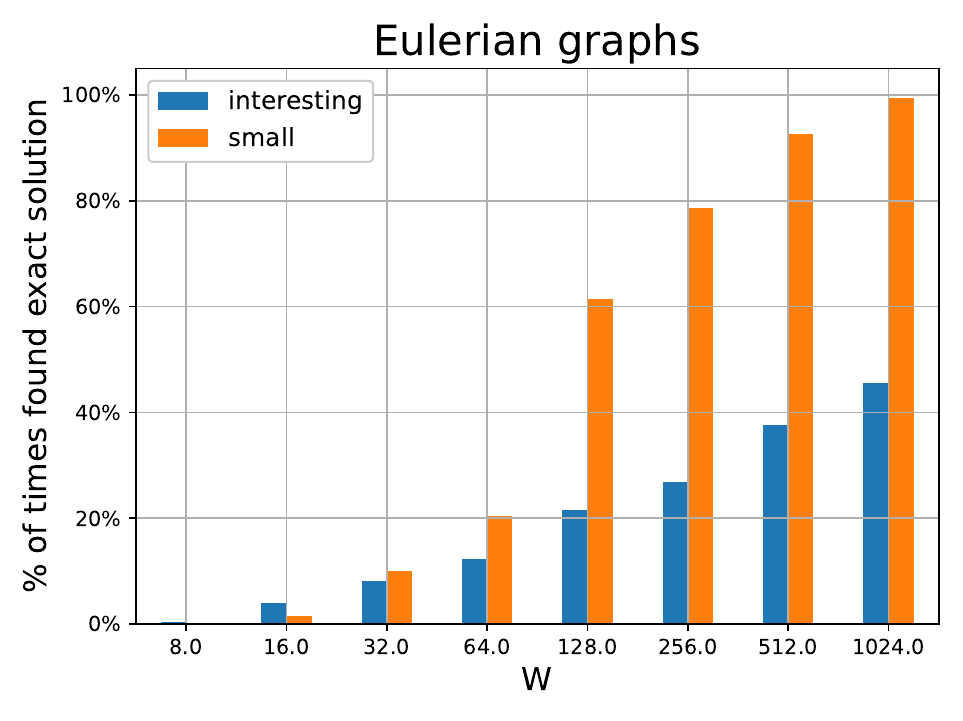}
    \hfil
    \includegraphics[width=0.48\textwidth]{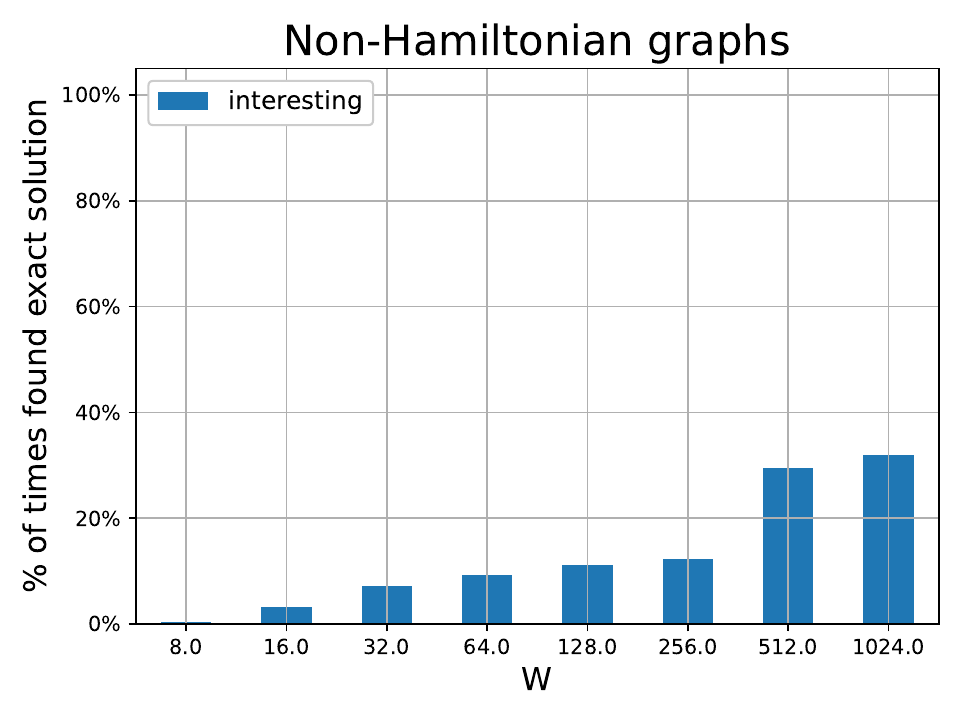}
    \hfil
    \includegraphics[width=0.48\textwidth]{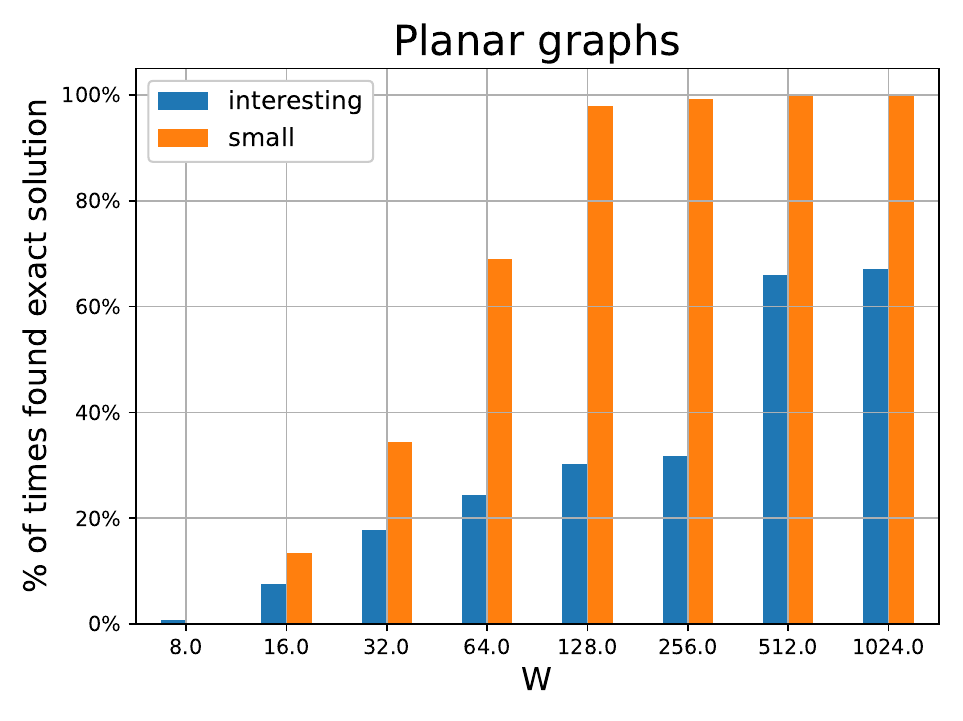}
    \caption{The probability that our heuristic algorithm proves optimality of a constructed solution for graphs of specific classes, in function of $W$.}
    \label{fig:classes_exactness}
\end{figure}

We also executed our algorithm on seven graph classes: bipartite graphs, claw-free graphs, cubic graphs, Eulerian graphs, non-Hamiltonian graphs, planar graphs and trees. Graphs in the first six classes can have arbitrarily large treewidth \citep{cygan2015parameterized}. We show the results of our experiments on these graph classes in Figs. \ref{fig:classes_data}, \ref{fig:classes_time}, \ref{fig:classes_quality} and \ref{fig:classes_exactness}. These show many similar patterns. 

\begin{itemize}
    \item Generally, the width of the tree decomposition increases as the number of vertices increases. The curves often show a linear trend, but with several outliers. The non-Hamiltonian and Planar graphs are an exception, for which the width is smaller for larger graphs. This is due to the use of the \textit{interesting} graphs from House of Graphs, for which no formal definition exists.
    
    \begin{figure}
        \centering
        \includegraphics[width=0.48\textwidth]{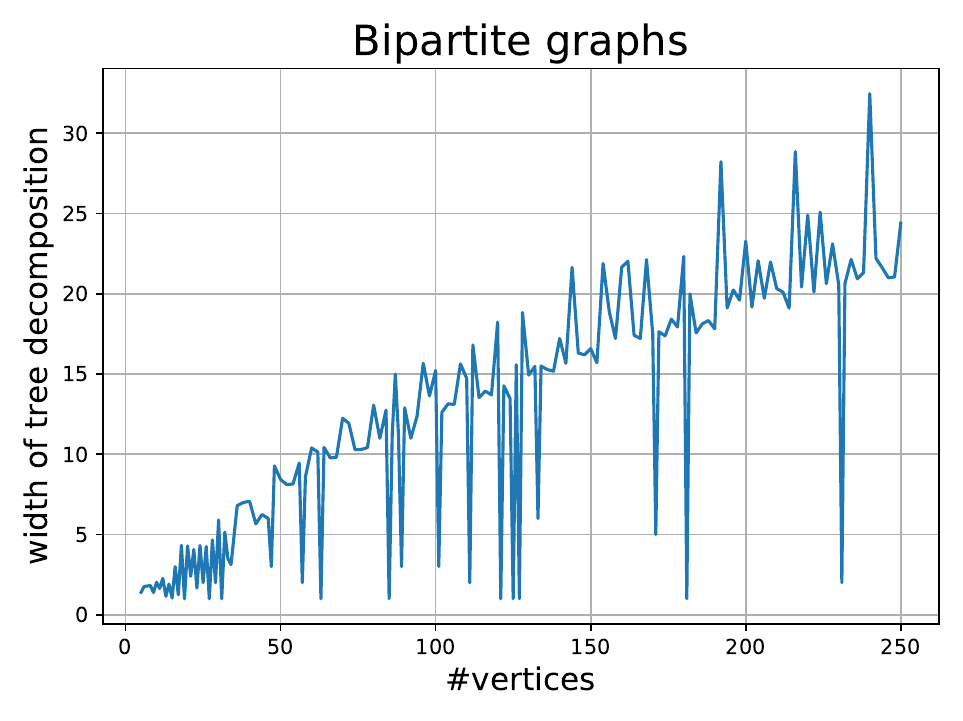}
        \hfil
        \includegraphics[width=0.48\textwidth]{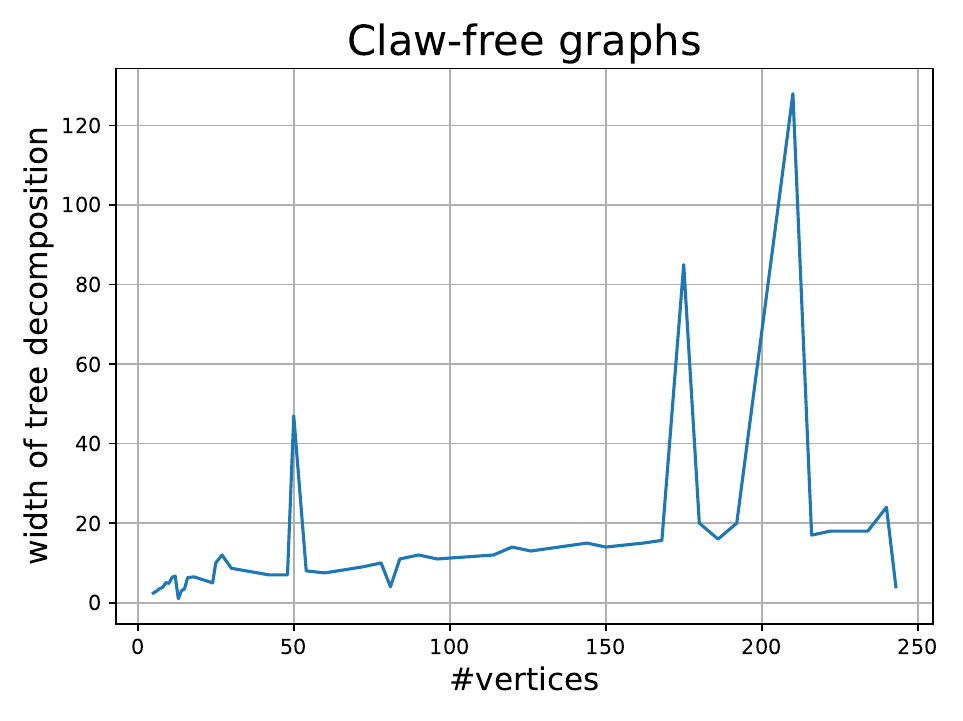}
        \hfil
        \includegraphics[width=0.48\textwidth]{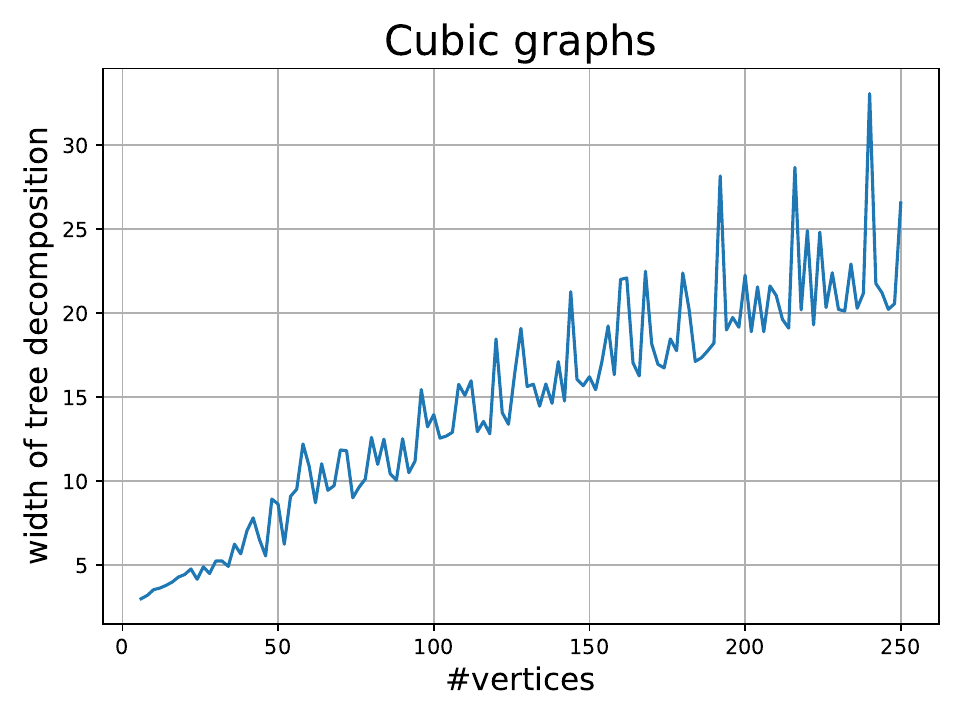}
        \hfil
        \includegraphics[width=0.48\textwidth]{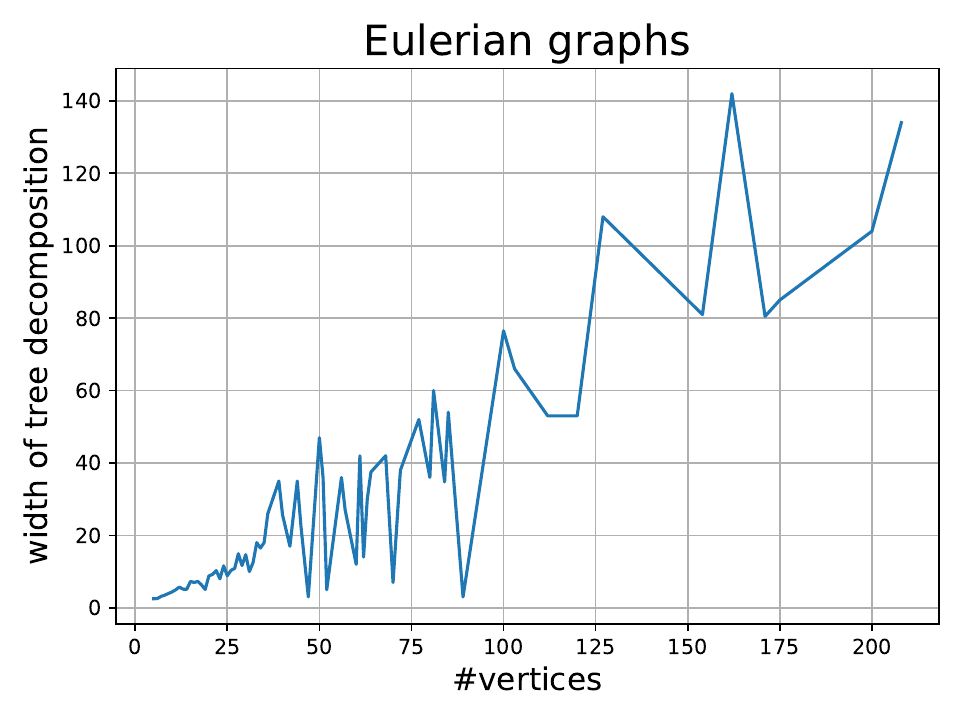}
        \hfil
        \includegraphics[width=0.48\textwidth]{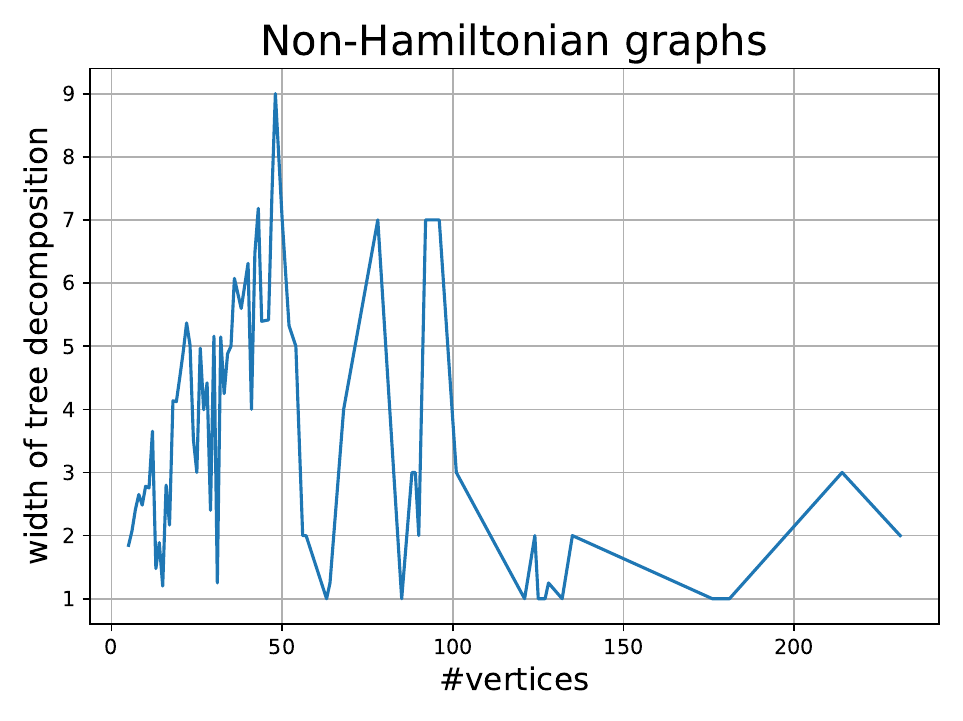}
        \hfil
        \includegraphics[width=0.48\textwidth]{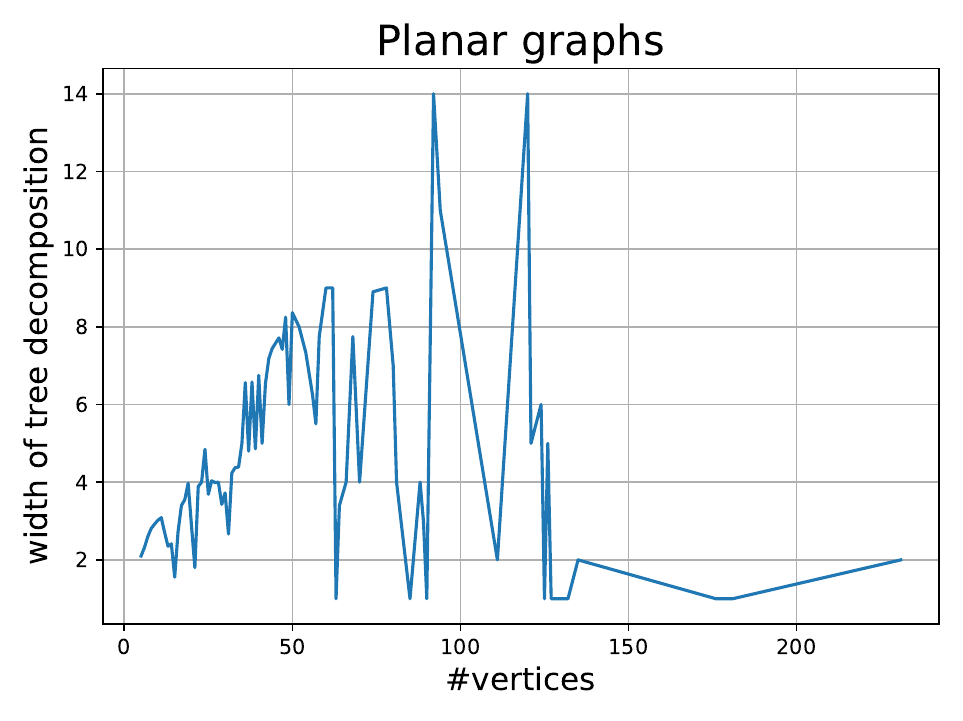}
        \caption{The size of tree decompositions for the graphs of specific classes.}
        \label{fig:classes_data}
    \end{figure}
    
    \item The required time to construct a colouring is strongly correlated with the width of the tree decomposition. This indicates that our algorithm is sensitive towards a varying treewidth. If the width increases, then the running time will increase as well. Conversely, our algorithm can more efficiently solve instances with a smaller treewidth. 
    
    \item Greedy-MHV and Growth-MHV generally perform better than our algorithm if the width of the tree decomposition is large. However, our algorithm results in a higher quality colouring for instances with smaller treewidth, for example the larger non-Hamiltonian and planar graphs. 
    
    \begin{figure}
        \centering
        \includegraphics[width=0.48\textwidth]{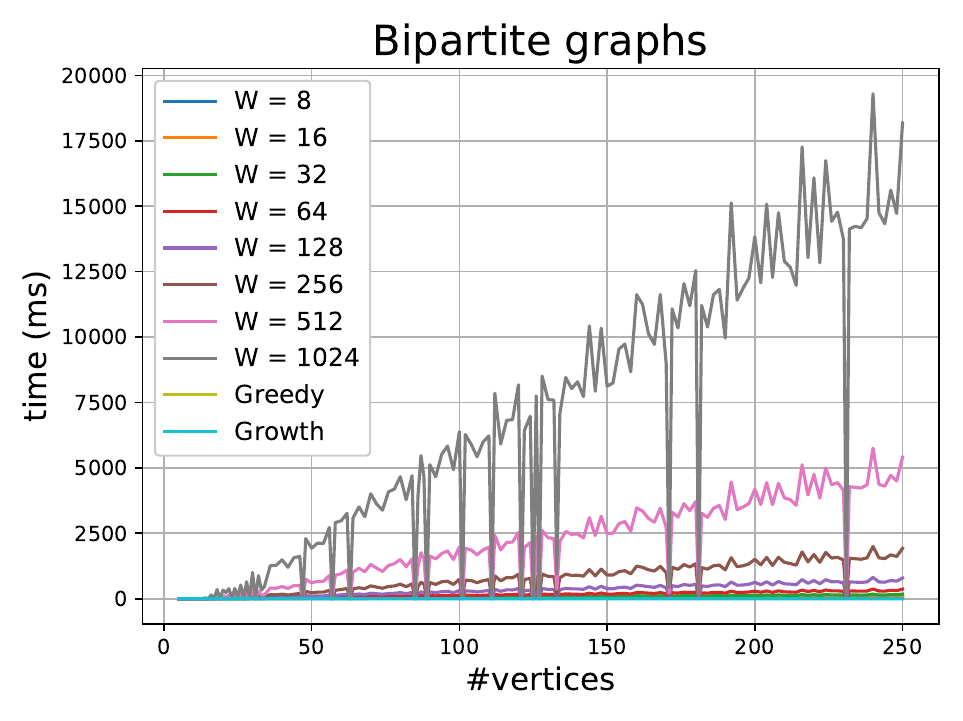}
        \hfil
        \includegraphics[width=0.48\textwidth]{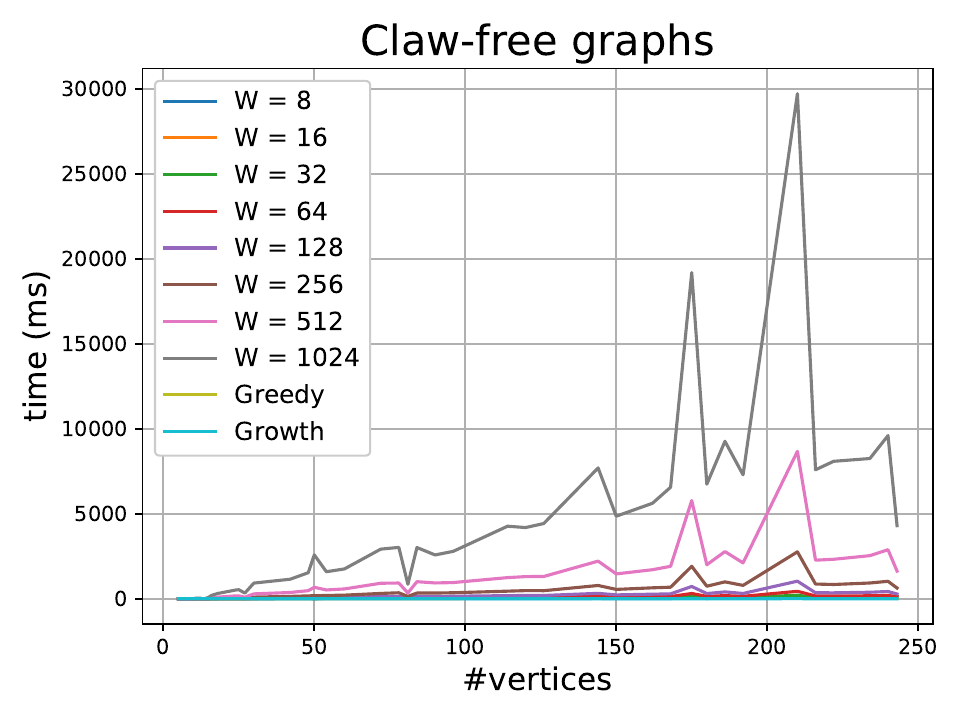}
        \hfil
        \includegraphics[width=0.48\textwidth]{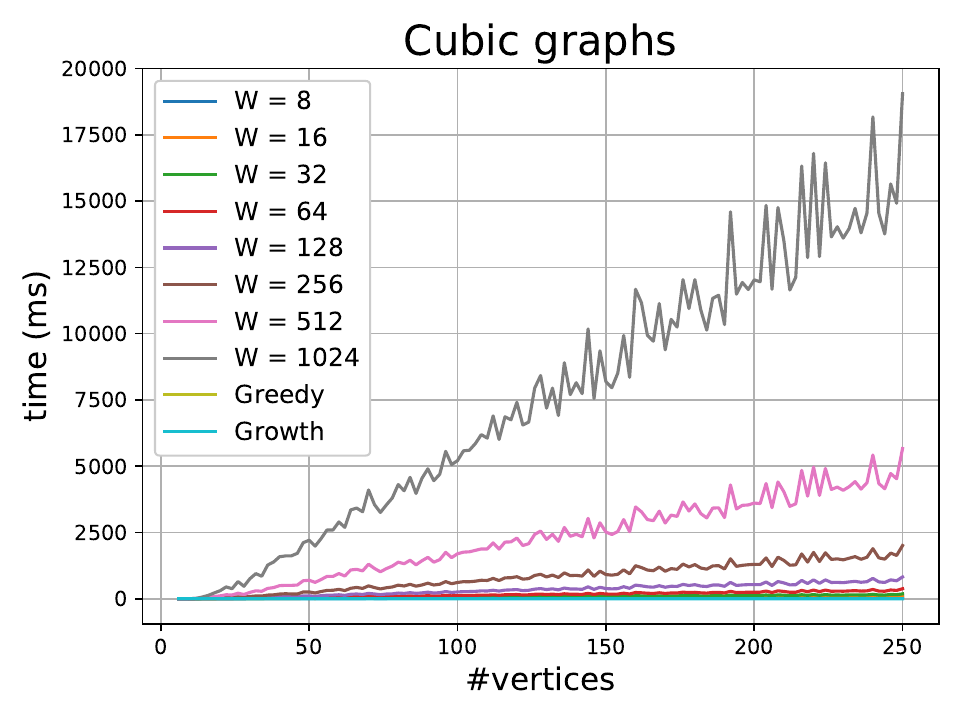}
        \hfil
        \includegraphics[width=0.48\textwidth]{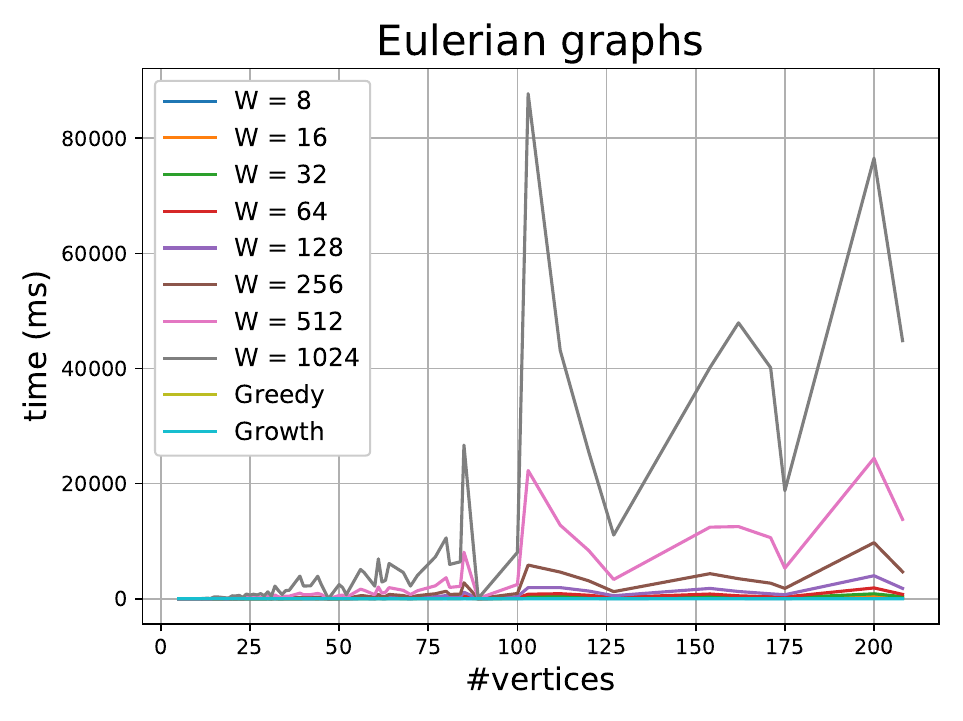}
        \hfil
        \includegraphics[width=0.48\textwidth]{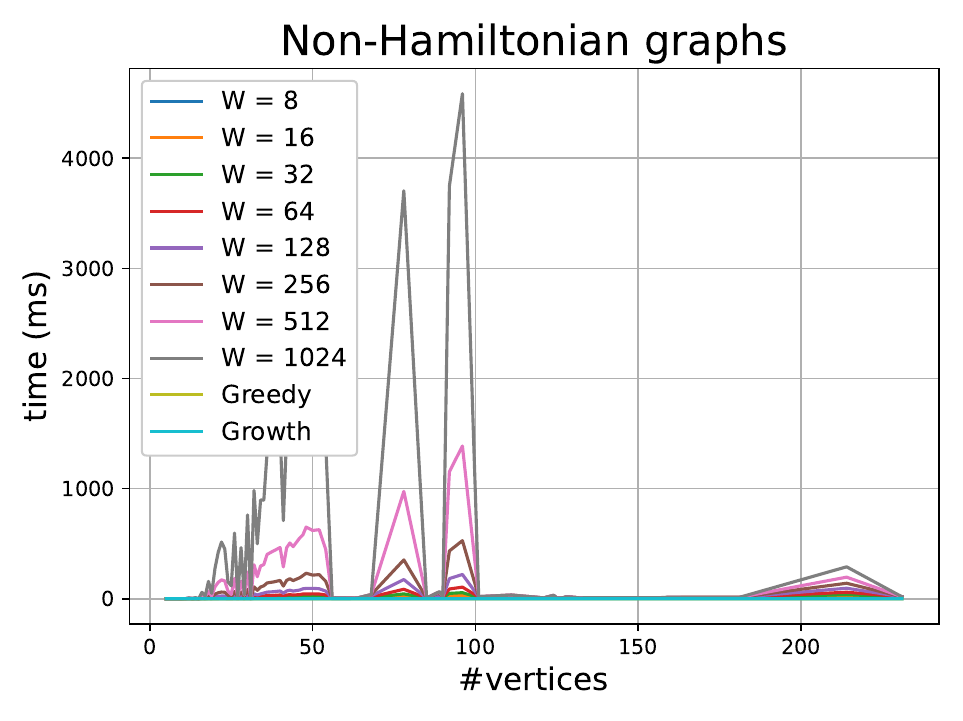}
        \hfil
        \includegraphics[width=0.48\textwidth]{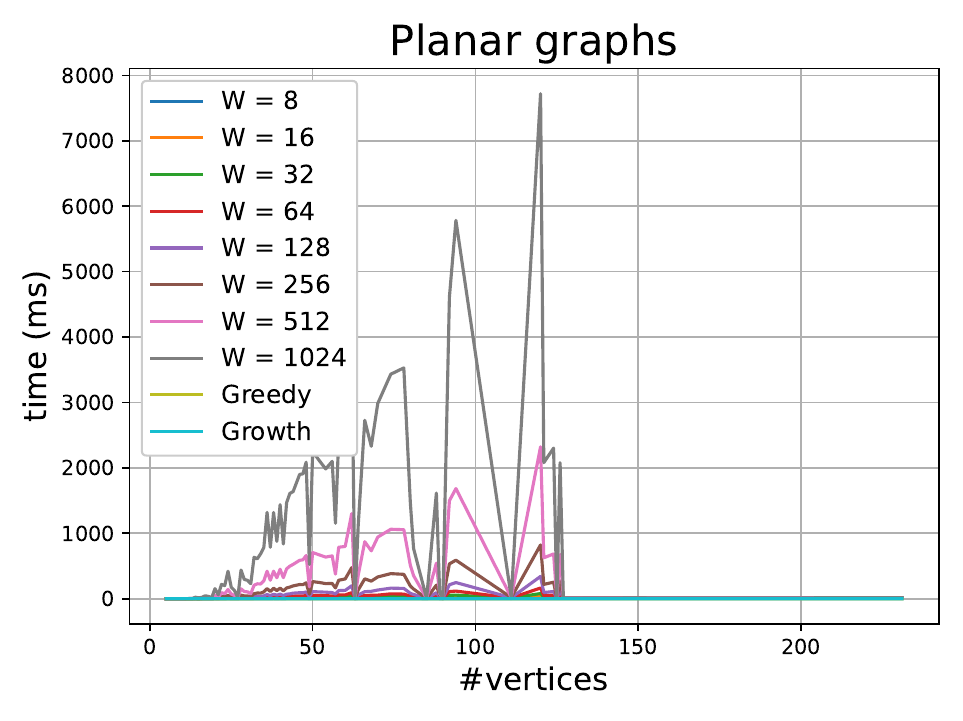}
        \caption{The mean running time of our heuristic algorithm, Greedy-MHV and Growth-MHV for graphs of specific classes, in function of the number of vertices in the graph. Only the running times of our algorithm with $W \leq 64$ are shown. Running times for larger $W$ behave similarly, but are larger and only make the figures unclear.}
        \label{fig:classes_time}
    \end{figure}

    \item For instances with a larger treewidth, our algorithm constructs colourings with fewer happy vertices, but so do Greedy-MHV and Growth-MHV. This could indicate that such instances have an inherent structure that results in fewer happy vertices. \cite{lewis2019finding} concluded that dense graphs tend to have fewer happy vertices, which corresponds to graphs with large treewidth. In this regard, we also refer to Section 7.7 in \cite{cygan2015parameterized}, where a win/win approach is sketched for algorithms using tree decompositions: either an instance has a small treewidth and can be solved efficiently, or the instance has a large treewidth which results in a complicated structure and lower quality solutions. 
    
    \begin{figure}
        \centering
        \includegraphics[width=0.48\textwidth]{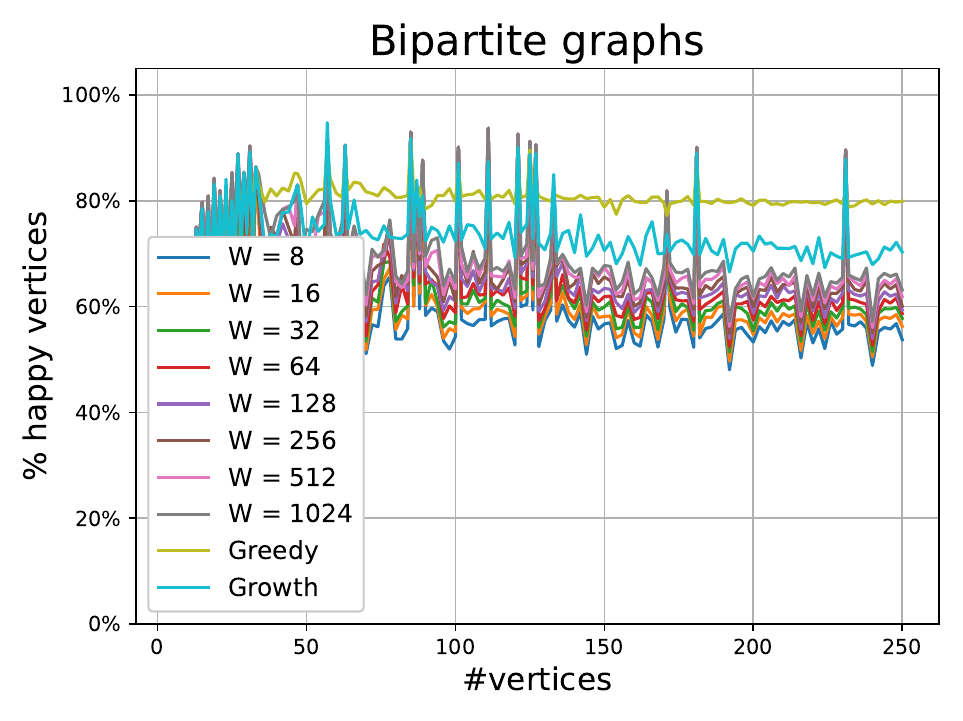}
        \hfil
        \includegraphics[width=0.48\textwidth]{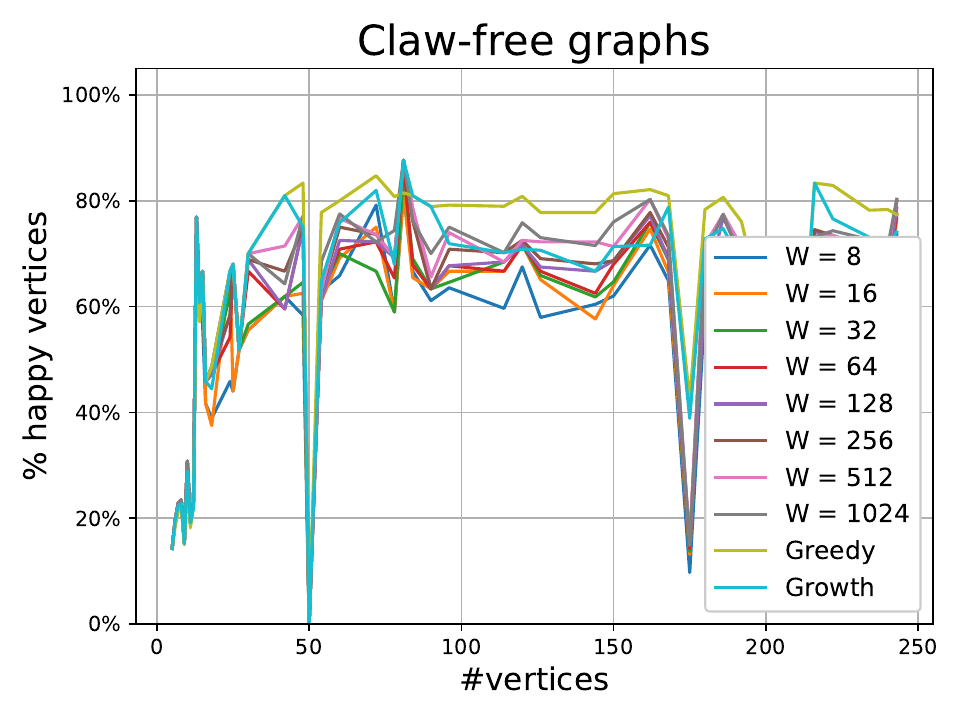}
        \hfil
        \includegraphics[width=0.48\textwidth]{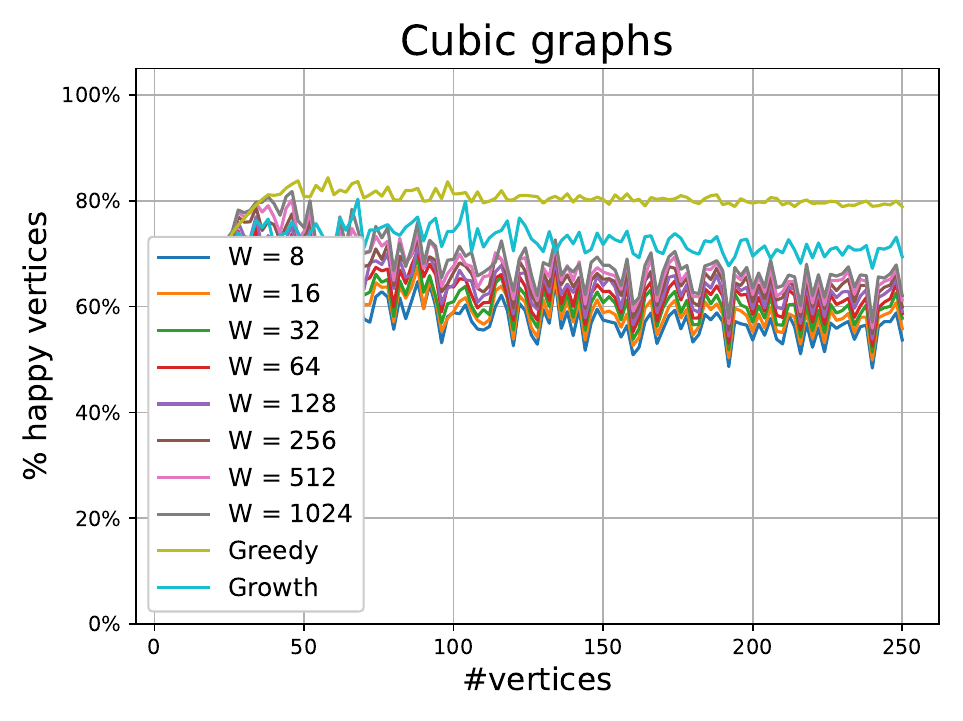}
        \hfil
        \includegraphics[width=0.48\textwidth]{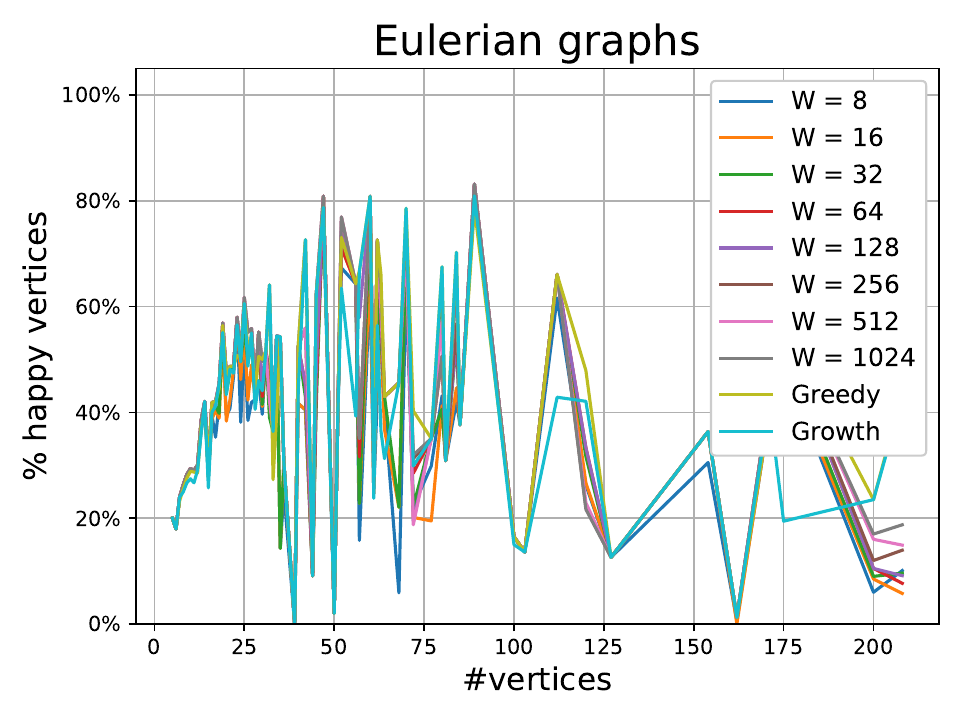}
        \hfil
        \includegraphics[width=0.48\textwidth]{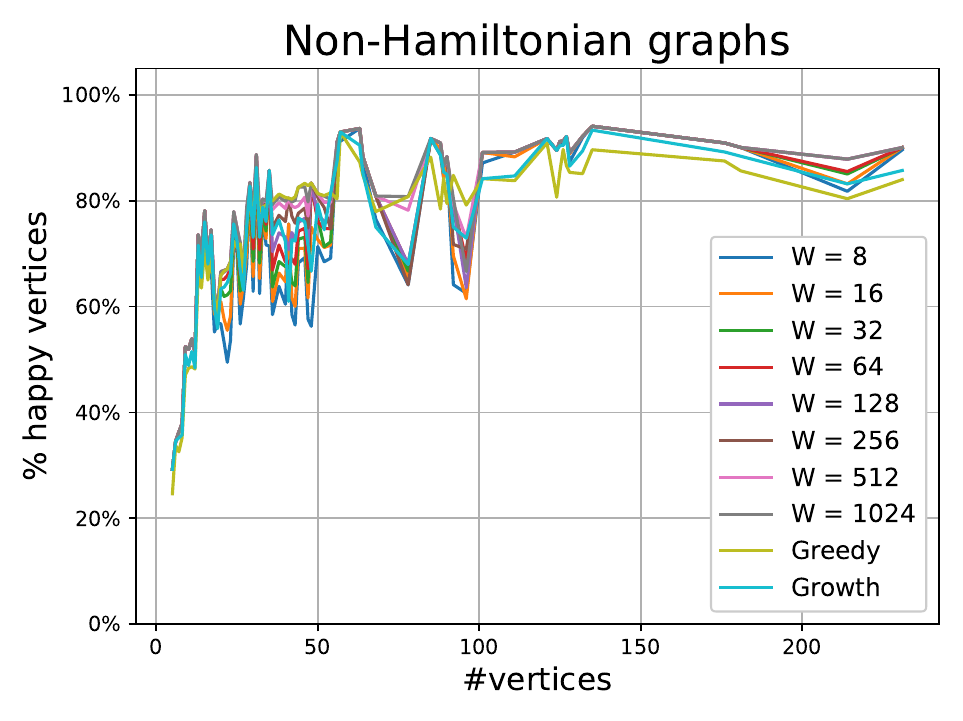}
        \hfil
        \includegraphics[width=0.48\textwidth]{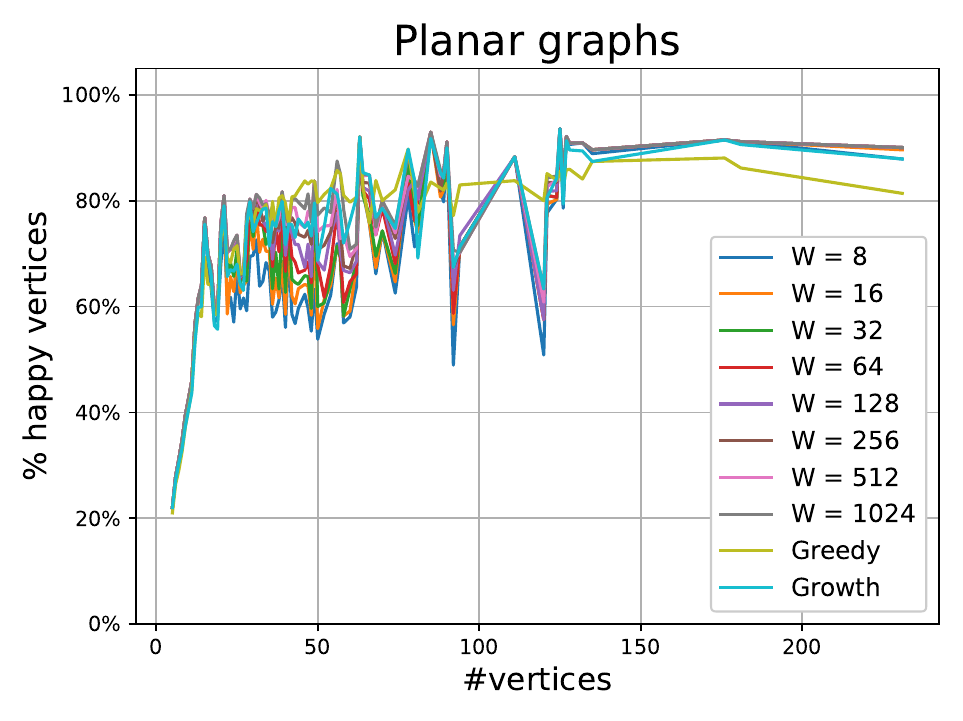}
        \caption{The quality of the final solution of our heuristic algorithm compared to Greedy-MHV and Growth-MHV for graphs of specific classes, in function of the number of vertices in the graph.}
        \label{fig:classes_quality}
    \end{figure}

    \item The algorithm proves optimality of the constructed colouring often for certain classes (e.g., claw-free graphs), but has trouble doing so for other classes (e.g., cubic graphs). This, again, originates from the width of the tree decomposition. If the width is smaller, then the bound in Equation \ref{eq:exactness_guarantee} is smaller, and our algorithm more likely proves optimality of the solution.  
\end{itemize}

\begin{figure}
    \centering
    \includegraphics[width=0.48\textwidth]{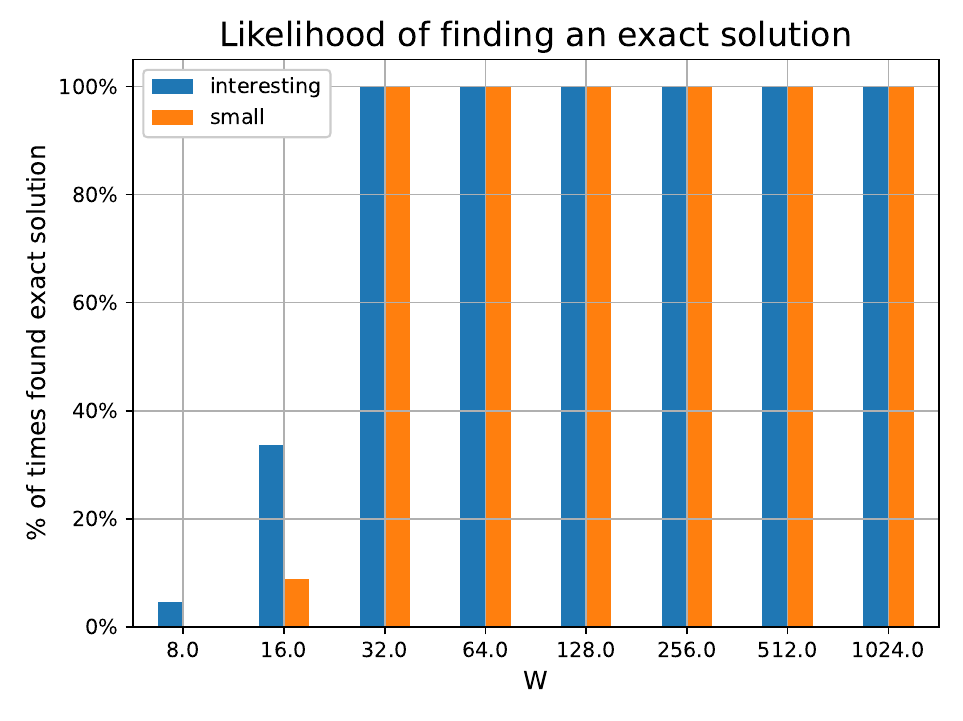}
    \hfil
    \includegraphics[width=0.48\textwidth]{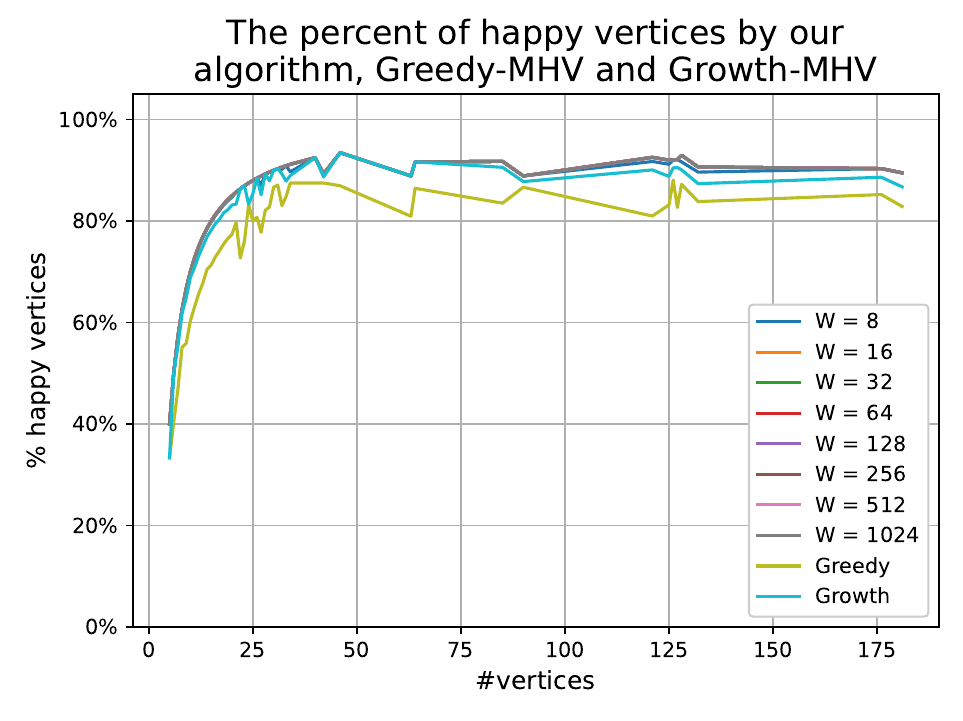}
    \hfil
    \includegraphics[width=0.48\textwidth]{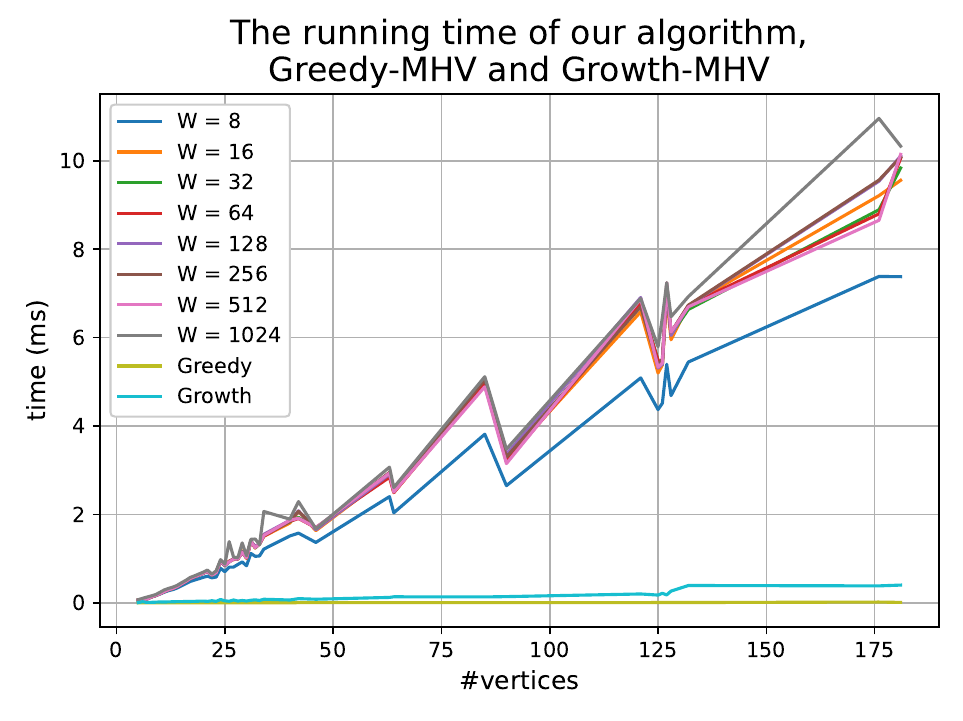}
    \caption{The performance of our algorithm when the graph in the MHV-instance is a tree.}
    \label{fig:trees}
\end{figure}

Trees on the other hand always have a treewidth equal to 1, independent of the number of vertices in the tree. For that reason we discuss the performance of our algorithm on trees separately. The results of our algorithm on trees are shown in Fig. \ref{fig:trees}. FlowCutter constructed an optimal tree decomposition for all instances, with treewidth 1. Applying Equation \ref{eq:exactness_guarantee}, if $(2k)^{w+1} = (2\cdot3)^{1 + 1} = 36 \leq W$, our algorithm computes an optimal solution for the given dataset (recall that we used 3 colours). Fig. \ref{fig:trees} shows that, indeed, this results in optimal solutions. Additionally all algorithm executions with $W = 32$ and some with $W \in \{8, 16\}$ resulted in an optimal colouring. Also note that more \texttt{interesting}-graphs are solved exactly for small $W$, but this is due to the randomness in the initial colouring since all instances have a treewidth of 1. 

The fact that many instances have been solved exactly results in more happy vertices. For $W \in \{8,16\}$, our algorithm often can not prove optimality of a solution, but these configurations still have more happy vertices than Greedy-MHV and Growth-MHV. Especially Greedy-MHV struggles to find an optimal number of happy vertices. Growth-MHV does perform better, but can not provide any exactness guarantee as our algorithm. Additionally, we see that our algorithm does not require much more computational resources, only several milliseconds. These observations make our algorithm most appropriate when the graph is known to be a tree. 

\section{Conclusions and future work} \label{sec:conclusions}

We proposed a new methodology to use tree decompositions in heuristic algorithms, which offers a dynamic trade-off between the exactness of the solutions and the runtime of the algorithm using only a single parameter. In particular, we implemented our methodology for the MHV problem and make the source code publicly available on \url{https://github.com/LouisCarpentier42/HeuristicAlgorithmsUsingTreeDecompositions}. Our approach more efficiently computes optimal solutions than the exact algorithm for graphs of bounded treewidth. The algorithm also constructs higher quality colourings than Greedy-MHV and Growth-MHV if at least 40\% of the vertices are initially coloured. However, our heuristic has problems scaling to larger instances with a larger treewidth, due to Equation~\ref{eq:exactness_guarantee}. Hence, the heuristic is most suitable for instances for which the treewidth is not too large. In fact, our algorithm can guarantee optimality of a constructed colouring, which is not possible with existing heuristic methods. This comes at a runtime cost, however.   

There are multiple possibilities to extend our heuristic algorithm. If only few tuples are constructed for each node in the tree decomposition, then these are potentially very similar. It would be interesting to validate a diversity promotion mechanism. As discussed earlier, each node in the tree decomposition may be composed of a huge search space. If the algorithm can explore more areas of this space, instead of focusing on only a single area, then more promising partial solutions may be computed, which will result in higher quality solutions. 

Several kernelisation techniques exist for the MHV problem \citep{PeetersFlorian2020OvhM,gao2018kernelization,agrawal2020parameterized}. A reduced graph can potentially lead to a completely different tree decomposition, which our algorithm could handle more easily. It would be interesting to see the effect of these techniques on the performance of our algorithm. 

We provide a new approach for using tree decompositions in algorithms. It would be interesting to use a similar approach for other graph optimisation problems. The existing literature of exact algorithms for graphs of bounded treewidth is extensive. \cite{cygan2015parameterized} cover numerous such algorithms for several optimisation problems. Each one of these can be modified by developing appropriate heuristic algorithms for handling the different types of nodes in a tree decomposition. Hence, such a methodology could result in a repertoire of new heuristic algorithms.

\textbf{Acknowledgments.} The research of Jan Goedgebeur was supported by Internal Funds of KU Leuven. Jorik Jooken is supported by a Postdocoral Fellowship of the Research Foundation Flanders (FWO) with contract number 1222524N. We gratefully acknowledge the support provided by the ORDinL project (FWO-SBO S007318N, Data Driven Logistics, 1/1/2018 - 31/12/2021). This research also received funding from the Flemish Government under the “Onderzoeksprogramma Artificiële Intelligentie (AI) Vlaanderen” programme. The computational resources and services used in this work were provided by the VSC (Flemish Supercomputer Center), funded by the Research Foundation - Flanders (FWO) and the Flemish Government - department EWI.

\bibliographystyle{unsrtnat}


\end{document}